\documentclass[onecolumn,secnumarabic,amssymb, nobibnotes, aps, prd]{revtex4-1}
\usepackage{epstopdf}
\usepackage{ulem}
\usepackage{amssymb}
\usepackage{natbib}
\usepackage{times}
\usepackage{graphicx}
\usepackage[usenames,dvipsnames]{pstricks}
 \usepackage{psfrag}
                                                                                        
\usepackage[colorlinks=true,linkcolor=blue,citecolor=blue]{hyperref}

\def\araa{ARA\&A}
\def\apj{ApJ}
\def\apjl{ApJ}
\def\apjs{ApJS}
\def\apss{Ap\&SS}
\def\aap{A\&A}
\def\jcap{J. Cosmology Astropart. Phys.}
\def\mnras{MNRAS}
\def\na{New A}
\def\prd{Phys.~Rev.~D}
\def\pasj{PASJ}
\def\nat{Nature}
\def\physrep{Phys.~Rep.}

\newcommand{\be}{\begin{equation}}
\newcommand{\ee}{\end{equation}}
\newcommand{\bary}{\begin{eqnarray}}
\newcommand{\eary}{\end{eqnarray}}
\newcommand{\en}{E_\nu}

\begin{document}
\title{Could a plasma in quasi-thermal equilibrium be associated to\\ the "orphan" TeV flares ?}
\author{N. Fraija}%
\email{Luc Binette-Fundaci\'on UNAM Fellow. nifraija@astro.unam.mx}
\affiliation{Instituto de Astronom\' ia, Universidad Nacional Aut\'onoma de M\'exico, Circuito Exterior, C.U., A. Postal 70-264, 04510 M\'exico D.F., M\'exico}
%
%
\begin{abstract}
TeV $\gamma$-ray detections in flaring states without  activity in X-rays from blazars have attracted much attention  due  to  the irregularity of these "orphan" flares.  Although the synchrotron self-Compton model has been very successful in explaining  the spectral energy distribution and  spectral variability of these sources, it has not been able to describe  these atypical flaring events. On the other hand,  an electron-positron pair plasma at the base of the AGN jet  was proposed as the mechanism of bulk acceleration of relativistic outflows.  This plasma in quasi-themal equilibrium called Wein fireball emits radiation at  MeV-peak energies serving as target of accelerated protons.  In this work we describe the "orphan" TeV flares presented in blazars 1ES 1959+650 and Mrk421 assuming geometrical considerations in the jet and  evoking the interactions of Fermi-accelerated protons and MeV-peak target photons coming from the Wein fireball.   After describing successfully these "orphan" TeV flares, we correlate the TeV $\gamma$-ray, neutrino and UHECR fluxes through p$\gamma$ interactions and calculate the number of  high-energy neutrinos and UHECRs expected  in IceCube/AMANDA and TA experiment, respectively. In addition, thermal MeV neutrinos produced mainly through electron-positron annihilation at the Wein fireball  will be able to propagate through it.  By considering two- (solar, atmospheric and accelerator parameters) and three-neutrino mixing, we study the resonant oscillations and estimate the neutrino flavor ratios as well as the number of thermal neutrinos expected on Earth.  
\end{abstract}

\maketitle



\section{Introduction}
\begin{figure}
\vspace{0.5cm}
{\centering
\resizebox*{1.2\textwidth}{0.38\textheight}
{\includegraphics{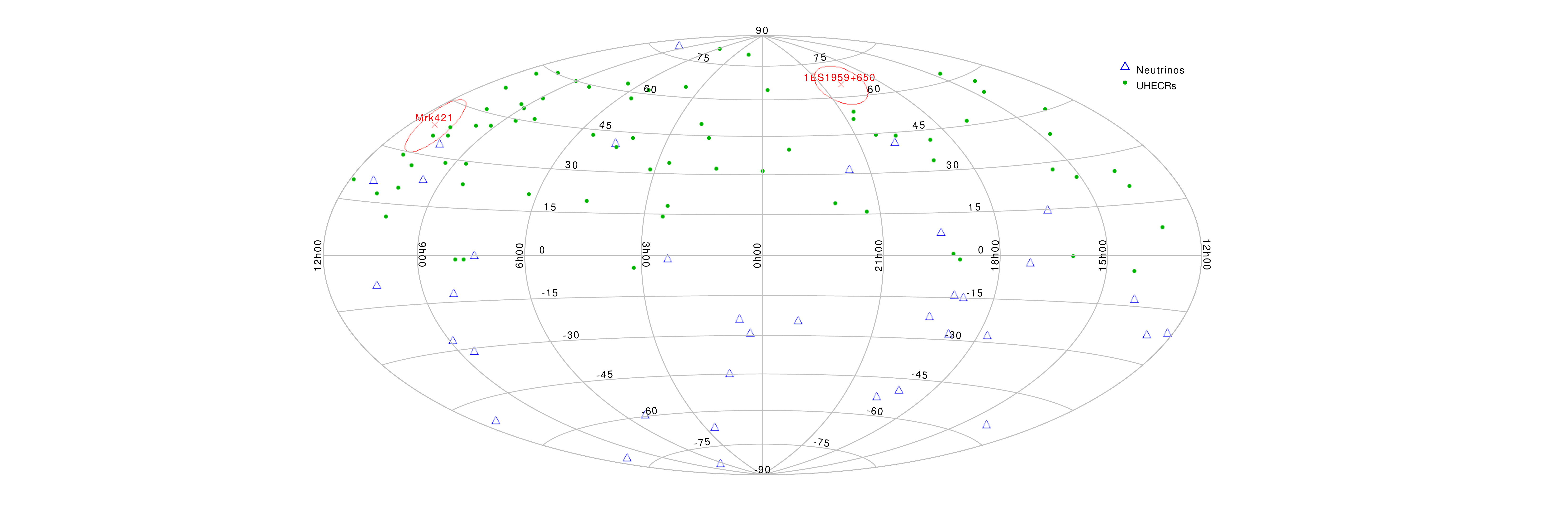}}
}
\caption{Sky-map with the location of the blazars  Mrk 421 and 1ES 1959+650 (marked with red color),  72 UHECRs collected by TA experiments (green  points) and  35 VHE neutrinos detected by IceCube (blue triangles). The red  circles  around the  blazars  have a size of  7$^\circ$.}
\label{skymap}
\end{figure}

%
Flares observed in very-high-energy (VHE) $\gamma$-rays with absence of high activity in X-rays, are very difficult to reconcile with the standard synchrotron self-Compton (SSC) although it has been very successful in explaining the spectral energy distribution (SED)  of blazars \citep{1996ApJ...461..657B, 1998ApJ...509..608T, 1998MNRAS.301..451G}.   Although most of the flaring activities occur almost simultaneously with TeV $\gamma$-ray and X-ray fluxes, observations of  1ES 1959+650  \citep{2003ApJ...583L...9H,2004ApJ...601..151K,2005ApJ...621..181D}  and Mrk421 \citep{2005ApJ...630..130B,2011ApJ...738...25A} have exhibited VHE $\gamma$-ray flares without their counterparts in X-rays, called "orphan" flares.\\
Leptonic and hadronic models have been developed to explain orphan flares.  A leptonic model based on geometrical considerations about the jet has been explored to reconcile the SSC model \citep{2006ApJ...651..113K} whereas  hadronic models where accelerated protons interact with both external photons  generated by electron synchrotron radiation \citep{2005ApJ...621..176B} and SSC photons at the low-energy tail \citep{2013PhRvD..87j3015S} have been performed to explain this anomalous behavior in these blazars. Based on these models HE neutrino emission has been studied by \citet{2005ApJ...630..186R} and \citet{2005APh....23..537H}.  In particular for the blazar 1ES 1959+650,  Halzen  and Hooper (2005) based on proton-proton (pp) and proton-photon (p$\gamma$) interactions estimated the number of events expected in Antarctic Muon And Neutrino Detector Array (AMANDA). They found that the neutrino rates were 1.8 ($10^{-3}$) events for pp (p$\gamma$) interactions. \\
It has been  widely  suggested that relativistic jets of active galactic nuclei (AGN) contain electron-positron pairs produced from accretion disks \citep{1998Natur.395..457W,1996ApJ...463..305K,2006astro.ph..3772K}.   Also electron-positron pair plasma has been proposed as a mechanism of bulk acceleration of relativistic outflows. In  gamma ray burst (GRB) jets, this plasma "fireball" formed inside the initial scale $\sim$10$^7$ cm is made of photons, a small amount of baryons  and $e^{\pm}$ pairs  in thermal equilibrium at some MeV \citep{2004IJMPA..19.2385Z}.   However, in AGN jets the fireball cannot be formed because the characteristic size is too large ($3r_g\sim 10^{14}$ cm) in comparison with GRB jets.  Some authors found that if the pair plasma is expected to be optically thin to absorption but thick to scattering, the "Wein fireball" could exist, even though for the size and luminosity of AGNs \citep{1999PhR...314..575P, 2004ApJ...601...78I,2002ApJ...565..163I}.  Afterward,  simulations with protons inside this plasma were performed by \citet{2007ApJ...655..762A,2009ApJ...690L..81A}.\\
At the initial stage of the Wein fireball, thermal neutrinos will be mainly created by electron-positron annihilation ($e^++e^-\to Z\to \nu_j+\bar{\nu}_j$). By considering  a small amount of baryons, neutrinos could also be generated by processes of positron capture on neutrons ($e^++n  \to p+  \bar{\nu_e}$), electron capture on protons   ($e^-+p  \to n+  \nu_e$) and  nucleon-nucleon bremsstrahlung ($NN\to NN+\nu_j+\bar{\nu}_j$) for $j=e,\nu,\tau$.  Taking into account that the temperature of Wein fireball is relativistic \citep{2004ApJ...601...78I,2002ApJ...565..163I}, then neutrinos of 1 - 5 MeV can be produced and fractions of them will be able to go through this plasma.   As known, the neutrino properties are modified when they propagate through a thermal medium, and although neutrino cannot couple directly to the magnetic field, its effect can be experimented through coupling to charged particles in the medium \citep{1951PhRv...82..664S}.  The resonance conversion of neutrino from one flavor to another due to the medium effect, known as Mikheyev-Smirnov-Wolfenstein effect \citep{1978PhRvD..17.2369W}, has been widely studied in the GRB fireball \citep{2009PhRvD..80c3009S, 2009JCAP...11..024S,2014ApJ...787..140F}.\\
Telescope Array (TA) experiment reported the arrival of 72 ultra-high-energy cosmic rays (UHECRs) above 57 EeV with a statistical significance of 5.1$\sigma$.  These events correspond to the period from 2008 May 11  to 2013 May 4.  Assuming  the error reported by TA experiment  in  the reconstructed directions, some UHECRs might be associated to the position of Mrk421 \citep{2014arXiv1411.7354F}. In addition, IceCube collaboration reported the detection of 37 extraterrestrial neutrinos at 4$\sigma$ level above 30 TeV  \citep{2013arXiv1311.5238I, 2013arXiv1304.5356I}, although none of them located in the direction of neither 1ES 1959+650 nor Mrk421, as shown in fig. \ref{skymap}.\\
Because TeV $\gamma$-ray "orphan" flares are very difficult to reconcile with SSC model, in this work we introduce a hadronic model by means of p$\gamma$ interactions to explain these atypical TeV flares registered in blazars 1ES 1959+650 and  Mrk421. In this model,  we consider some geometrical assumptions of the jet  and  the interactions between the MeV-peak photons coming from the Wein fireball  and relativistic  protons accelerated at the emitting region.   Then,  we correlate the TeV $\gamma$-ray, neutrino and UHECR fluxes to calculate the number of HE neutrino and UHECR events. In addition, we study the resonance oscillations of thermal MeV neutrinos.   The paper is arranged as follows. In Section 2 we show the dynamic model of the radiation coming from Wein fireball and its interactions with the protons accelerated at the emitting region. In section 3 we study the emission, production and oscillation of neutrinos.  In Section 4 we discuss the mechanisms for accelerating UHECRs and also estimate the number of these events expected  in the TA experiment, supposing that the proton spectrum is extended up to energies greater than $57$ EeV.  In Section 5 we describe the TeV orphan flares of the blazars 1ES 1959+650 and Mrk 421 and give a discussion on our results; a brief summary is given in section 6. We hereafter use primes (unprimes) to define the quantities in a comoving (observer) frame, natural units  (c=$\hbar$=k=1) and redshifts z$\simeq$ 0.
\section{Orphan TeV $\gamma$-ray emission}
Different hadronic models have been considered to explain TeV $\gamma$-ray observations presented in blazars \citep{1993A&A...269...67M, 2000ApJ...534..109S, 2002MNRAS.332..215A, 2000NewA....5..377A}.  In those models,  SEDs are described in terms of co-accelerated electrons and protons at the emitting region.  In this hadronic model, we describe the TeV $\gamma$-ray emission through $\pi^0$ decay products generated in the interactions of accelerated protons and seed photons coming from the Wein fireball, as shown in fig. \ref{picture}.  
\subsection{MeV radiation from the Wein fireball}
The Wein fireball connects the base of the jet with the black hole (BH). We assume that at the initial state, it is formed by e$^\pm$ pairs with photons inside the initial scale $r_o=2r_g=4{\rm GM}$,  being ${\rm G}$ the gravitational constant and ${\rm M}$ the BH mass. The initial temperature can be defined through microscopic processes at the base of the jet (Compton scattering, $\gamma\gamma$ pair production, etc).    At the first state, photons inside the Wein fireball are at relativistic temperature. The internal energy starts to be converted into kinetic energy  and the Wein fireball begins to  expand. As a result of this expansion, temperature decreases and bulk Lorentz factor increases at the first state.  The initial optical depth is \citep{2002ApJ...565..163I,2004ApJ...601...78I}
\be\label{opt}
\tau_o \simeq \frac{n_{e,o}\,\sigma_T\,r_{o}}{\Gamma_{W,o}}\,,
\ee
where $\sigma_T=6.65\times 10^{-25}\,{\rm cm}^2$ is the Thompson cross section, $\Gamma_{W,o}=1/\sqrt{1-\beta^2_{W,o}}$ is the initial Lorentz factor of the plasma and $n_{e,o}$ is the initial electron density which is given by
\be\label{elec_den}
n_{e,o}=\frac{1}{4\sigma_TG_NM} \frac{1}{\Gamma^2_{W,o}\,\beta_{W,o} \langle \gamma_{e,o}\rangle} \left(\frac{m_p}{m_e}\right) \left(\frac{r_g}{r_o}\right)^2\left(\frac{L_j}{L_{Edd}}\right)\,, 
\ee
where $L_j$ is the total luminosity of the jet, $L_{Edd}=2\pi m_pr_g/\sigma_T$ is the Eddington luminosity, $\langle\gamma_{e,o}\rangle=K_3(1/\theta_o)/K_2(1/\theta_o)-\theta_o$ is the average Lorentz factor of electron thermal velocity, $\theta_o=T_o/m_e$ is the initial temperature normalized to electron mass (m$_e$), $m_p$ is the proton mass and $K_i$ is the modified Bessel function of integral order.  For a steady and spherical flow, the conservation equations of energy  and momentum can be written  as
\bary
\frac{1}{r^2}\frac{d}{dr}[r^2(\rho_T+P_T)\Gamma^2_W\beta_W]=0\label{3}\,,\\
\frac{1}{r^2}\frac{d}{dr}[r^2(\rho_T+P_T)\Gamma^2_W\beta^2_W]+\frac{dP_T}{dr} =0\,.
\eary
Here $\rho_T$ is the total energy density of pairs ($\rho_e=2m_en_e\langle\gamma_e\rangle$) and photons ($\rho_\gamma=3m_en_\gamma \theta$) and $P_T$ is the total pressure of pairs ($P_e=2m_en_e\theta$) and photons ($P_\gamma=m_en_\gamma\theta$).   The number density of electrons and photons in the Wein equilibrium, and the number conservation equations are given by \citep{1984MNRAS.209..175S}
\be\label{ratio_den}
\frac{n_e}{n_\gamma}=\frac{K_2(1/\theta)}{2\theta^2}\equiv f(\theta)\,,
\ee
and
\bary
\frac{1}{r^2}\frac{d}{dr}(r^2n_e\,\Gamma_W\beta_W)=\dot{n}_e\,,\\
\frac{1}{r^2}\frac{d}{dr}(r^2n_\gamma\,\Gamma_W\beta_W)=\dot{n}_\gamma\,,\label{7}
\eary
respectively.  Taking into account the momentum conservation law {\small $\frac{1}{\Gamma} \frac{d\Gamma}{dr}+\frac{1}{(\rho_T+P_T)}\frac{dP_T}{dr}=0$},  eqs. from (\ref{3}) to (\ref{7}) and  following  to \citet{2004ApJ...601...78I,2002ApJ...565..163I},  it is possible to write the evolution of the Lorentz factor, the temperature and the total number density of photons and pairs as
\be\label{Gamma_W}
\Gamma_W=\Gamma_{0,W}\frac{r}{r_0},
\ee
\be\label{theta}
\theta=\theta_0\frac{r_0}{r},
\ee
and
\be\label{number_total}
n_\gamma+2n_e=3n_{e,0}\left(\frac{r_0}{r}\right)^3\,,
\ee
respectively.  Eqs. ($\ref{Gamma_W}$), (\ref{theta}) and (\ref{number_total}) represent the evolution of the Wein fireball in the optical thick regime.  As the Wein fireball expands, the photon density, optical thickness and temperature decrease.  At a certain radius,  flow becomes optically thin  ($\tau=1$), then photon emission will be radiated away.  Defining this radius as the photosphere radius  (r=r$_{ph}$), then from eqs. (\ref{opt}), (\ref{elec_den}), (\ref{ratio_den}) and (\ref{number_total}), the photon density,  radius, temperature and Lorentz factor at the photosphere can be written as
\be\label{den_ph}
n_\gamma=\frac{\theta_0}{\theta_{ph}(1+2f(\theta_{ph}))}\frac{3 L_\gamma}{4\pi r^2_{ph}\,\epsilon\,\Gamma^2_{W,ph}\,\beta_{W,ph}\,m_e\,\langle \gamma_{e,o}\rangle}\,,
\ee
\be\label{r_ph}
r_{ph}\simeq \tau^{1/3}_o\,r_o\,,
\ee
\be\label{theta_ph}
\theta_{ph}=\tau_o^{-1/3}\,\theta_o\,,
\ee
and
\be\label{Gamma_ph}
\Gamma_{W,ph}=\tau^{1/3}_o\,\Gamma_{W,o}\,,
\ee
respectively.  Here $L_\gamma=\epsilon L_j$ is the luminosity radiated at the photosphere and $\epsilon$ is a parameter $0 \leq \epsilon\leq 1$.  One can see that  $n_\gamma$,  r$_{ph}$, $\theta_{ph}$ and $\Gamma_{ph}$ only depend on the initial conditions ($\theta_o$, $r_o$, $\Gamma_{W,o}$ and $L_j/L_{Edd}$).  The numerical results exhibit a radiation centered around  5 MeV   \citep{2004ApJ...601...78I}.  Hence, the output spectrum of simulated photons could be described by
\be\label{esp_phot}
\frac{d n_\gamma (\epsilon_\gamma)}{d\epsilon_\gamma} \propto
\cases{
( \epsilon_\gamma)^{-\beta_l},                                                                   & ${\rm if}\,\,\, \epsilon_\gamma < \epsilon_{\gamma b}$,\cr
(\epsilon_{\gamma b})^{-\beta_l+\beta_h} (\epsilon_\gamma)^{-\beta_h}, & ${\rm if}\,\,\, \epsilon_\gamma \ge \epsilon_{\gamma b}$,\cr
}
\ee
where $\beta_h\simeq 2$, $\beta_l\simeq 1$  \citep{2004APh....20..591F} and the peak energy around $\epsilon_{\gamma b} \simeq$ 5 MeV.   Also we can define the optical thickness to pair creation around the peak as  \citep{1997iagn.book.....P}
\be\label{opt_gg}
\tau_{\gamma\gamma}\simeq  \frac{\sigma_T\,\epsilon_{\gamma b}\,r_{ph}}{4\,m_e\,\Gamma_{W,ph}}\,n_\gamma\,,
\ee
where we have taken into account that the cross section of pair production reaches a maximum value close to the Thomson cross section. Additionally, it is very important to say that in the optically thick regimen, the angular distribution of MeV photons is almost isotropic, whereas in the optically thin regime it has a skew distribution. Specifically at a distance $r \gg r_g$, the outgoing photons are distributed in the range $0^\circ \leq\phi_{ph}\leq 60^\circ$ \citep{2004ApJ...601...78I}.  
\subsection{Geometrical considerations and assumptions}
\begin{figure}
\vspace{0.5cm}
{\centering
\resizebox*{0.5\textwidth}{0.4\textheight}
{\includegraphics{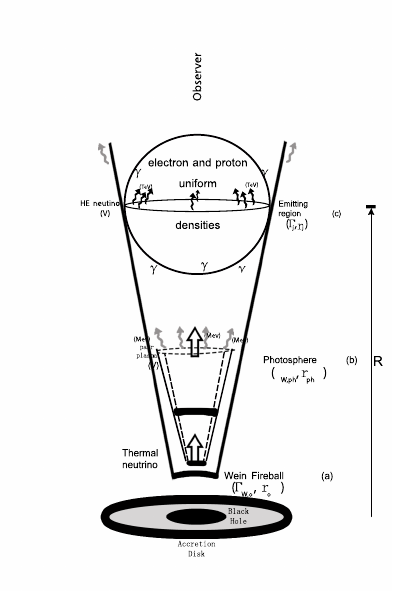}}
}
{\centering
\resizebox*{0.5\textwidth}{0.4\textheight}
{\includegraphics{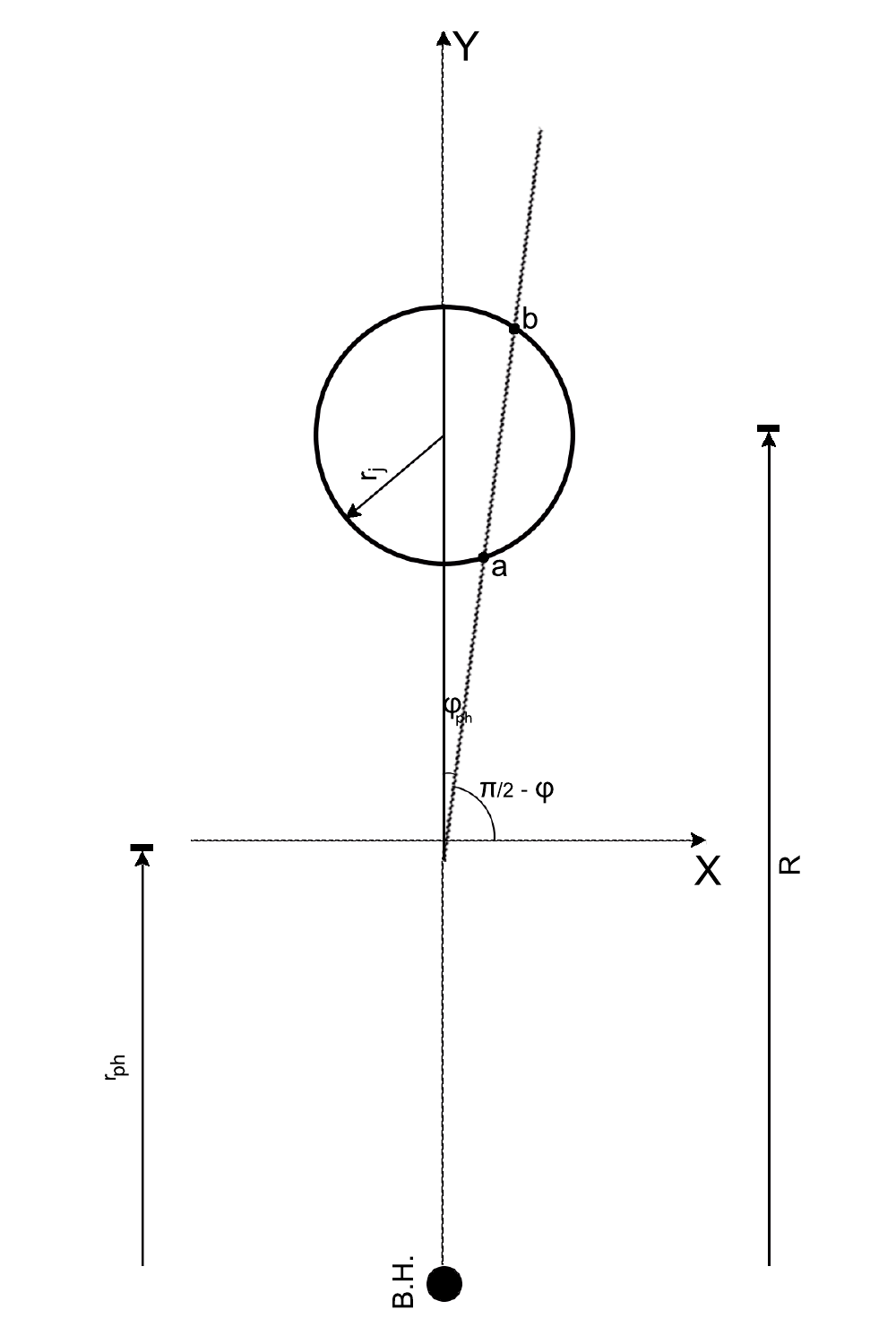}}
}
\caption{Outline of the jet and geometrical considerations. In the  figure  above is showed the evolution of the Wein fireball in three stages:   Initial state,   photosphere and emitting region and in the figure below is showed the geometrical diagram of a circle (emitting region) and a straight line (photon path).}
\label{picture}
\end{figure}

We consider a spherical emitting region with a uniform particle density and radius $r_j$, located at a distance $R$ from the BH  and moving at relativistic speed with bulk Lorentz factor $\Gamma_j$, as shown in fig. \ref{picture} (above) \citep{1996MNRAS.280...67G, 2002ApJ...564...86B}.  Seed photons coming from the photosphere of the 'Wein' fireball will interact with Fermi-accelerated  electrons and protons injected in the emitting region.  Taking into account  $R\gg r_{ph}$,  these photons (with an angular distribution;  \citet{2004ApJ...601...78I}) will arrive and go through the emitting region following  different paths with an angle ($\phi_{ph}$);  longer paths around the center and shorter ones as the trajectories get farther away from the center.  To estimate the distances of any path, we find the common points of the circle, $(y-y'_0)^2+(x-x'_0)^2=r_j^2$, and the straight line, $y-y_0=m\,(x-x_0)$, through their intersections (points $a$ and $b$)  (see fig. \ref{picture} (below)).   As shown in fig. \ref{picture} (below), we assume $y'_0=R-r_{ph}$, $x'_0=0=x_0=y_0=0$ and $m=\pi/2-\phi_{ph}$. Solving this equation system, we obtain that  the distance between two points ($a$ and $b$) as a function of $\phi_{ph}$ can be written as 
\be\label{d}
d=\sqrt{\frac{4(R-r_{ph})^2 - 4(1+\tan\phi^2_{ph})\,\left[(R-r_{ph})^2  -r_j^2 \right]}{1+\tan\phi^2_{ph}}}\,.
\ee
Photons coming from the 'Wein' fireball will be able to go or not through the emitting region depending on their paths (distances) and the mean free path $\lambda_{\gamma,e}=1/(\sigma_T\,n_e)$  inside of it.  For instance, photospheric photons going through the emitting region will be absorbed in it if the paths are longer than the mean free path ($d>\lambda_{\gamma,e}$); otherwise, ($d<\lambda_{\gamma,e}$), they will be transmitted.  Defining the optical depth as a function of these two quantities (d and $\lambda_{\gamma,e}$) 
{\small
\bary
\tau&=& \frac{d}{\lambda_{\gamma,e}} \\
&=&2\sigma_T\,n_e\,\sqrt{\frac{(R-r_{ph})^2 - (1+\tan\phi^2_{ph})\,\left[(R-r_{ph})^2  -r_j^2 \right]}{1+\tan\phi^2_{ph}}} \,,\nonumber
\eary
}
it is possible to relate this angle with the electron particle density.  Taking into account  that a medium is said to be optically thick or opaque when $\tau > 1$ and optically thin or transparent  when $\tau < 1$, we write the electron particle density for this transition ($\tau=1$) as 
{\small
\be\label{den_opt}
n_e=\frac{1}{2\sigma_T}\,\sqrt{\frac{1+\tan\phi^2_{ph}}{(R-r_{ph})^2 - (1+\tan\phi^2_{ph})\,\left[(R-r_{ph})^2  -r_j^2 \right]}}\,.
\ee
}
The previous equation gives the values of electron density for which  photospheric photons going through the emitting region with a particular angle could be transmitted or absorbed.
\subsection{P$\gamma$ interactions}
Once emitted, the MeV-peak photons also interact with  protons accelerated at the emitting region.  Assuming a baryon content in this region \citep{2011ApJ...736..131A,2000ApJ...534..109S,2004MNRAS.349..336K, 2012ApJ...753...40F, 2014ApJ...783...44F},  accelerated protons lose their energies by electromagnetic and hadronic interactions.  Electromagnetic interaction such as proton synchrotron radiation and inverse Compton will not be considered here, we only assume that protons will be cooled down  by p$\gamma$ interactions.  The optical depth  of this process is 
\be
\tau_{p\gamma}\simeq \frac{r_j\, n_{\gamma} \sigma_{p\gamma}}{\Gamma_j},
\label{opt_pg}
\ee
where the photon density ($n_{\gamma}$) is given by eq. (\ref{den_ph}) and $\sigma_{p\gamma}$ is the cross section for p$\gamma$ interactions.   The photopion process p$\gamma\to N\pi$ has a threshold photon energy $\epsilon_{th}=m_\pi + m^2_\pi/2m_p$ where the neutral and charged pion mass are $m_{\pi^0}=135$ MeV and $m_{\pi^0}=139.6$ MeV, respectively \citep{der09}.   The two main contributions to the total photopion cross section at low energies come from the resonance production and direct production.\\
{\bf Resonance Production ($\Delta^+$).}  The photopion production is given through the resonances $\Delta^+(1.232)$,  $\Delta^+(1.700)$,  $\Delta^+(1.905)$ and $\Delta^+(1.950)$ where the mass $m_{\Delta^+}$ and Lorentzian width $\Gamma_{\Delta^+}$ for all resonances are reported by \citet{2000CoPhC.124..290M}.   The $\pi^0$ threshold is $\simeq 145$ MeV, and only the $p\gamma \longrightarrow\Delta^{+}\longrightarrow p\pi^0$ is cinematically allowed.\\ 
{\bf Direct Production.} This channel exhibits the non resonant contribution to direct two-body channels consisting of outgoing charged pions. This channel includes the reaction $p\gamma\to n\pi^+$, $p\gamma\to \Delta^{++}\pi^-$, and $p\gamma\to \Delta\pi^0$. For the $\pi^+$ threshold,  the direct pion channel  p$\gamma\to n\pi^+$ is dominant for an energy between 0.150 GeV and 0.25 GeV.\\ 
Other contributions  to the total photopion cross section (multi pion production and diffraction) are only important at high energies.
\subsubsection{$\pi^0$ decay products}
Neutral pion decays into two photons, $\pi^0\rightarrow \gamma\gamma$, and each photon carries less than $15\%$  of the proton's energy $\epsilon_\gamma \simeq \xi_{\pi^0}/2 E_p=0.15\,E_p$.  The $\pi^0$ cooling time scale for this process is \citep{1968PhRvL..21.1016S, PhysRevLett.78.2292}
\be\label{tpg}
{t'}^{-1}_{\pi^0}=\frac{1}{2\,\Gamma^2_p}\int\,d\epsilon\,\sigma_\pi(\epsilon)\,\xi_{\pi^0}\,\epsilon\int dx\, x^{-2}\, \frac{dn_\gamma}{d\epsilon_\gamma} (\epsilon_\gamma=x)
,\ee
where $\Gamma_p$ is the proton Lorentz factor, $\sigma_\pi(\epsilon_\gamma)\simeq\sigma_{p\gamma}$ is the cross section of pion production. The target photon spectrum  $dn_\gamma/d\epsilon_\gamma$ is given by eq. (\ref{esp_phot}) which is normalized through $\int^\infty_0\,\epsilon_\gamma (dn_\gamma/d\epsilon_\gamma) d\epsilon_\gamma= U_\gamma\simeq \epsilon_{\gamma b}\,n_\gamma/2\Gamma^2_{W,ph}$.  The threshold for p$\gamma$ interaction is computed through  the minimum proton energy, which can be written as
\bary
 E_{p,min}&=&\Gamma_j\,\Gamma_{W,ph}\,\frac{(m^2_{\pi^0}+2m_pm_{\pi^0})}{2\epsilon_{\gamma b}\,(1-\cos\phi)}\cr
&\simeq& 0.14\, {\rm TeV}\, \frac{\Gamma_j\,\Gamma_{W,ph}\,\epsilon^{-1}_{\gamma b, MeV}}{(1-\cos\phi)}\,,
\eary
where $\phi$ is the angle of this interaction.  The photopion production efficiency $f_{\pi^0}= \frac{t'_d}{t'_{\pi^0}}$ can be defined through the dynamical ($t'_d\simeq\,r_j/\Gamma_j$) and photopion time scales \citep{2013avhe.book..225D, 2014arXiv1403.4089M}
{\small
\bary\label{f_pi}
f_{\pi^0}&=& \frac{t'_d}{t'_{\pi^0}}\simeq \frac{r^2_{ph}\,n_\gamma\,\sigma_{p\gamma}\,\xi_{\pi^0}\,\Delta \epsilon'_{peak}}{4\,\Gamma_j\,\Gamma_{W,ph}\,r_j\,\epsilon'_{peak}}\times\cases{\left(\frac{\epsilon_{\pi^0,\gamma,c}}{\epsilon_{0}}\right)^{-1} \left(\frac{\epsilon_{\pi^0,\gamma}}{\epsilon_{0}}\right)       &  $\epsilon_{\pi^0,\gamma} < \epsilon_{\pi^0,\gamma,c}$\cr\nonumber
1                                                                                      &  $\epsilon_{\pi^0,\gamma,c} < \epsilon_{\pi^0,\gamma}$\,,\cr
}
\eary
}
where the break photopion energy  is given by
\be
\epsilon_{\pi^0,\gamma,c}\simeq0.25\, \Gamma_j\,\Gamma_{W,ph}\,\xi_{\pi^0}\,(m_{\Delta^+}^2-m_p^2) {\epsilon_{\gamma b}}^{-1}\,,
\label{pgamma}
\ee
$\epsilon'_{peak}$ and $\Delta \epsilon'_{peak}$ correspond to the energy and the width around the resonance of  total cross section, respectively.  Taking into account the proton spectrum as a simple power law
\be\label{Ap}
\left( \frac{dN}{dE}\right)_p=A_p\,\left(\frac{E_p}{{\rm GeV}}\right)^{-\alpha},
\ee
the photopion production efficiency ($f_{\pi^0}$), and  $f_{\pi^0}\,E_p\,(dN/dE)_p\,dE_p=\epsilon_{\pi^0,\gamma}\,(dN/d\epsilon)_{\pi^0,\gamma}\,d\epsilon_{\pi^0,\gamma}$, then  we can write the photopion spectrum as
{\small
\bary\label{pgammam}
\left(\epsilon^2\,\frac{dN}{d\epsilon}\right)_{\pi^0,\gamma} &=&A_{p,\gamma}\times\cases{
\left(\frac{\epsilon_{\pi^0,\gamma,c}}{\epsilon_{0}}\right)^{-1} \left(\frac{\epsilon_{\pi^0,\gamma}}{\epsilon_{0}}\right)^{-\alpha+3}          & $\epsilon_{\pi^0,\gamma} < \epsilon_{\pi^0,\gamma,c}$\cr\nonumber
\left(\frac{\epsilon_{\pi^0,\gamma}}{\epsilon_{0}}\right)^{-\alpha+2}                                                                                       &   $\epsilon_{\pi^0,\gamma,c} < \epsilon_{\pi^0,\gamma}$,\cr
}
\eary
}
\noindent with the  proportionality constant  given by
\be\label{Apg}
A_{p,\gamma}\simeq \frac{3r^2_{ph}\,n_\gamma\,\epsilon^2_0\,\sigma_{p\gamma}\,\left(\frac{2}{\xi_{\pi^0}}\right)^{1-\alpha}\,\Delta \epsilon'_{peak}}{4\,\Gamma_j\,\Gamma_{W,ph}\,r_j\,\epsilon'_{peak}   } \,A_p\,.
\ee
From eqs. (\ref{Ap}) and (\ref{Apg}), the proton luminosity {\small $L_p\simeq 4\pi D^2_z F_p= 4\pi D^2_z E^2_p \left(\frac{dN}{dE}\right)_p$} can be written as
\be\label{lum}
L_p\simeq \frac{8\,\pi\,\Gamma_j\,\Gamma_{W,ph}\,r_j\, D^2_z\, \left(\frac{2}{\xi_{\pi^0}}\right)^{\alpha-1} \,\epsilon'_{peak}}{(\alpha-2)\,r^2_{ph}\,n_\gamma\,\sigma_{p\gamma}\Delta \epsilon'_{peak}}\,A_{p,\gamma}\,\biggl(\frac{E_p}{GeV}\biggr)^{2-\alpha},
\ee
where $A_{p,\gamma}$ is obtained through the photopion spectrum (eq. \ref{Apg}).
\subsubsection{$\pi^\pm$ decay products}
Muons and positrons/electrons are produced through charged pion decay products ($\pi^{\pm}\rightarrow \mu^{\pm}+ \nu_{\mu}/\bar{\nu}_{\mu} \rightarrow e^{\pm}+\nu_{\mu}/\bar{\nu}_{\mu}+\bar{\nu}_{\mu}/\nu_{\mu}+\nu_{e}/\bar{\nu}_{e}$).  Although muon's lifetime $t'_{\mu^+,dec}=\frac{E'_{\mu^+}}{m_{\mu}}\,\tau_{\mu^+}$ is  very short with $\tau_{\mu^+}=$ 2.2 $\mu$s and m$_\mu=105.7$ MeV,  muons  could be rapidly  accelerated for a short period of time in the presence of a magnetic field ($B'$) and radiate photons by synchrotron emission $\epsilon'_{\gamma}=\frac{3\pi q_e\,B'}{8\,m_{\mu}^3}\,E_\mu^{'2}$ \citep{1998PhRvD..58l3005R,2000NewA....5..377A,1993A&A...269...67M, 2011ApJ...736..131A}.  After muons decay, positrons/electrons could radiate photons $\epsilon'_{\gamma}=\frac{3\pi q_e\,B'}{8\,m_e^3}\,E_e^{'2}$ at the same place. Therefore, muons and positrons/electrons  cool down in accordance with the cooling time scale characteristic, $t'_{syn,i}=6\pi m_{i}^4/(\sigma_T\,m^2_e\,B'^2\, E'_{i})$  up to a maximum acceleration time scale given by  $t'_{syn,max}=16\,E'_{i}/(3\,q_e\,B')$. Here $q_e$ is the  elementary charge and the subindex  $\rm {i}$ is for $\mu$ and $e$. Hence, the break and maximum photon energies in the observed frame are
\bary\label{ene_ph}
\epsilon_{\gamma,c} &=& \frac{m^5_i}{m^5_e} \epsilon_{\gamma,c-e}\,,\cr
\epsilon_{\gamma, max} &=& \frac{m_i}{m_e}\epsilon_{\gamma, max-e}\,,
\eary
where we have taken into account the Lorentz factor ratios $\gamma_{i}=m^2_i/m^2_e\,\gamma_{e}$. Assuming that  Fermi-accelerated muons and electrons/positrons within a volume $\simeq 4\pi r_j^3/3$  with energies $\gamma_i\,m_i$ and maximum radiation powers $P_ {\nu,max,i} \simeq \frac{dE/dt}{\epsilon_\gamma(\gamma_i)}$  are  well described by broken power laws $N_i(\gamma_i)$:  $\gamma_i^{-\alpha}$ for $\gamma_i < \gamma_{i,b}$ and $\gamma_{i,b} \gamma_i^{-(\alpha+1)}$ for $\gamma_{i,b} \leq  \gamma_i<\gamma_{i,max}$, then the observed synchrotron spectrum can be written as \citep{1994hea2.book.....L,  1986rpa..book.....R}
{\small
\bary\label{espsynm}
\left(\epsilon^2\,\frac{dN}{d\epsilon}\right)_{syn,\gamma}&=& A_{syn,\gamma-i} \times\cases{
\left(\frac{\epsilon_{\gamma,c}}{\epsilon_0}\right)^{-1/2} \left(\frac{\epsilon_\gamma}{\epsilon_0}\right)^{-(\alpha-3)/2} &  $\epsilon_\gamma < \epsilon_{\gamma,c}$,\cr\nonumber
\left(\frac{\epsilon_\gamma}{\epsilon_0}\right)^{-(\alpha-2)/2}           &  $\epsilon_{\gamma,c} < \epsilon_\gamma < \epsilon_{\gamma,max}$\,,\cr
}
\eary
}
with
\be
A_{syn,\gamma-i}= \frac{4\,\sigma_T\,m^4_i}{27\pi^2 q_e\,m_e^3}r^3_j\,D^{-2}_z\,\Gamma_j\,\epsilon_{\gamma,c-e}\,B'\,N_i.
\ee
Here $N_i$ is the number of radiating muons and/or positrons/electrons.
\section{Neutrino Emission}
Although in the current model there are multiple places where neutrinos with different energies could be generated,  only we are going to consider the thermal neutrinos created at the initial stage of the Wein fireball and the HE neutrinos generated by p$\gamma$ interactions at the emitting region (see fig. \ref{picture} (above)).
\subsection{Thermal Neutrinos}
At the initial stage of the Wein fireball, thermal neutrinos are created by electron-positron annihilation ($e^++e^-\to Z\to \nu_j+\bar{\nu}_j$). By considering a small amount of baryons, neutrinos could be generated by processes of positron capture on neutrons ($e^++n  \to p+  \bar{\nu_e}$), electron capture on protons   ($e^-+p  \to n+  \nu_e$) and  nucleon-nucleon bremsstrahlung ($NN\to NN+\nu_j+\bar{\nu}_j$) for $j=e,\nu,\tau$.  Electron antineutrinos can be detected indirectly (in water Cherenkov detector)  through the interactions with the positrons created by the inverse neutron decay processes ($\bar{\nu}_e + p \to n +e^+$).   
The expected event rate can be estimated by 
\be\label{rate}
N_{ev}=T\,\rho_N\,N_A\,V_w  \int_{E_{\bar{\nu}_e}}  \sigma^{\bar{\nu}_ep}_{cc}(E_\nu) \left(\frac{dN}{dE}\right)_\nu\,dE_\nu\,,
\ee
\noindent where $N_A=6.022\times 10^{23}$ g$^{-1}$ is the Avogadro's number, $\rho_N=2/18\, {\rm g\, cm^{-3}}$ is the nucleons density in water \citep{2004mnpa.book.....M}, $V_{w} $ is the volume of the detector,  $ \sigma^{\bar{\nu}_ep}_{cc}\simeq 9\times 10^{-44}\,E^2_{\bar{\nu}_e}/MeV^2 \, {\rm cm^2}$  is the cross section \citep{1989neas.book.....B,2007fnpa.book.....G}, $T$ is the observation time  and (${dN/dE})_\nu$ is the neutrino spectrum.   Taking into account the relation between the luminosity $L_\nu$ and neutrino flux $F_\nu$, $L_\nu=4\pi D^2_z  F_\nu(<E>)=4\pi D^2_z   (E^2 dN/dE)_\nu$, then the number of events expected will be
\bary\label{ther_neu}
N_{ev}&\simeq&\frac{T}{<E_{\bar{\nu}_e}>}V_{w} N_A\,  \rho_N  \sigma^{\bar{\nu}_ep}_{cc} <E_{\bar{\nu}_e}>^2 \left(\frac{dN}{dE}\right)_{\bar{\nu}_e} \cr
&\simeq&\frac{T}{4\pi D^2_z <E_{\bar{\nu}_e}>}V_{w} N_A\,  \rho_N  \sigma^{\bar{\nu}_ep}_{cc}\,L_{\bar{\nu}_e}\,.
\eary
\noindent Here we have averaged over the electron antineutrino energy.  After thermal neutrinos are produced, they oscillate firstly in matter (due to Wein plasma) and secondly in vacuum on their path to Earth.
%
%
\subsubsection{Neutrino effective potential} 
As known, neutrino properties get modified when they propagate in a heat bath. A massless neutrino acquires an effective mass and undergoes an effective potential in the background.  Because electron neutrino ($\nu_e$) interacts with electrons via both neutral and charged currents (CC), and muon/tau ($\nu_\mu/\nu_\tau$) neutrinos interact only via the neutral current (NC), $\nu_e$ experiments a different effective potential in comparison with $\nu_\mu$ and $\nu_\tau$.  This would induce a coherent effect in which maximal conversion of $\nu_e$ into $\nu_\mu$ ($\nu_\tau$) takes place even for a small intrinsic mixing angle \citep{1978PhRvD..17.2369W}.  On the other hand, although neutrino cannot couple directly to the magnetic field, its effect which is entangled with the matter can be undergone by means of coupling to charged particles in the medium.  Recently, \citet{2014ApJ...787..140F} derived the neutrino self-energy and effective potential up to order $m_W^{-4}$ at strong, moderate and weak magnetic field approximation as a function of  temperature, chemical potential ($\mu$) and neutrino energy (E$_\nu$) for moving neutrinos along the magnetic field. In this approach, we will use  the neutrino effective potential in the weak field approximation, which is given by
{\small
\bary\label{Veffw}
V_{eff}&=&\frac{\sqrt2\,G_F\,m_e^3}{\pi^2}\biggl[\sum^{\infty}_{l=0}(-1)^l\sinh\alpha_l \biggl\{\left(2+ \frac{m^2_e}{m^2_W}\left(3+4\frac{E^2_\nu}{m^2_W}\right)\right)\times\left(\frac{K_0(\sigma_l)}{\sigma_l}+2\frac{K_1(\sigma_l)}{\sigma^2_l} \right) - 2\left(1+ \frac{m^2_e}{m^2_W} \right)\frac{B}{B_c} K_1(\sigma_l) \biggr\}\nonumber\\ 
&&-4\frac{m^2_e}{m^2_W}\,\frac{E_\nu}{m_e}\sum^\infty_{l=0}(-1)^l\cosh\alpha_l \biggl\{\left(\frac{2}{\sigma^2_l}-\frac{B}{4B_c}\right)K_0(\sigma_l)+\left(1+\frac{4}{\sigma^2_l}\right)\frac{K_1(\sigma_l)}{\sigma_l} \biggr\} \biggr]\,. \nonumber\\
\eary
}
Taking into account  the condition that the plasma has equal number of electrons and positrons ($N_e-\bar{N}_e=0$), then the neutrino effective potential is reduced to
{\small
\bary \label{Veffw_N0}
V_{eff}&=&-\frac{4\sqrt2\,G_F\,m_e^4 E_\nu}{\pi^2\,m^2_W}\sum^\infty_{l=0}(-1)^l \biggl\{\left(\frac{2}{\sigma^2_l}-\frac{B}{4B_c}\right)K_0(\sigma_l)+\left(1+\frac{4}{\sigma^2_l}\right)\frac{K_1(\sigma_l)}{\sigma_l} \biggr\} \,, \nonumber\\
\eary
}
where  K$_i$ is once again  the modified Bessel function of integral order i,  $G_F=\sqrt 2g^2/8m_W^2$ is the Fermi coupling constant,  $B_c=m^2_e/e$ is the critical magnetic field, m$_W$ is the W-boson mass,  $\alpha_l=(l+1)\mu/(\theta_o\,m_e) $ and $\sigma_l=(l+1)/\theta_o$.\\
\subsubsection{Resonant oscillations}
When neutrino oscillations occur in matter, a resonance could take place that would dramatically enhance the flavor mixing and lead to a maximal conversion from one neutrino flavor to another.  This resonance depends on the effective potential and neutrino oscillation parameters.  The equations that determine the neutrino evolution in matter for two and  three flavors are related as follows \citep{2014ApJ...787..140F,2014MNRAS.442..239F}.
\paragraph{Two-Neutrino Mixing.}
The evolution equation for neutrinos that propagate in the  medium  ($\nu_e\leftrightarrow \nu_{\mu, \tau}$) is given by \citep{2014MNRAS.437.2187F}
\be
i
{\pmatrix {\dot{\nu}_{e} \cr \dot{\nu}_{\mu}\cr}}
={\pmatrix
{V_{eff}-\frac{\delta m^2}{2E_\nu} \cos 2\psi & \frac{\delta m^2}{4E_\nu}\sin 2\psi \cr
\frac{\delta m^2}{4E_\nu}\sin 2\psi  & 0\cr}}
{\pmatrix
{\nu_{e} \cr \nu_{\mu}\cr}},
\ee
where $\delta m^2$ is the mass difference,  $\psi$ is the neutrino mixing angle and $V_{eff}$ is the neutrino effective potential (eqs. \ref{Veffw} and \ref{Veffw_N0}).  The oscillation length for the neutrino is given by
\be
l_{osc}=\frac{4\pi\,E_\nu}{\delta m^2\sqrt{\cos^2 2\psi (1-\frac{2E_\nu\,V_{eff}}{\delta m^2 \cos 2\psi})^2+\sin^2 2\psi}}\,,
\label{osclength}
\ee
and the conversion probability  by
\be
P_{\nu_e\rightarrow {\nu_{\mu}{(\nu_\tau)}}}(t) = 
\frac{ \delta m^4  \sin^2 2\psi}{4\,\omega^2\,E^2_\nu}\sin^2\left (\frac{\omega t}{2}\right
),
\label{prob}
\ee
with {\small $\omega=\sqrt{(V_{eff}- (\delta m^2/2E_\nu) \cos 2\psi)^2+ (\delta m^4/4E^2_\nu) \sin^2 2\psi}$}.   Taking into account the resonance condition
\bary
V_{eff}&=&5\times 10^{-7}\, eV\,\frac{\delta m^2_{eV}}{E_{\nu,MeV}}\,\cos2\psi\,,
\label{reso2}
\eary
then the resonance length ($l_{res}$) can be written as
\be
l_{res}=\frac{4\pi\,E_\nu}{\delta m^2 \sin 2\psi}.
\ee
The best fit values of the two neutrino mixing  are: \textbf{Solar Neutrinos}:  $\delta m^2=(5.6^{+1.9}_{-1.4})\times 10^{-5}\,{\rm eV^2}$ and $\tan^2\psi=0.427^{+0.033}_{-0.029}$\citep{aha11},  \textbf{Atmospheric Neutrinos}: $\delta m^2=(2.1^{+0.9}_{-0.4})\times 10^{-3}\,{\rm eV^2}$ and $\sin^22\psi=1.0^{+0.00}_{-0.07}$ \citep{abe11a} and   \textbf{Accelerator Neutrinos}: $\delta m^2  \approx  0.5\, {\rm eV^2}$ and $\sin^22\psi\sim 0.0049$  \citep{ath96, ath98}.

\paragraph{Three-Neutrino Mixing.}
In the three-flavor framework, the evolution equation of the neutrino system in the matter can be written as
\be
i\frac{d\vec{\nu}}{dt}=H\vec{\nu},
\ee
where the state vector is $\vec{\nu}\equiv(\nu_e,\nu_\mu,\nu_\tau)^T$, the effective Hamiltonian is $H=U\cdot H^d_0\cdot U^\dagger+diag(V_{eff},0,0)$ with  $H^d_0=\frac{1}{2E_\nu}diag(-\delta m^2_{21},0,\delta m^2_{32})$,  the neutrino effective potential $V_{eff}$ is defined by eqs. (\ref{Veffw}) and (\ref{Veffw_N0}) and $U$ is the three neutrino mixing matrix \citep{gon03,akh04,gon08,gon11}.
The oscillation length for the neutrino is given by
\be
l_{osc}=\frac{4\pi E_{\nu}/\delta m^2_{32}}{\sqrt{\cos^2 2\psi_{13} (1-\frac{2 E_{\nu} V_e}{\delta m^2_{32} \cos 2\psi_{13}}
    )^2+\sin^2 2\psi_{13}}}\,.
\label{osclength}
\ee
The resonance condition and resonance length are
\be\label{reso3}
V_{eff}-5\times 10^{-7}\frac{\delta m^2_{32,eV}}{E_{\nu,MeV}}\,\cos2\psi_{13}=0\,,
\ee
and
\be
l_{res}=\frac{4\pi E_{\nu}/\delta m^2_{32}}{\sin 2\psi_{13}},
\ee
respectively.  In addition to the resonance condition, the dynamics of this transition from one flavor to another must be determined by adiabatic conversion which is given by 
\bary
\kappa_{res}&\equiv & \frac{2}{\pi}
\left ( \frac{\delta m^2_{32}}{2 E_\nu} \sin 2\psi_{13}\right )^2   \left | \frac{dV_{eff}}{dr} \right |^{-1} \ge 1\,.\nonumber\\
&& \frac{2}{\pi}\left ( \frac{\delta m^2_{32}}{2 E_\nu} \sin 2\psi_{13}\right )^2\,\left(\frac{\theta_o\,r_o}{\theta^2} \right)      \left | \frac{\partial V_{eff}}{\partial \theta}\right |^{-1} \ge 1\,.\nonumber\\
\label{adbcon}
\eary
Combining solar, atmospheric, reactor and accelerator parameters, \textbf{the best fit values of the three neutrino mixing  are}, {\small ${\rm for}\,\,\sin^2_{13} < 0.053: \Delta m_{21}^2= (7.41^{+0.21}_{-0.19})\times 10^{-5}\,{\rm eV^2}$ and $\tan^2\psi_{12}=0.446^{+0.030}_{-0.029}$ and, {\rm for}\,\,$\sin^2_{13} < 0.04: \Delta m_{23}^2=(2.1^{+0.5}_{-0.2})\times 10^{-3}\,{\rm eV^2}$ and  $\sin^2\psi_{23}=0.50^{+0.083}_{-0.093}$}, \citep{aha11,wen10}.
\paragraph{Neutrino flavor ratio expected on Earth.}
The probability for a neutrino to oscillate from a flavor state $\alpha$ to another flavor state $\beta$ in its path (from the source to  Earth) is 
{\small
\bary
P_{\nu_\alpha\to\nu_\beta} &=&\delta_{\alpha\beta}-4 \sum_{j>i}\,U_{\alpha i}U_{\beta i}U_{\alpha j}U_{\beta i}\,\sin^2\biggl(\frac{\delta m^2_{ij}\, L}{4\, E_\nu}   \biggr)\,,
\eary
}
where $U_{ij}$ are the elements of  the three-neutrino mixing matrix \citep{gon03,akh04,gon08,gon11}. Using the set of oscillation parameters \citep{aha11,wen10}, the mixing matrix can be written as
\be
U =
{\pmatrix
{
0.817    &  0.545     &     0.191\cr
 -0.505  & 0.513      &	 0.694\cr
 0.280   &  -0.663    &     0.694\cr
}}\,.
\ee
Additionally, averaging the term sin  $\sim 0.5$ \citep{lea95} for distances longer than the solar system, the neutrino flavor vector at source ($\nu_e$, $\nu_\mu$, $\nu_\tau$)$_{source}$ and Earth  ($\nu_e$, $\nu_\mu$, $\nu_\tau$)$_{Earth}$ are related through  the probability matrix given by
\be
{\pmatrix
{
\nu_e   \cr
\nu_\mu   \cr
\nu_\tau   \cr
}_{Earth}}
=
{\pmatrix
{
0.534	  & 0.266	  & 0.200\cr
 0.266	  & 0.367	  &  0.368\cr
 0.200	  & 0.368	  & 0.432\cr
}}
{\pmatrix
{
\nu_e   \cr
\nu_\mu   \cr
\nu_\tau   \cr
}_{source}}\,.
\label{matrixosc}
\ee
\subsection{High-energy neutrinos}
\begin{figure}
\vspace{0.5cm}
{\centering
\resizebox*{1.\textwidth}{0.6\textheight}
{\includegraphics{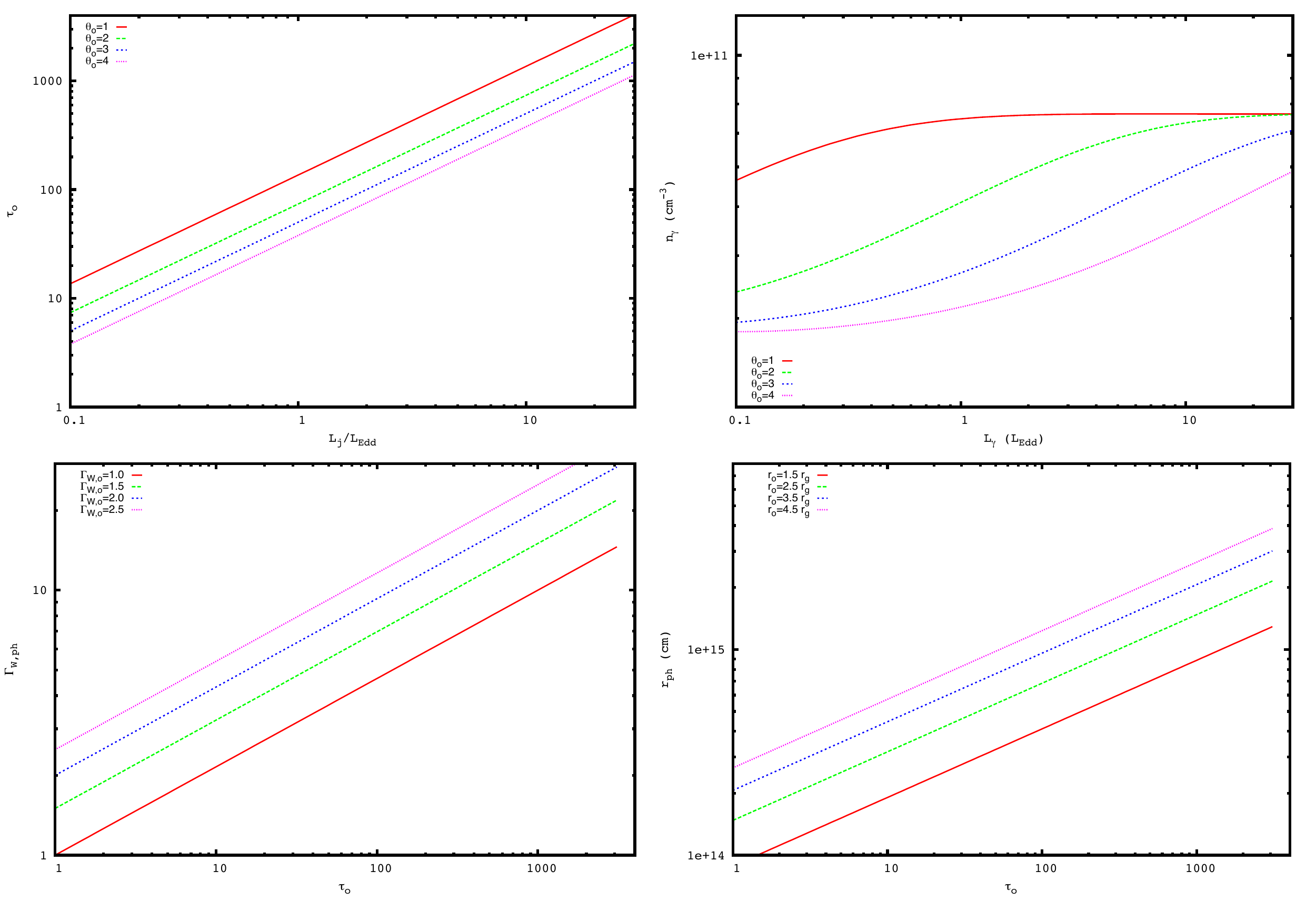}}
}
\caption{Figures above: Initial optical depth ($\tau_o$)  as a function of $L_j/L_{Edd}$ (left panel) and photon density $n_{\gamma}$ as a function of L$_\gamma$ (right panel) are plotted. In both figures we use $\theta_0$ for 1, 2 3 and 4.  Figures below: Lorentz factor ($\Gamma_{W,ph}$)   (left panel)  and photospheric radius ($r_{ph}$) (right panel) as a function of optical depth ($\tau_o$). }
\label{fireball_plot}
\end{figure}

HE neutrinos are created in p$\gamma$ interactions through n$\pi^+$ channel.  The  charged pion decays into leptons and neutrinos, $\pi^{\pm}\rightarrow e^{\pm}+\nu_{\mu}/\bar{\nu}_{\mu}+\bar{\nu}_{\mu}/\nu_{\mu}+\nu_{e}/\bar{\nu}_{e}$. 
Assuming that TeV flares can be described as $\pi^0$ decay products, we can estimate the number of neutrinos associated to these flares \citep{2000APh....13....1A, 2001Natur.410..441A, 2001NuPhS..91..423A}.  For $p\gamma$ interactions, the neutrino flux, dN$_\nu/d\en=A_{\nu} \,\en^{-\alpha_\nu}$, is related  with  the photopion flux by  \citep[see, e.g.] [and reference therein]{2007Ap&SS.309..407H,2005APh....23..537H}
\be
\int \left(\frac{dN}{dE}\right)_\nu\,\en\,d\en=\frac14\int \left(\frac{dN}{d\epsilon}\right)_{\pi^0,\gamma}\,\epsilon_{\pi^0,\gamma}\,d\epsilon_{\pi^0,\gamma}\,.
\ee
Assuming that the spectral indices of neutrino and photopion spectra are similar  $\alpha\simeq \alpha_\nu$ \citep{2008PhR...458..173B}, taking into account that  each neutrino  carries $\sim$ 5\%  of the  proton energy ($\en\simeq1/20\,E_p$) \citep{Halzen:2013bta} and also from eq. (\ref{pgammam}),  we can write the normalization factors  of HE neutrino and photopion   as 
\be\label{Anu}
A_{\nu}\simeq\frac14A_{p,\gamma}\,\left (10\,\xi_{\pi^0}\right)^{-\alpha+2}\, {\rm TeV}^{-2},
\ee
with A$_{p,\gamma}$ given by Eq. (\ref{Apg}).  Therefore,  we could infer the number of events expected through
\be
N_{ev}\approx T \rho_{ice}\,N_A\, V_i \int_{E_{th}}^\infty   \sigma_{\nu N}(\en)\, \left(\frac{dN}{dE}\right)_\nu\,d\en,
\label{evneu1}
\ee
where E$_{th}$ is the threshold energy, $T$ corresponds to the observation time of the flare \citep{2005APh....23..537H},  $ \sigma_{\nu N}(\en)=6.78\times 10^{-35}(\en/TeV)^{0.363}$ cm$^2$ is the charged current cross section \citep{1998PhRvD..58i3009G}, $\rho_{ice}$=0.9 g cm$^{-3}$ is the density of the ice and V$_i$ is the effective volume of detector, then the expected number of neutrinos inferred from this flare is
{\small
\be\label{numneu}
N_{ev} \approx  \frac{T \rho_{ice}\,N_A\, V_i}{\alpha-1.363}\,A_{\nu}\,(6.78\times 10^{-35}\,{\rm cm^2})\left(\frac{E_{\nu,th}}{{\rm TeV}}\right)^{\beta}\,{\rm TeV},
\ee
with the power index $\beta=-\alpha$+1.363 and $A_\nu$ given by eq. (\ref{Anu}).
\section{Ultra-high-energy cosmic rays}
It has been suggested that TeV $\gamma$-ray observations from low-redshift sources could be good candidates for studying UHECRs \citep{2012ApJ...749...63M}. Also special features in these $\gamma$-ray observations coming from blazars favor acceleration of UHECRs in blazars \citep{2012ApJ...745..196R}.  In the current  model we consider that the proton spectrum is extended up to $\sim 10^{20}$ eV energies and  based on this assumption, we  calculate the  number of events expected in TA experiment.
\subsection{Hillas Condition}
By considering that the BH has the power to accelerate particles  up to UHEs by means of Fermi processes,  protons accelerated in the emitting region are limited by the Hillas condition \citep{1984ARA&A..22..425H}. Although this requirement is a necessary condition and acceleration of UHECRs in AGN jets \citep{2012ApJ...749...63M,2012ApJ...745..196R,2010ApJ...719..459J},  it is far from trivial (see e.g., Lemoine \& Waxman 2009 for a more detailed energetics limit \citep{2009JCAP...11..009L}). The Hillas criterion says that the maximum proton energy achieved is   
\be\label{Emax}
E_{p,max}=e\,B'\,r_j\,\Gamma_j\,,
\ee
where $B'$ is  the strength of the magnetic field.  Alternatively, during flaring intervals for which the apparent luminosity can achieve L$^*\approx 10^{47}$ erg s$^{-1}$ and from the equipartition magnetic field $\epsilon_B$, the maximum energy of UHECRs can be derived and written as \citep{2009NJPh...11f5016D,2014ApJ...783...44F}
\begin{equation}
E_{max}\approx 1.0\times10^{21}\,\frac{e}{\Phi}\frac{\sqrt{\epsilon_B\,L^*/10^{47}\, erg\, s^{-1}}}{\Gamma_j}\,eV,
\end{equation}
\noindent where  $\Phi \simeq 1$ is the acceleration efficiency factor. \\
\subsection{Deflections}
The magnetic fields in the Universe play important roles because UHECRs are deflected by them.   UHECRs traveling from source to Earth are randomly deviated by galactic (B$_G$) and extragalactic (B$_{EG}$) magnetic fields. By considering a quasi-constant and homogeneous magnetic fields,  the deflection angle due to the B$_G$ is
\be\label{thet_G}
\psi_G\simeq 3.8^{\circ}\left(\frac{E_{p,th}}{57 EeV}\right)^{-1} \int^{L_G}_0  | \frac{dl}{{\rm kpc}}\times \frac{B_G}{4\,{\rm \mu G}} |\,,
\ee
and  due to B$_{EG}$  can be written as \citep{1997ApJ...479..290S}
\be\label{thet_EG}
\psi_{EG}\simeq 4^{\circ}\left(\frac{E_{p,th}}{57 EeV}\right)^{-1} \left( \frac{B_{EG}}{1\,{\rm nG}} \right)\,\left(\frac{L_{EG}}{100\, {\rm Mpc}}\right)^{1/2}\,\left(\frac{l_c}{1\, {\rm Mpc}}\right)^{1/2}\,,   
\ee
where L$_G$ corresponds to the distance of our Galaxy (20 kpc) and $l_c$ is the coherence length.   Due to the strength of extragalactic ($B_{EG}\simeq$ 1 nG) and galactic ($B_{G}\simeq$ 4 $\mu$G) magnetic fields,   UHECRs are deflected ($\psi_{EG}\simeq 4^{\circ}$ and $\psi_G\simeq 3.8^{\circ}$) between the true direction to the source, and the observed arrival direction, respectively.  Evaluation of these deflection angles  links the transient UHECR sources with the  high-energy neutrino and $\gamma$-ray emission. Regarding eqs. (\ref{thet_G}) and (\ref{thet_EG}), it is reasonable to associate UHECRs lying within  5$^\circ$ of a source.

\subsection{Expected number of events}
TA experiment located in Millard Country, Utah,  is made of  a scintillator surface detector (SD) array and three fluorescence detector (FD) stations \citep{2012NIMPA.689...87A}. With an area of $\sim$ 700 m$^2$,  it was designed to study UHECRs with energies above 57 EeV.  To estimate the number of UHECRs associated to "orphan" flares,  we take into account the TA  exposure, which  for a point source is given by $\Xi\,t_{op}\, \omega(\delta_s)/\Omega$, where $\Xi\,t_{op}=(5)\,7\times10^2\,\rm km^2\,yr$, $t_{op} $ is the total operational time (from 2008 May 11 and 2013 May 4),  $\omega(\delta_s)$ is an exposure correction factor for the declination of the source \citep{2001APh....14..271S} and $\Omega\simeq\pi$.  The expected number of UHECRs above an energy $E_{p,th}$ yields
\be
N_{\tiny UHECR}= F_r\, ({\rm  TA\, Expos.})\times \,N_p, 
\label{num}
\ee
where F$_r$ is the fraction of propagating cosmic rays that survives over a distance $>$ D$_z$ \citep{2011ARA&A..49..119K} and  $N_p$ is calculated from the proton spectrum extended up to energies higher than $E_{p,th}$ (eq. \ref{Ap}).   The expected number can be written as
\bary
N_{\tiny UHECR}=F_r\,\frac{\Xi\,t_{op}\, \omega(\delta_s)\,(\alpha-2)}{4\pi\,(\alpha-1)\Omega\,d^2_z\,E_{p,th}}\,L_p\,,
\label{nUHE1}
\eary
where  $L_p$ is obtained from the TeV $\gamma$-ray observations of the flaring activities (eq. \ref{lum}).\\
\section{Application to 1ES 1959+650 and Mrk 421}
Following \citet{2002ApJ...565..163I,2004ApJ...601...78I}, we have considered the dynamics of a plasma made of  $e^\pm$ pairs and photons, and generated by the Wein equilibrium \citep{1984MNRAS.209..175S}. We have obtained the evolution equations of the bulk Lorentz factor (eq. \ref{Gamma_W}), temperature (eq. \ref{theta}) and density of photons and pairs (eq. \ref{number_total}).   In particular, we have computed the radius (eq. $\ref{r_ph}$), temperature (eq. $\ref{theta_ph}$),  Lorentz factor (eq. $\ref{Gamma_ph}$) and the photon density (eq. $\ref{den_ph}$)  at the photosphere as a function of  the initial conditions ($r_o$, $\Gamma_{W,o}$, $\theta_o$ and  $L_j/L_{Edd}$).    By considering the values of initial radius $r_o=2r_g$ and Lorentz factor $\Gamma_{W,o}=(3/2)^{1/2}$, we plot (see fig. \ref{fireball_plot}) the initial optical depth (left-hand figure above) and  photon density (right-hand figure above) as a function of  $L_j/L_{Edd}$ and  $L_\gamma$, respectively, and the Lorentz factor (left-hand figure below) and radius (right-hand figure below) at the photosphere as a function of initial optical depth. From the panels above we take into account  the range of values 0.1$\,\leq L_j/L_{Edd}\leq$ 30 and  0.1$L_{Edd}\leq L_\gamma \leq 30\, L_{Edd}$ for $\theta_o$ = 1, 2, 3 to 4, and in particular for $\theta=4$ we obtain that the initial optical depth and photon density  lie in the range of 3.8 $\leq\tau_o\leq$ 990 and $1.8 \times  10^{10} {\rm cm}^{-3} \leq n_{\gamma} \leq 6.75 \times 10^{10} {\rm cm}^{-3}$,  respectively.  In the left-hand figure, we can observe that $\tau_o$ behaves as an increasing function with $L_j/L_{Edd}$ and a decreasing function with $\theta_o$ whereas in the right-hand figure,  photon density is an increasing function of $L_\gamma$ and a decreasing function of $\theta_o$. From the panels below, we consider the initial optical depth in the range $1\leq \tau_0\leq 3\times 10^3$ to find the values of the Lorentz factors and radius in the photosphere for the initial conditions  $\Gamma_{W,o}$=1.0, 1.5, 2.0, and 2.5 and   $r_o$=$1.5r_g$, $2.5 r_g$, $3.5 r_g$ and $4.5 r_g$.  One can see that both the Lorenz factor and radius are increasing functions of initial optical depth and the initial conditions ($\Gamma_{W,0}$ and $r_o$). In particular for $\Gamma_{W,o}$=2, the Lorentz factor lies in the range 1$\leq \Gamma_{W,ph}\leq $ 30 and for $r_o=3.5 r_g$ the photosphere radius at $10^{14}  {\rm cm} \leq r_{ph}\leq  10^{15.6}  {\rm cm}$.\\   
Once the MeV-peak photons are released from the photosphere, they will arrive and go through the emitting region following different paths.  By considering that each path (see fig. \ref{sketch}) has a length given by eq. (\ref{d}), we plot the distance of these paths as a function of angle for three values of R = (8, 10 and 12 $\times \,10^{16}$ cm). In this figure, one can see that the length of the paths are shorter when angle increases, achieving a maximum value around 30$^\circ$. For instance,  taking into account $R= 12\times 10^{16}$ cm, a photon arriving with an angle $\phi_{ph}=25^{\circ}$ only has to go through a distance of $d= 2\times 10^{16}$ cm.   Regarding the mean free path of photons in the emitting region, we compute the optical depth as a function of election particle density and the angle of the photon's paths.  Taking the condition  $\tau=1$, we obtain  eq. (\ref{den_opt}) and plot the electron density as a function of angle $\phi_{ph}$, as shown in fig \ref{optical}. From this figure, we can observe two zones: the transparent and opaque zones.   For instance, a value of electron density higher than  $n_e \geq 0.12\times 10^8$ cm$^{-3}$ for $\phi_{ph}\leq 10^\circ$, is optically opaque, otherwise it is transparent. Therefore, MeV-peak photons coming from the 'Wein' fireball could be absorbed around the center and be transmitted with angles higher than 25$^\circ$. These photons (out of the line of view) would interact  in their wake with the Fermi-accelerated protons (through p$\gamma$ interactions) producing  $\pi^0$ decay products.  These TeV $\gamma$-ray photons  are resealed  isotropically whereas MeV-peak photons only are directed  out of the line of view (more than 25$^\circ$).  In this approach multi TeV photons can be imaged by the observer, but not the MeV-peak photons.\\
%
%
We have introduced a hadronic model  invoking  the p$\gamma$ interactions to interpret the "orphan" flares at TeV energies as $\pi^0$ decay products and to estimate the neutrinos and UHECRs from these events.  Using the developed Monte-Carlo event generator SOPHIA (Simulations Of Photo Hadronic Interactions in Astrophysics; \citet{2000CoPhC.124..290M}),   we obtain the cross sections for the resonance production ($p\gamma \longrightarrow\Delta^{+}\longrightarrow p\pi^0$) and the direct production ($p\gamma\to n\pi^+$), as shown in fig. \ref{cross}. From the left-hand panel, one can see that the most important contribution to the cross section comes from $\Delta^+(1.232)$ with $\sigma_{p\gamma}\simeq$ 281 $\mu$barn for $\epsilon'_{peak}\simeq$0.31 GeV. From the  right-hand panel,  we can observe that the value of the cross section ($p\gamma \longrightarrow n\pi^+$ ) at $\epsilon'_{peak}\simeq$0.31 GeV is $\sigma_{p\gamma}\simeq$ 254 $\mu$barn.  In both cases the width around $\epsilon'_{peak}$ is $\Delta \epsilon'_{peak}\simeq 0.21$ GeV. The previous values of the $p\gamma$ interaction parameters will be used to compute the TeV $\gamma$-rays from $\pi^0$ decay products and HE neutrinos from $\pi^+$ decay products.  On the other hand, we obtain the TeV $\gamma$-ray  spectrum generated by this process (eq. \ref{pgammam}), which depends on: the parameters of the proton ($A_p$ and $\alpha$) and seed photon ($\epsilon_{\gamma b}$ and  $n_\gamma$) spectrum,  the size of the emission region,  the p$\gamma$ cross section as well as the Lorentz factors of accelerated protons and target photons.  Using the photopion model and the method of Chi-square ($\chi^2$) minimization proposed by \citet{1997NIMPA.389...81B}, we find the values of proportionality constant ($A_{p,\gamma}$) and spectral index ($\alpha$) through the fitting parameters; [0] and [1], respectively, that best describe the photopion spectrum (eq \ref{pgammam}) \citep{2014ApJ...783...44F,  2014MNRAS.441.1209F}
{\small
\bary
\label{pgamma_fit}
\left(\epsilon^2\,\frac{dN}{d\epsilon}\right)_{\pi^0,\gamma}= [0]\hspace{6cm}\\
\times\cases{
\left(\frac{\epsilon_{\pi^0,\gamma,c}}{\epsilon_{0}}\right)^{-1} \left(\frac{\epsilon_{\pi^0,\gamma}}{\epsilon_{0}}\right)^{-[1]+3} &  $ \epsilon_{\pi^0,\gamma}< \epsilon_{\pi^0,\gamma,c}$\,,\cr\nonumber
\left(\frac{\epsilon_{\pi^0,\gamma}}{\epsilon_{0}}\right)^{-[1]+2}                                                                                        &   $\epsilon_{\pi^0,\gamma,c} < \epsilon_{\pi^0,\gamma}$\,.
}
\eary
}
It is important to say that the photopion spectrum can undergo pair production on their way to the observer which will reduce this observed spectrum. In this case the proportionality proton constant ($A_p$) increases to reproduce the observed TeV photon flux.   Based on the values of the dynamics of the "Wein" fireball and those values reported after describing  the multiwavelength campaigns for 1ES 1959+650 (May 2006; \citet{2008ApJ...679.1029T}) and Mrk 421 (from 2009 January 19 to 2009 June 1; \citet{2011ApJ...736..131A}), we will analyze the flares presented in both blazars.  We consider the target photons at break energy  $\epsilon_{\gamma b}$= 5 MeV \citep{1999PhR...314..575P, 2004ApJ...601...78I,2002ApJ...565..163I},  the values of normalization energy $\epsilon_0$=1 TeV and Lorentz factors $\Gamma_W$= 8, 12 and 18.\\
\subsection{1ES 1959+650}
The blazar 1ES 1959+650, localized at a distance of $D_z$=210 Mpc (z=0.047), is  one of the best studied high broad-line Lacertae objects (HBL) \citep{2006A&A...455..773V,2014ApJ...797...89A}.   This BL Lac object has a central  BH mass of $\sim2\times 10^8 M_{\odot}$ \citep{2003ApJ...595..624F}. It was first detected in X-rays during the Slew Survey made by Einstein satellite's Imaging Proportional Counter \citep{1992ApJS...80..257E}. Based on the X-ray/radio versus X-ray/optical color diagram, the source was classified as a BL Lac object by Schachter et al. 1993 \citep{1993ApJ...412..541S}.  1ES 1959+650 was further detected in X-rays with ROSAT and BeppoSAX \citep{2002A&A...383..410B}.  In particular, a multiwavelength campaign on 1ES 1959+650 performed by Suzaku and Swift satellites and MAGIC telescope  in 2006 May and other historical data showed the usual double-peak structure on its SED, with the first peak at an energy of $\simeq$ 1 keV and the second one,  at hundreds of GeV.  These observations  with a higher flux level in the first peak than the second one were modeled by the one-zone SSC \citep{2008ApJ...679.1029T}. 
In 2002 May, the X-ray flux increased significantly and both the Whipple \citep{2003ApJ...583L...9H} and High Energy Gamma Ray Astronomy (HEGRA) \citep{2003A&A...406L...9A} collaborations subsequently confirmed an increasing $\gamma$-ray flux as well.   An interesting aspect of the source activity in 2002 was the observation of a so-called "orphan" flare, recorded on June 4 by the Whipple Collaboration \citep{2004ApJ...601..151K,2005ApJ...621..181D} which was observed in $\gamma$-rays without any increased activity in other wavelength simultaneously.   Within the limited statistics, a pair of hours of observations showed the same spectral shape, the fit of a single power law $dN/dE=N_0 \times (E/TeV)^{-\alpha}$ to the differential energy spectrum results in $N_0=(7.4\pm 1.3_{stat}\pm 0.9_{sys}) \times 10^{-11}$ TeV$^{-1}$ cm$^{-2}$ s$^{-1}$  and a spectral index of $\alpha=2.83\pm 0.14_{stat}\pm 0.08_{sys}$.   Whereas the flaring activities recorded during  the month of May were well described by means of SSC emission, the "orphan flare" registered on June 4 has been interpreted in different contexts.\\ 
Using eq. (\ref{pgamma_fit}), we found the best set of parameters for p$\gamma$ interaction, as shown in Table 1.  In addition, we plot the SED with the fitted parameters (see fig. \ref{fit_1es}). 
\begin{center}\renewcommand{\arraystretch}{1}\addtolength{\tabcolsep}{-1pt}\label{fit_1es}
\begin{tabular}{ l c c c}
 \hline \hline
 \scriptsize{} &\scriptsize{Parameter} & \scriptsize{Symbol} & \scriptsize{Value}  \\
\hline
\hline
$\Gamma_{W,ph}$ = 8 & & &\\ 
\hline
\hline
\scriptsize{Proportionality constant} ($10^{-10}\,{\rm erg\,cm^{-2}s^{-1}}$)  &\scriptsize{[0]}                    &   \scriptsize{$A_{p,\gamma} $}      &  \scriptsize{ $1.381\pm 0.231$} \\
\scriptsize{Spectral index}                                                                 &\scriptsize{[1]}                    &   \scriptsize{$\alpha$}                     & \scriptsize{2.716$\pm$ 0.081} \\
\scriptsize{Chi-square/NDF}                                                              &                                          &    \scriptsize{$ \chi^2/{\rm NDF}$}   &  \scriptsize{ $12.84/6.0$} \\
\hline
\hline
$\Gamma_{W,ph}$ = 12 & & &\\ 
\hline
\hline
\scriptsize{Proportionality constant} ($10^{-10}\,{\rm erg\,cm^{-2}s^{-1}}$)  &\scriptsize{[0]}                    &   \scriptsize{$ A_{p,\gamma} $}      &  \scriptsize{ $1.581\pm 0.268$} \\
\scriptsize{Spectral index}                                                                 &\scriptsize{[1]}                    &   \scriptsize{$\alpha$}                     & \scriptsize{2.767$\pm$ 0.056} \\
\scriptsize{Chi-square/NDF}                                                              &                                          &    \scriptsize{$ \chi^2/{\rm NDF}$}   &  \scriptsize{ $12.45/6.0$} \\
\hline
\hline
$\Gamma_{W,ph}$ = 18 & & &\\ 
\hline
\hline
\scriptsize{Proportionality constant} ($10^{-10}\,{\rm erg\,cm^{-2}s^{-1}}$)  &\scriptsize{[0]}                    &   \scriptsize{$ A_{p,\gamma} $}      &  \scriptsize{ $1.853\pm 0.358$} \\
\scriptsize{Spectral index}                                                                 &\scriptsize{[1]}                    &   \scriptsize{$\alpha$}                     & \scriptsize{2.834$\pm$ 0.096} \\
\scriptsize{Chi-square/NDF}                                                              &                                          &    \scriptsize{$ \chi^2/{\rm NDF}$}   &  \scriptsize{ $14.66/6.0$} \\
\hline
\end{tabular}
\end{center}
\begin{center}
\scriptsize{\textbf{Table 1. Parameters obtained after fitting the TeV flare present in 1ES 1959+650 with the $\pi^0$ spectrum.}} \\
\end{center}
From this table, we can see that  for $\Gamma_j=\Gamma_W$=18,  the value of spectral index is equal to that obtained by \citet{2003A&A...406L...9A}. Taking into account  the values used by \citet{2008ApJ...679.1029T} to describe the average SED reasonably well: $\Gamma_j=18$, $r_j= 4.66 \times 10^{16}$ cm and B= $0.25$ G,  the initial conditions of the Wein fireball: temperature $\theta_o=1$, initial radius  $r_o=4\, r_g$ and Lorentz factor $\Gamma_{W,o}=1.2$  and  considering $\Gamma_{W,ph}= \Gamma_j=18$ we obtained the table  2.  The number of HE neutrinos reported in this table was computed with the effective volume of AMANDA detector ($V_{eff}\simeq 10^{-3} {\rm km}^3$) \citep{2003PhRvD..67a2003A}, therefore this number corresponds to those events that could have been expected in this detector.\\
\begin{center}\renewcommand{\arraystretch}{0.5}\addtolength{\tabcolsep}{5pt}
\begin{tabular}{ l c c}
\hline
\hline
\scriptsize{Lorentz factor} &\scriptsize{$\Gamma_{W,ph}=\Gamma_j$}& \scriptsize{18} \\ 
\hline
\hline
\scriptsize{Initial optical depth}  & \scriptsize{$\tau_o$} &\scriptsize{$3375.0$}         \\
\scriptsize{Photospheric radius} ($10^{15}\,{\rm cm}$)  & \scriptsize{$r_{ph}$} &\scriptsize{$6.01$}         \\
\scriptsize{Photospheric temperature}  & \scriptsize{$\theta_{ph}$} &\scriptsize{$0.07$}          \\
\scriptsize{Ratio of total and Eddington luminosity}  & \scriptsize{$L_j/L_{Edd}$} &\scriptsize{$47.34$}         \\
\scriptsize{Photon density} ($10^{11}\,{\rm cm^{-3}}$)  & \scriptsize{$\eta_\gamma$} &\scriptsize{$6.26$}         \\
\scriptsize{Proton luminosity} ($10^{47}\,{\rm erg/s}$)  & \scriptsize{$L_p$} &\scriptsize{$1.17$}         \\
\scriptsize{Maximum escaping proton energy ($10^{20}\,{\rm eV}$)} & \scriptsize{$E_{p,max}$}&\scriptsize{$2.07$}  \\
\scriptsize{Number of UHECRs} & \scriptsize{$N_{UHECR}$}&\scriptsize{$-$}  \\
\scriptsize{Number of HE neutrinos} ($10^{-5}$) & \scriptsize{$N_{ev}$}&\scriptsize{$4.39$}  \\
\hline
\end{tabular}
\end{center}
\begin{center}
\scriptsize{\textbf{Table 2.  Calculated quantities with our model after the fit of the TeV $\gamma$-ray flare in 1ES 1959+650.}}\label{res_1es} \\
\end{center}
From the previous analysis it is important to highlight that this model required a super-Eddington luminosity  $L_j/L_{Edd}=47.34$.   Additionally,  as has been pointed out muons and secondary positrons/electrons are created as charged pion decay products in p$\gamma$ interactions.  We look into the contribution of radiation released by these charged particles.  First of all as shown in fig. \ref{sync_1es},  we calculate and plot the time cooling scales as a function of muon (left-hand figure above) and positron/electron (right-hand figure above) energies. Comparing the time scales of synchrotron radiation and life-time of muons, one can see that synchrotron time scale is much longer than lifetime scale for any muon energy, therefore muons lose negligible energy before they decay. Additionally, comparing both figures above one can see that muons cool down much slower than electrons/positrons, for instance for muons and electrons/positrons with energies $E'_{\mu}=E'_{e^\pm}=$10 TeV the cooling times are $4.43 \times 10^{13}$ s and  $7.95  \times 10^{3}$ s, respectively.   Furthermore,  we calculate the positron/electron synchrotron emission  at  break (199.40 eV)  and maximum (22.09 keV) energies and then plot (figure below) the proportionality constant of the electron/positron spectrum ($A_{syn,\gamma-e^\pm}$) as a function of number density of positrons/electrons ($N_{e^\pm}$).  In this figure,  one can see that for a number density of positrons/electrons 10 cm$^{-3}$,  the  electron/positron synchrotron spectrum  would be $A_{syn,\gamma-e^\pm}\simeq10^{-11}$ erg cm$^{-2}$ s$^{-1}$. The SSC flux can be roughly obtained though the Compton parameter $Y\simeq U_e/U_B $ is $A_{ssc,\gamma-e^\pm}\sim Y A_{syn,\gamma-e^\pm}$, where  electron and magnetic energy densities are $U_e=m_e N_e$ and $U_B=B^2/8\pi$, respectively. Hence, if we consider the number density of electrons/positrons  of 10 cm$^{-3}$ and magnetic field B= 0.25 G,  then the SSC flux is  $A_{ssc,\gamma-e^\pm}\sim 10^{-14}$ erg cm$^{-2}$ s$^{-1}$.\\ 
On the other hand,  target photons from the Wein fireball could also be up-scattered by electrons accelerated  at emitting region.  In this case and taking into account once again the values of electron Lorentz factors reported by \citet{2008ApJ...679.1029T}, ($\gamma_{e,b}=5.7\times 10^4$ and $\gamma_{e, max}=6.0\times 10^5$), the energies of scattered photons can be calculated by $\epsilon^{ic}_{\gamma,i}\simeq  2\gamma^2_{e,i} \epsilon_{\gamma,b}$, with subindex $i$ for break and max.  Considering the values of target photons $\epsilon_{\gamma b} \simeq$ 5 MeV, we obtained that the energies of scattered photons are not in the energy range considered during the flare ($\epsilon^{ic}_{\gamma,break} \simeq3.24\times 10^{16}$ eV and  $\epsilon^{ic}_{\gamma,max}\simeq3.60\times 10^{18}$ eV).
\subsection{Mrk 421}
At a distance of 134.1 Mpc, the BL Lac object Mrk 421 (z=0.03, \citealp{2009ApJ...691L..13D}) with a central BH mass of $\sim 2\times 10^8 M_{\odot}$ \citep{2003ApJ...583..134B}  is one of the closest and brightest sources in the extragalactic X-ray/TeV sky.  This object has been widely studied from radio to $\gamma$-ray  at GeV - TeV energy range. Its SED presents the double-humped shape typical of blazars, with a low energy peak at energies $\simeq$ 1 keV \citep{2008ApJ...677..906F} and the second peak in the VHE regime. As for the other blazars, the low energy peak is interpreted as synchrotron radiation from relativistic electrons within the jet; the existing observations do not allow for an unambiguous discrimination between leptonic and hadronic mechanisms as responsible for the HE emission. For instance,   \citet{2011ApJ...736..131A} found that both a leptonic scenario (inverse Compton scattering of low energy synchrotron photons) and a hadronic scenario (synchrotron radiations of accelerated protons) are able to describe the Mrk 421 SED reasonably well, implying comparable jet powers but very different characteristics for the blazar emission site. \\
Although flaring activities at TeV $\gamma$-ray \citep{1992Natur.358..477P,1996Natur.383..319G,2011ApJ...738...25A, 2002ApJ...575L...9K, 2007ApJ...663..125A, 2005A&A...437...95A, 2002A&A...393...89A, 2003A&A...410..813A,2007ApJ...662..199C, 2014ApJ...782..110A} with  X-ray counterpart \citep{2010PASJ...62L..55I,2009ApJ...691L..13D, 2012ApJ...759...84N, 2009ATel.2292....1K, 2008ATel.1574....1C,2006ATel..848....1L,2013ATel.4974....1B}  have been usually detected, atypical  TeV flares  without  X-ray counterpart have occurred  \citep{2005ApJ...630..130B,2011ApJ...743...62A}.  Firstly, at around  MJD 53033.4 a TeV flare was present when the X-ray flux was low. This X-ray flux seems to have peaked 1.5 days before the  $\gamma$-ray flux, and secondly,  the higher TeV flux during the night of MJD 54589.21 was not accompanied by simultaneous X-ray activity. \\
Using eq. (\ref{pgamma_fit}), we found the best set of parameters for p$\gamma$ interaction, as shown in table 3.  In addition, we plot the broadband SED with the fitted parameters (see fig. \ref{fit_mrk}). 
\begin{center}\renewcommand{\arraystretch}{0.8}\addtolength{\tabcolsep}{5pt}
\begin{tabular}{ l c c c c}
 \hline \hline
 \scriptsize{} &\scriptsize{Param.} & \scriptsize{Symbol} & \scriptsize{Flare 1} & \scriptsize{Flare 2} \\
\hline
\hline
$\Gamma_{W,ph}$ = 8 & & &\\ 
\hline
\hline
\scriptsize{P. constant} ($10^{-10}\,{\rm erg\,cm^{-2}s^{-1}}$)  &\scriptsize{[0]}                    &   \scriptsize{$ A_{p,\gamma} $}      &  \scriptsize{ $5.663\pm 0.479$} &  \scriptsize{ $3.179\pm 0.995$}\\
\scriptsize{Spectral index}                                                                 &\scriptsize{[1]}                    &   \scriptsize{$\alpha$}                     & \scriptsize{3.082$\pm$ 0.091} & \scriptsize{2.786$\pm$ 0.045} \\
\scriptsize{Chi-square/NDF}                                                              &                                          &    \scriptsize{$ \chi^2/{\rm NDF}$}   &  \scriptsize{ $6.711/6.0$} &  \scriptsize{ $33.31/10.0$} \\
\hline
\hline
$\Gamma_{W,ph}$ = 12 & & &\\ 
\hline
\hline
\scriptsize{P. constant} ($10^{-10}\,{\rm erg\,cm^{-2}s^{-1}}$)  &\scriptsize{[0]}                    &   \scriptsize{$ A_{p,\gamma} $}      &  \scriptsize{ $6.138\pm 0.535$} &  \scriptsize{ $4.078\pm 0.132$} \\
\scriptsize{Spectral index}                                                                 &\scriptsize{[1]}                    &   \scriptsize{$\alpha$}                     & \scriptsize{3.143$\pm$ 0.095} & \scriptsize{3.018$\pm$ 0.053} \\
\scriptsize{Chi-square/NDF}                                                              &                                          &    \scriptsize{$ \chi^2/{\rm NDF}$}   &  \scriptsize{ $5.24/6.0$} &  \scriptsize{ $36.15/10.0$} \\
\hline
\hline
$\Gamma_{W,ph}$ = 18 & & &\\ 
\hline
\hline
\scriptsize{P. constant} ($10^{-10}\,{\rm erg\,cm^{-2}s^{-1}}$)  &\scriptsize{[0]}                    &   \scriptsize{$ A_{p,\gamma} $}      &  \scriptsize{ $8.810\pm 0.854$} &  \scriptsize{ $5.867\pm 0.189$} \\
\scriptsize{Spectral index}                                                                 &\scriptsize{[1]}                    &   \scriptsize{$\alpha$}                     & \scriptsize{3.412$\pm$ 0.111} & \scriptsize{3.272$\pm$ 0.053} \\
\scriptsize{Chi-square/NDF}                                                              &                                          &    \scriptsize{$ \chi^2/{\rm NDF}$}   &  \scriptsize{ $3.148/6.0$} &  \scriptsize{ $18.96/10.0$} \\
\hline
\end{tabular}
\end{center}
\begin{center}
\scriptsize{\textbf{Table 3. Parameters obtained after fitting the TeV flare presented in Mrk 421 with the $\pi^0$ spectrum.}}\label{fit_mrk}  \\
\end{center}
Requiring the values used by \citet{2011ApJ...736..131A}  to explain the whole spectrum: $\Gamma_j=12$, $r_j =3.11 \times 10^{16}$ cm and B = $50$ G, the initial conditions of Wein fireball: temperature $\theta_o=2$, initial radius  $r_o=1.5\, r_g$ and Lorentz factor $\Gamma_{W,o}=2$  and  considering $\Gamma_{W,ph}= \Gamma_j=12$ we obtained table 4.  The value of HE neutrinos  reported was calculated with the effective volume of Ice Cube ($V_{eff}\simeq 1\, {\rm km}^3$) \citep{2013arXiv1304.5356I}.\\
\begin{center}\renewcommand{\arraystretch}{0.8}\addtolength{\tabcolsep}{5pt}
\begin{tabular}{ l c c c}
\hline
\hline
\scriptsize{} & \scriptsize{} & \scriptsize{Flare 1} & \scriptsize{Flare 2} \\
\hline
\hline
\scriptsize{Lorentz factor} &\scriptsize{$\Gamma_{W,ph}=\Gamma_j$}& \scriptsize{12}& \scriptsize{12} \\ 
\hline
\hline
\scriptsize{Initial optical depth}  & \scriptsize{$\tau_o$} &\scriptsize{$216.0$} &\scriptsize{$216.0$}         \\
\scriptsize{Photospheric radius} ($10^{15}\,{\rm cm}$)  & \scriptsize{$r_{ph}$} &\scriptsize{$5.31$}&\scriptsize{$5.31$}         \\
\scriptsize{Photospheric temperature}  & \scriptsize{$\theta_{ph}$} &\scriptsize{$0.33$}&\scriptsize{$0.33$}          \\
\scriptsize{Ratio of total and Eddington luminosity}  & \scriptsize{$L_j/L_{Edd}$} &\scriptsize{$15.21$} &\scriptsize{$15.21$}         \\
\scriptsize{Photon density} ($10^{11}\,{\rm cm^{-3}}$)  & \scriptsize{$\eta_\gamma$} &\scriptsize{$0.85$}&\scriptsize{$0.85$}         \\
\scriptsize{Proton luminosity} ($10^{45}\,{\rm erg/s}$)  & \scriptsize{$L_p$} &\scriptsize{$2.86$} &\scriptsize{$8.89$}         \\
\scriptsize{Maximum escaping proton energy  ($10^{21}\,{\rm eV}$)} & \scriptsize{$E_{p,max}$}&\scriptsize{$ 18.47$}&\scriptsize{$18.47$}  \\
\scriptsize{Number of UHECRs ($10^{-3}$)} & \scriptsize{$N_{UHECR}$}&\scriptsize{$0.10$} &\scriptsize{$1.42$}  \\
\scriptsize{Number of HE neutrinos} & \scriptsize{$N_{ev}$}&\scriptsize{$0.12$}&\scriptsize{$0.09$}  \\
\hline
\end{tabular}
\end{center}
\begin{center}
\scriptsize{\textbf{Table 4.  Calculated quantities with our model after the fit of the TeV $\gamma$-ray flare in Mrk 421.}}\label{res_mrk} \\
\end{center}
As pointed out for the blazar 1ES 1959+650,  a super-Eddington luminosity  $L_j/L_{Edd}=15.21$ is necessary to describe the TeV flares for Mrk 421.   By considering again the values reported by  \citet{2011ApJ...736..131A}, we plot  (see fig. \ref{sync_mrk})   the time cooling scales of muon (left-hand figure above) and positron/electron (right-hand figure below), and the proportionality constant of the positron/electron spectrum ($A_{syn,\gamma-e^\pm}$) (figure below). Comparing the time scales of muons (synchrotron radiation and life-time), we can see that both scales start to be similar at E$'_\mu=10^{16}$ eV, then in this energy range muons could radiate before they decay.   From figures (\ref{fit_mrk}) and (\ref{sync_mrk}) and taking into account the break photon energies radiating by positron/electrons ($\epsilon_{\gamma,br}\simeq 73.64$ eV and $\epsilon_{\gamma,max}\simeq$ 13.09 keV), we can see that for the number density of positrons/electrons greater than $N_{e^\pm}\geq 10^{2}$ cm$^{-3}$ the contribution of the synchrotron flux to the SED could be noticeable ($\nu F_\nu \simeq 10^{-11}$ erg cm$^{-2}$ s$^{-1}$}).   The SSC flux for this case would be $A_{ssc,\gamma-e^\pm}\sim 10^{-17}$ erg cm$^{-2}$ s$^{-1}$.\\
Doing a similar analysis on 1ES 1959+650, electron accelerated in the emitting region could up-scatter photons from the Wein fireball. In this case, from the values of the electron Lorentz factors reported by  \citet{2011ApJ...736..131A}  ($\gamma_{e,min}=7\times 10^2$ and $\gamma_{e, max}=4.0\times 10^4$), photons in the energy range  $\sim$10 TeV $\leq\epsilon^{ic}_{\gamma} \leq$1.6 $\times 10^4$ TeV would be expected. Eventually, these TeV photons will undergo pair productions with the 5 MeV photons  in the Wein fireball or less energetic photons around. 
\subsection{Neutrino Oscillations}
\begin{figure}
\vspace{0.5cm}
{\centering
\resizebox*{1.03\textwidth}{0.7\textheight}
{\includegraphics{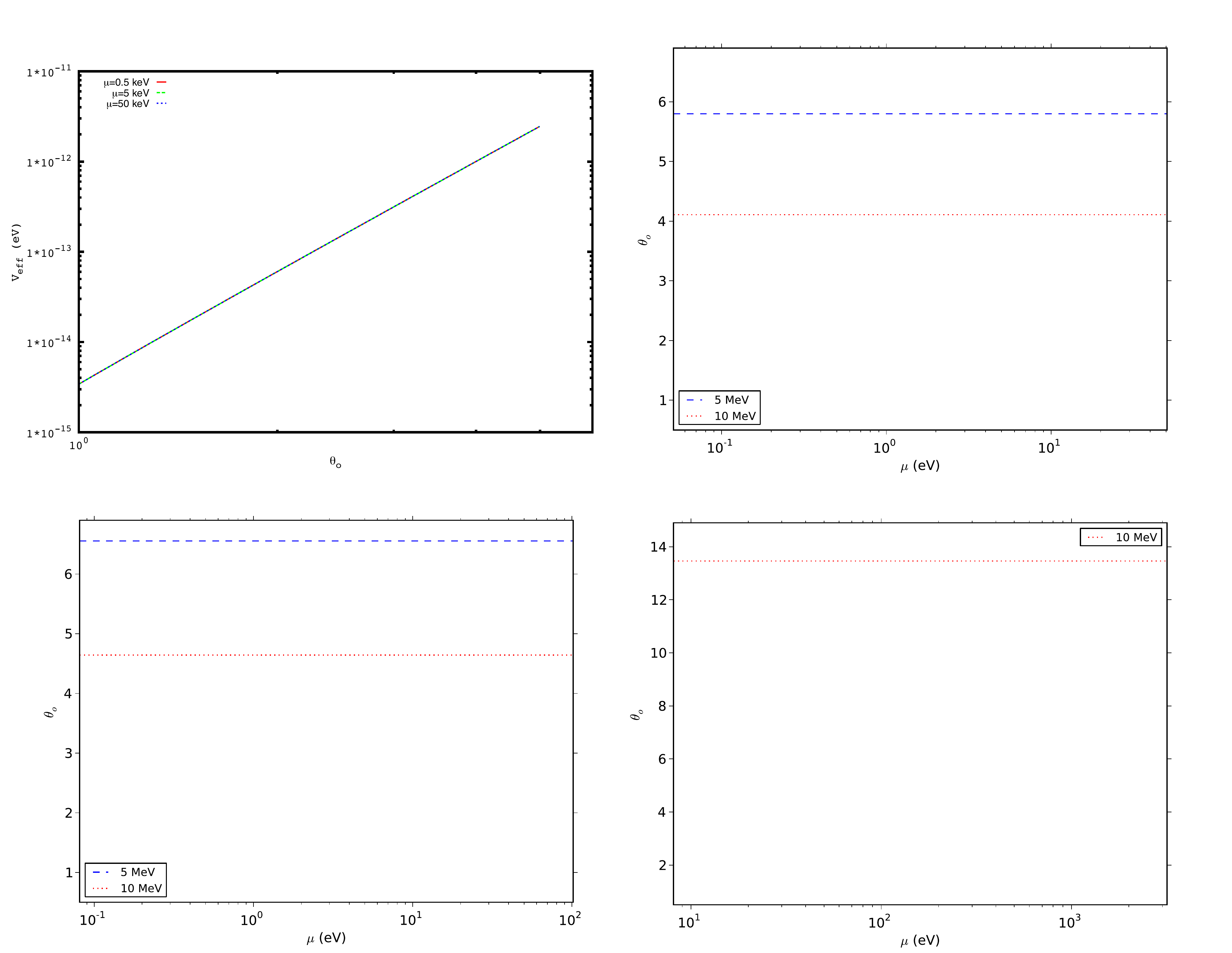}}
}
\caption{Neutrino effective potential in the weak magnetic field regime (left-hand figure above) and contour lines of normalized temperature ($\theta_o$), chemical potential ($\mu$) and  neutrino energy (E$_\nu$) for which the resonance condition is satisfied for $N_e - \bar{N}_e=0$. We have applied the neutrino effective potential with magnetic field  B=1 G and used the best fit values of the two; solar (right-hand figure above), atmospheric (left-hand figure below), and three (right-hand figure below) neutrino mixing.}
\label{res_osc_N0}
\end{figure}

\begin{figure}
\vspace{0.5cm}
{\centering
\resizebox*{1.03\textwidth}{0.7\textheight}
{\includegraphics{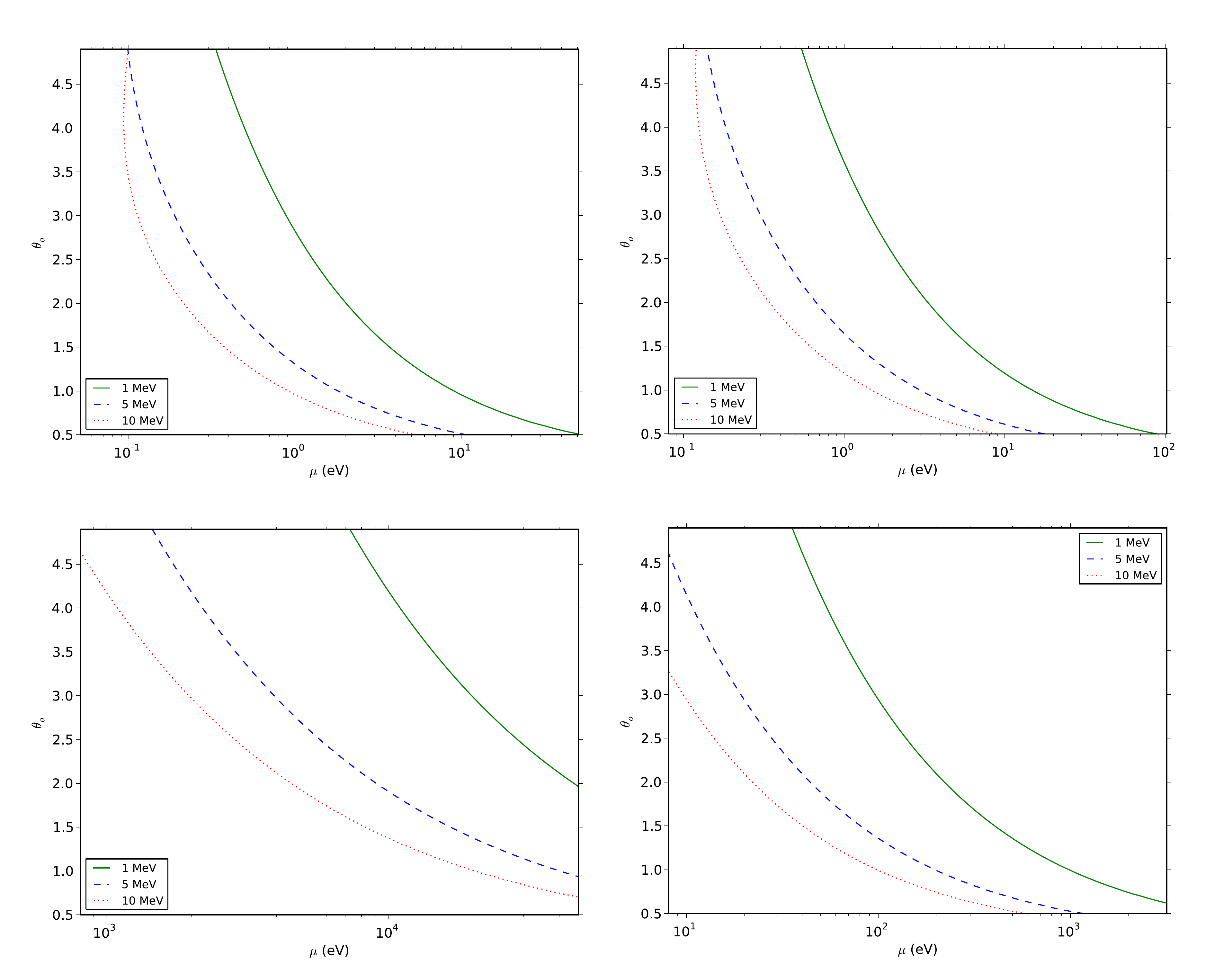}}
}
\caption{Contour lines of normalized temperature ($\theta_o$), chemical potential ($\mu$) and  neutrino energy (E$_\nu$) for which the resonance condition is satisfied. We have applied the neutrino effective potential with magnetic field  B=1 G and used the best fit values of the two; solar (left-hand figure above), atmospheric (right-hand figure above), and accelerator (left-hand figure below), and three (left-hand figure below) neutrino mixing.}
\label{res_osc}
\end{figure}

\begin{figure}
\vspace{0.5cm}
{\centering
\resizebox*{1.03\textwidth}{0.35\textheight}
{\includegraphics{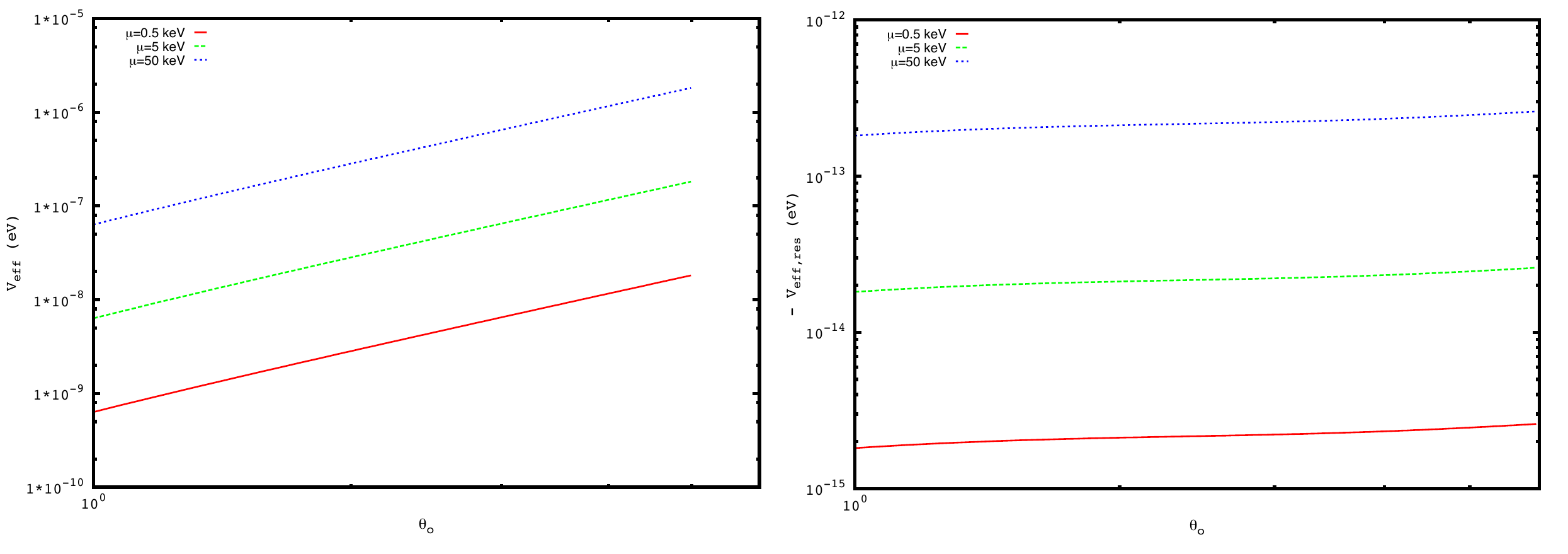}}
}
\caption{Neutrino effective potential in the weak magnetic field regime (left panel)  and the magnetic field contribution ($V_{eff} {\rm (B=1 G)}-V_{eff}{\rm  (B=0)}$) (right panel) as a function of temperature ($\theta_o$). We use a neutrino energy $E_\nu=1$ MeV and chemical potential $\mu=$ 0.5, 5 and 50 keV.}
\label{potencial}
\end{figure}

\begin{figure}
\vspace{0.5cm}
{\centering
\resizebox*{0.6\textwidth}{0.35\textheight}
{\includegraphics{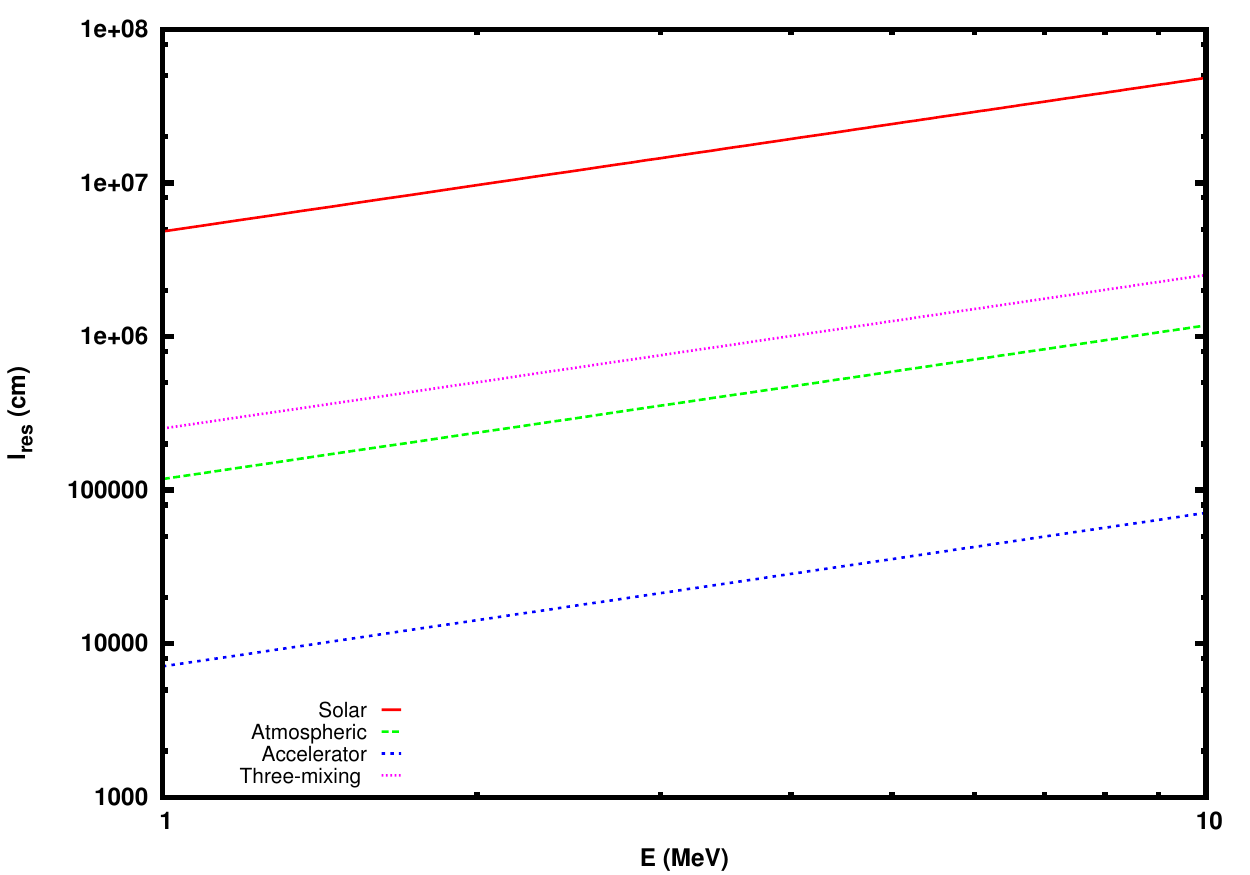}}
}
\caption{Resonance length as a function of neutrino energy for the best fit parameters of two- (solar, atmospheric and accelerator) and three-neutrino mixing.}
\label{reson}
\end{figure}

We have studied the active-active neutrino process in the Wein fireball. We have used the neutrino effective potential at the weak field limit obtained by \citet{2014ApJ...787..140F}.  It depends on temperature ($\theta$),  chemical potential ($\mu$), magnetic field ($B$) and neutrino energy (E$_\nu$) for $N_e=\bar{N}_e$ (eq. \ref{Veffw_N0}) and $N_e\neq\bar{N}_e$ (eq. \ref{Veffw}).  As shown in fig. \ref{res_osc_N0}, we plot the neutrino effective potential at the weak field limit for $N_e=\bar{N}_e$.  Taking into account the range of values of initial temperature 1 $\leq\theta_o\leq$ 4,   one can see that the neutrino effective potential for $\mu=0.5$, 5 and 50 keV lies in the range from $\simeq 10^{-9}$ eV to $\simeq 10^{-25}$ eV.  It can be seen that the neutrino effective potential is negative and then, due to its negativity ($V_{eff}<0$ eV), only anti-neutrinos can oscillate resonantly. From the resonance conditions (eqs. \ref{reso2} and \ref{reso3}),  we have obtained the contour plots of initial temperature and chemical potential  (fig. \ref{res_osc}).  Our analysis shows that chemical potential lies in the range 0.1 eV $\leq \mu\leq$ 50 eV for solar parameters, 0.09 eV$\leq \mu\leq$ 100 eV for atmospheric parameters,  and  10 eV $\leq \mu\leq$ $3.3\times 10^{3}$  eV for three-neutrino mixing.  By considering a CP-asymmetric $\gamma$ and $e^{\pm}$ fireball, where the excess of electrons could come from the electrons associated with the baryons within the fireball, we plot the neutrino effective potential at the weak field limit  (eq. \ref{Veffw})  (above) and  the magnetic field contribution (below), i.e. by subtracting the effective potential with B=0, as a function of temperature, as shown in fig. \ref{potencial}. Taking into account the range of values of initial temperature 1 $\leq\theta_o\leq$ 4,   one can see that the neutrino effective potential for $\mu=5$ keV lies in the range from $7.6\times 10^{-9}$ eV to $3\times 10^{-7}$ eV (left panel) and the magnetic field contribution is the opposite as compared to the plasma contribution for any value of $\mu$ (right panel). From fig. \ref{potencial} (left panel)  you can see that the neutrino effective potential is positive and then, due to its positivity ($V_{eff}>7.6\times 10^{-9}$ eV), neutrinos can oscillate resonantly.  From the resonance conditions (eqs. \ref{reso2} and \ref{reso3}),  we have obtained the contour plots of initial temperature and chemical potential  (fig. \ref{res_osc}).  Our analysis shows that chemical potential lies in the range 0.1 eV $\leq \mu\leq$ 50 eV for solar parameters, 0.19 eV$\leq \mu\leq$ 95 eV for atmospheric parameters, 0.8$\times 10^{3}$ eV $\leq \mu\leq$ $4.3\times 10^{4}$ eV for accelerator parameters and  8 eV $\leq \mu\leq$ $3.2\times 10^{3}$  eV for three-neutrino mixing. One can observe that initial temperature is a decreasing function of chemical potential which gradually increases as neutrino energy  decreases.  In addition, we have plotted the resonance lengths, as shown  in fig. \ref{reson}. As we can see the resonance lengths  lie in the range  $\simeq$ 10$^4$-10$^8$ cm for neutrino energy in the range 1 MeV $\leq$ E$_{\nu}$ $\leq$ 10 MeV.  One can observe that regardless the values of mixing parameters and neutrino energy, the resonance lengths are much less than $\sim r_g$, therefore neutrinos will oscillate  resonantly many times before leaving the Wein fireball.\\ 
Taking into account the neutrino effective potential (eq. \ref{Veffw}) and the vacuum oscillation effects between the source and Earth (eq. \ref{matrixosc}),  we estimate the flavor ratio expected on Earth for four neutrino energies  ($E_{\nu}=1$, 3, 5 and 10 MeV), as shown in table 5.  It is important to clarify that  we consider a small amount of baryons, then in addition to electron-positron annihilation we take into account inverse beta decay and nucleonic bremsstrahlung. For  the electron-positron annihilation and nucleonic bremsstrahlung processes,  neutrinos of all  flavors  are created with equal probability  whereas for the inverse beta decay process just electron neutrinos can be generated.
\begin{center}\renewcommand{\arraystretch}{1.}\addtolength{\tabcolsep}{5pt}
\begin{tabular}{ c c c c }
\hline
\hline
\scriptsize{Energy} & \scriptsize{$\phi_{\nu_e}:\phi_{\nu_\mu}:\phi_{\nu_\tau}$} & \scriptsize{$\phi_{\nu_e}:\phi_{\nu_\mu}:\phi_{\nu_\tau}$} & \scriptsize{$\phi_{\nu_e}:\phi_{\nu_\mu}:\phi_{\nu_\tau}$}  \\
\scriptsize{(MeV)} & \scriptsize{$r_o=1.0\times 10^{14}\,cm$} & \scriptsize{$r_o=5.0\times 10^{14}\,cm$} & \scriptsize{$r_o=1.0\times 10^{15}\,cm$} \\
\hline
\hline
\scriptsize{$\theta_o=2$} & & &\\
\scriptsize{1} & \scriptsize{$1.2383:0.8870:0.8747$}& \scriptsize{$1.2387: 0.8868: 0.8745$}&\scriptsize{$1.2383: 0.8703: 0.8747$} \\ 
\scriptsize{3} & \scriptsize{$1.2389:0.8868: 0.8744$}& \scriptsize{$1.2390: 0.8867: 0.8743$} &\scriptsize{$1.2388: 0.8868: 0.8744$}\\ 
\scriptsize{5} & \scriptsize{$1.2391:0.8866: 0.8743$}& \scriptsize{$1.2390: 0.8867: 0.8743$} &\scriptsize{$1.2389: 0.8867: 0.8744$}\\ 
\scriptsize{10} & \scriptsize{$1.2391:0.8866: 0.8743$}& \scriptsize{$1.2391: 0.8867: 0.8743$} &\scriptsize{$1.2391: 0.8866: 0.8743$}\\ 
\hline
\hline
\scriptsize{$\theta_o=4$} & & &\\
\scriptsize{1} & \scriptsize{$1.2389 :0.8867: 0.8744$}& \scriptsize{$1.2389: 0.8867: 0.8744$}&\scriptsize{$1.2391: 0.8867: 0.8743$} \\ 
\scriptsize{3}& \scriptsize{$1.2391 :0.8866: 0.8743$}& \scriptsize{$1.2391: 0.8866: 0.8743$}  &\scriptsize{$1.2391: 0.8864: 0.8743$}\\ 
\scriptsize{5} & \scriptsize{$1.2391 :0.8866: 0.8743$}& \scriptsize{$1.2391: 0.8866: 0.8743$}&\scriptsize{$1.2391: 0.8866: 0.8743$} \\ 
\scriptsize{10} & \scriptsize{$1.2391: 0.8866: 0.8743$}& \scriptsize{$1.2391: 0.8866: 0.8743$} &\scriptsize{$1.2391: 0.8866: 0.8743$}\\ 
\hline
\end{tabular}
\end{center}
\begin{center}
\scriptsize{\textbf{Table 5.  The flavor ratio of thermal neutrinos expected on Earth. For this calculation, we use the neutrino energies  E$_{\nu}$=1, 3, 5 and 10 MeV, initial radius $r_o=1.0\times 10^{14}$, $5.0\times 10^{14}$  and $1.0\times 10^{15}$ cm  and initial temperatures $\theta_o$= 2 and 4.}} \label{flaratio}  \\
\end{center}
In addition to the flavor ratio of thermal neutrinos,  we calculate the number of events that could be expected.   From eq. (\ref{ther_neu}) and taking into account the volume of Hyper-Kamiokande experiment $\sim 10^{5} {\rm m^{3}}$  \citep{2011arXiv1109.3262A},  the distance of the BL Lac objects considered $D_z=100$ Mpc, luminosity $L_\nu\simeq10^{47}$ erg/s and neutrino energy $E_\nu=3$ MeV, we would expected  $8.14\times 10^{-5}$ events. \\
\section{Summary and conclusions}
We have considered the target photons coming from the Wein fireball, the geometry of the emitting region and the protons accelerated at emitting region in order to explain through p$\gamma$ interactions the "orphan" TeV flares presented in 1ES1950+650 and Mrk421.  Taking into account the values reported by the semianalytical description of the Wein fireball dynamics \citep{2004ApJ...601...78I, 2002ApJ...565..163I} and required for the descriptions of the SEDs of 1ES1950+650 \citep{2008ApJ...679.1029T} and  Mrk421 \citep{2011ApJ...736..131A},  we have naturally interpreted the TeV "orphan" flare as $\pi^0$ decay products, as shown in figs.  \ref{fit_1es} and \ref{fit_mrk}.   Using the values of these parameters in our model for  $\Gamma_{W,ph}=\Gamma_j$ we have computed  optical depth ($\tau_o$), photospheric radius ($r_{ph}$), photospheric temperature ($\theta_{ph}$), ratio of total and Eddington Luminosity ($L_j/L_{Edd}$), photon density ($\eta_\gamma$), proton luminosity ($L_p$), number of UHECRs ($N_{UHECR}$), number of HE neutrino ($N_{ev}$), as shown in tables 2 and 4.  From these tables one can see that in both cases, this model which has been assumed a spherical symmetry  'Wein' fireball,  required a super-Eddington luminosity.   However, the required total luminosity becomes much smaller if the outflow is collimated within a small opening angle, as observed.  It is important to clarify that when the jet is collimated to the solid angle $\Omega_j$ \citep{1999Natur.401..891J, Sauty:2001wx,2014arXiv1410.0679K}, the dynamics will not be much different from the spherical 'Wein' fireball and then the total luminosity of the jet becomes smaller by a factor $\Omega_j/4\pi$.  For a typical value of solid angle $\Omega\simeq 10^{-2}$, the required total luminosity can be much smaller than the Eddington Luminosity.  Future simulations will be needed to solve the collimation problem \citep{2004ApJ...601...78I,2002ApJ...565..163I}. \\
We have showed that protons in the emitting region  could be accelerated up to $2.08 \times 10^{20}$ eV and $1.84\times 10^{22}$ eV for 1ES 1959+650 and Mrk 421, respectively, then assuming that proton spectrum can be extended at energies greater than $57$ EeV and  taking into account the GZK limit \citep{1966PhRvL..16..748G,1966JETPL...4...78Z} we estimated the number of UHECRs expected from Mrk421 in TA experiment.  Taking into account that 30\%  of proton fraction will survive to a distance of 134 Mpc \citep{2011ARA&A..49..119K},  UHECRs  cannot be expected during this flare as shown in the table 4.\\
In addition, correlating the $\gamma$-ray and neutrino fluxes we have computed the number of neutrinos expected on AMANDA and IceCube, for 1ES 1959+650 and Mrk 421, respectively. In both cases the number of neutrinos would have been  $4\times 10^{-5}$ and 0.09 for 1ES 1959+650 and Mrk 421, which is in agreement with the non-detection of HE neutrinos in the direction of 1ES 1959+650 and Mrk 421 (see. fig. \ref{skymap}).\\
We have studied the active-active neutrino oscillation processes for two- and three-neutrino mixing in the Wein fireball. We have studied the resonance condition for a neutral plasma (eq. \ref{Veffw_N0}) and CP-asymmetric plasma  (eq. \ref{Veffw}).  We found that for a neutral plasma the chemical potential lies in the range 0.1 eV $\leq \mu\leq$ 50 eV for solar parameters, 0.09 eV$\leq \mu\leq$ 100 eV for atmospheric parameters,  and  10 eV $\leq \mu\leq$ $3.3\times 10^{3}$  eV for three-neutrino mixing and for a charged-asymmetric plasma the chemical potential lies in the range 0.1 eV $\leq \mu\leq$ 50 eV for solar parameters, 0.19 eV$\leq \mu\leq$ 95 eV for atmospheric parameters, 0.8$\times 10^{3}$ eV $\leq \mu\leq$ $4.3\times 10^{4}$ eV for accelerator parameters and  8 eV $\leq \mu\leq$ $3.2\times 10^{3}$  eV for three-neutrino mixing. Considering the neutrino oscillations in a CP-asymmetric Wein fireball and in vacuum, we have computed the flavor ratio expected on Earth, as shown in table 5.   In addition, considering a volume of the Hyper-Kamiokande experiment \citep{2011arXiv1109.3262A}  we have computed that the number of thermal neutrinos expected is of the order of $10^{-5}$. However, if the source is of the order of Mpc with a luminosity $\geq 10^{47}$ erg/sec, a few events could be expected on this experiment.\\
We have proposed that the observations of "orphan" TeV flares would be directly related with the numbers of thermal and HE neutrinos, UHECR events and flavor thermal neutrino ratios,  as calculated in tables 2, 4 and 5, respectively. Although the numbers of these events were less than one, improvements in the sensibility of detectors for cosmic rays as well as  new techniques for detecting neutrino oscillations will allow us  in the near future to confirm our estimates.\\
 Recently, Murase et al. (2014) studied high-energy neutrino emission  in the  inner jets of radio-loud AGN \citep{2014arXiv1403.4089M}.   Considering the photohadronic interactions the authors explored the effects of external photon fields (broadline and dust radiation)  on neutrino production in blazars.  They found that the diffuse neutrino intensity from radio-loud AGN is dominated by blazars with $\gamma$-ray luminosity of $\gtrsim 10^{48} {\rm erg s^{-1}}$, and the arrival directions of neutrinos in the energy range $\sim$ 1 - 100 PeV  correlate with the luminous blazars detected by Fermi.  In our model we studied  high-energy neutrino emission correlated with the observations of "orphan" TeV flares presented in blazars  1ES1950+650 and Mrk421,  when these atypical observations are naturally explained through the photohadronic interactions between the Wein fireball photon field and the protons accelerated at emitting region.  We found that although neutrinos in the energy range of $\sim$ 1- 100 TeV are generated together with these atypical observations, a $\gamma$-ray luminosity of  $\gtrsim 10^{47}$ erg/sec is necessary to expect more than one event in IceCube observatory.  In both models one can see  similar conclusions, although the unusual TeV $\gamma$-ray and neutrino correlations have been only present in two blazars.  In a forthcoming paper we will present a more detailed neutrino estimation from the Wein fireball model.\\ 
It is important to say that although VHE (E $\geq$ 100 GeV) $\gamma$-rays coming from these pion decay products can be attenuated through pair production with low energy photons from the diffuse extragalactic photon fields in the ultraviolet (UV) to far-infrared (FIR) (commonly referred to as extragalactic background light ; EBL), the attenuation of the TeV photon flux  for 1ES1959+650 and Mrk 421 sources which have a small redshift (z $\ll$ 0.2) is low \citep{2008IJMPD..17.1515R}.\\
Finally,  correlations of TeV $\gamma$- and X-ray fluxes in flaring activities  by  observatories like High Altitude Water Cherenkov (HAWC) which will have a sensitivity of $5 \times 10^{-13}\, {\rm cm}^{-2} {\rm s}^{-1}$ ($\sim$ 50 mCrab) above 2 TeV \citep{2013APh....50...26A} and satellites like  Nuclear Spectroscopy Telescope array (NuSTAR) with a sensitivity of  2$\times 10^{15}$ erg cm$^{-2} {\rm s}^{-1}$ (1$\times 10^{14}$ erg cm$^{-2} s^{-1}$) for an energy range  6 - 10 keV (10 - 30 keV)\citep{2013ApJ...770..103H} and Swif-XRT with a sensitivity of  3$\times 10^{13}$ erg cm$^{-2} s^{-1}$ in the range 0.3 - 10 keV\citep{2014ApJS..210....8E} will provide information on the amount of hadronic acceleration in blazars as well as HE neutrino fluxes.  On the other hand,  taking into account the energy range of both the 37 events detected by IceCube  \citep{2013arXiv1311.5238I, 2013arXiv1304.5356I} and operation of HAWC (100 GeV - 100 TeV), and considering the relation between the neutrinos and photons from p$\gamma$ interactions $E_\nu\simeq E_\gamma/2$,   HAWC will provide unparalleled  sensitivity to co-relate the TeV flares and HE neutrinos. In addition, it is significant to say that  NuSTAR and Swift-XRT could put lower limits on the synchrotron flux from secondary electrons/positrons produced by $\pi^+$ decay products thus confirming our model, as shown in figs. \ref{sync_1es} and \ref{sync_mrk}.\\
Further observations are needed to understand this atypical TeV flaring activity.
\section*{Acknowledgements}
We thank to Bing Zhang, Francis Halzen, Barbara Patricelli, Antonio Marinelli, Tyce DeYoung and William Lee  for useful discussions, Matthias Beilicke for sharing with us the data, Anita M$\ddot{\rm u}$cke and Antonio Galv\'an for helping us to use the SOPHIA program and the TOPCAT team for the useful sky-map tools.  This work was supported by Luc Binette scholarship and the projects IG100414 and Conacyt 101958.
%

\begin{thebibliography}{124}%
\makeatletter
\providecommand \@ifxundefined [1]{%
 \@ifx{#1\undefined}
}%
\providecommand \@ifnum [1]{%
 \ifnum #1\expandafter \@firstoftwo
 \else \expandafter \@secondoftwo
 \fi
}%
\providecommand \@ifx [1]{%
 \ifx #1\expandafter \@firstoftwo
 \else \expandafter \@secondoftwo
 \fi
}%
\providecommand \natexlab [1]{#1}%
\providecommand \enquote  [1]{``#1''}%
\providecommand \bibnamefont  [1]{#1}%
\providecommand \bibfnamefont [1]{#1}%
\providecommand \citenamefont [1]{#1}%
\providecommand \href@noop [0]{\@secondoftwo}%
\providecommand \href [0]{\begingroup \@sanitize@url \@href}%
\providecommand \@href[1]{\@@startlink{#1}\@@href}%
\providecommand \@@href[1]{\endgroup#1\@@endlink}%
\providecommand \@sanitize@url [0]{\catcode `\\12\catcode `\$12\catcode
  `\&12\catcode `\#12\catcode `\^12\catcode `\_12\catcode `\%12\relax}%
\providecommand \@@startlink[1]{}%
\providecommand \@@endlink[0]{}%
\providecommand \url  [0]{\begingroup\@sanitize@url \@url }%
\providecommand \@url [1]{\endgroup\@href {#1}{\urlprefix }}%
\providecommand \urlprefix  [0]{URL }%
\providecommand \Eprint [0]{\href }%
\providecommand \doibase [0]{http://dx.doi.org/}%
\providecommand \selectlanguage [0]{\@gobble}%
\providecommand \bibinfo  [0]{\@secondoftwo}%
\providecommand \bibfield  [0]{\@secondoftwo}%
\providecommand \translation [1]{[#1]}%
\providecommand \BibitemOpen [0]{}%
\providecommand \bibitemStop [0]{}%
\providecommand \bibitemNoStop [0]{.\EOS\space}%
\providecommand \EOS [0]{\spacefactor3000\relax}%
\providecommand \BibitemShut  [1]{\csname bibitem#1\endcsname}%
\let\auto@bib@innerbib\@empty
\bibitem [{\citenamefont {{Bloom}}\ and\ \citenamefont
  {{Marscher}}(1996)}]{1996ApJ...461..657B}%
  \BibitemOpen
  \bibfield  {author} {\bibinfo {author} {\bibfnamefont {S.~D.}\ \bibnamefont
  {{Bloom}}}\ and\ \bibinfo {author} {\bibfnamefont {A.~P.}\ \bibnamefont
  {{Marscher}}},\ }\href {\doibase 10.1086/177092} {\bibfield  {journal}
  {\bibinfo  {journal} {\apj}\ }\textbf {\bibinfo {volume} {461}},\ \bibinfo
  {pages} {657} (\bibinfo {year} {1996})}\BibitemShut {NoStop}%
\bibitem [{\citenamefont {{Tavecchio}}\ \emph {et~al.}(1998)\citenamefont
  {{Tavecchio}}, \citenamefont {{Maraschi}},\ and\ \citenamefont
  {{Ghisellini}}}]{1998ApJ...509..608T}%
  \BibitemOpen
  \bibfield  {author} {\bibinfo {author} {\bibfnamefont {F.}~\bibnamefont
  {{Tavecchio}}}, \bibinfo {author} {\bibfnamefont {L.}~\bibnamefont
  {{Maraschi}}}, \ and\ \bibinfo {author} {\bibfnamefont {G.}~\bibnamefont
  {{Ghisellini}}},\ }\href {\doibase 10.1086/306526} {\bibfield  {journal}
  {\bibinfo  {journal} {\apj}\ }\textbf {\bibinfo {volume} {509}},\ \bibinfo
  {pages} {608} (\bibinfo {year} {1998})},\ \Eprint
  {http://arxiv.org/abs/arXiv:astro-ph/9809051} {arXiv:astro-ph/9809051}
  \BibitemShut {NoStop}%
\bibitem [{\citenamefont {{Ghisellini}}\ \emph {et~al.}(1998)\citenamefont
  {{Ghisellini}}, \citenamefont {{Celotti}}, \citenamefont {{Fossati}},
  \citenamefont {{Maraschi}},\ and\ \citenamefont
  {{Comastri}}}]{1998MNRAS.301..451G}%
  \BibitemOpen
  \bibfield  {author} {\bibinfo {author} {\bibfnamefont {G.}~\bibnamefont
  {{Ghisellini}}}, \bibinfo {author} {\bibfnamefont {A.}~\bibnamefont
  {{Celotti}}}, \bibinfo {author} {\bibfnamefont {G.}~\bibnamefont
  {{Fossati}}}, \bibinfo {author} {\bibfnamefont {L.}~\bibnamefont
  {{Maraschi}}}, \ and\ \bibinfo {author} {\bibfnamefont {A.}~\bibnamefont
  {{Comastri}}},\ }\href {\doibase 10.1046/j.1365-8711.1998.02032.x} {\bibfield
   {journal} {\bibinfo  {journal} {\mnras}\ }\textbf {\bibinfo {volume}
  {301}},\ \bibinfo {pages} {451} (\bibinfo {year} {1998})},\ \Eprint
  {http://arxiv.org/abs/arXiv:astro-ph/9807317} {arXiv:astro-ph/9807317}
  \BibitemShut {NoStop}%
\bibitem [{\citenamefont {{Holder}}\ and\ \citenamefont
  {et~al.}(2003)}]{2003ApJ...583L...9H}%
  \BibitemOpen
  \bibfield  {author} {\bibinfo {author} {\bibfnamefont {J.}~\bibnamefont
  {{Holder}}}\ and\ \bibinfo {author} {\bibnamefont {et~al.}},\ }\href
  {\doibase 10.1086/367816} {\bibfield  {journal} {\bibinfo  {journal} {\apjl}\
  }\textbf {\bibinfo {volume} {583}},\ \bibinfo {pages} {L9} (\bibinfo {year}
  {2003})},\ \Eprint {http://arxiv.org/abs/arXiv:astro-ph/0212170}
  {arXiv:astro-ph/0212170} \BibitemShut {NoStop}%
\bibitem [{\citenamefont {{Krawczynski}}\ and\ \citenamefont
  {et~al.}(2004)}]{2004ApJ...601..151K}%
  \BibitemOpen
  \bibfield  {author} {\bibinfo {author} {\bibfnamefont {H.}~\bibnamefont
  {{Krawczynski}}}\ and\ \bibinfo {author} {\bibnamefont {et~al.}},\ }\href
  {\doibase 10.1086/380393} {\bibfield  {journal} {\bibinfo  {journal} {\apj}\
  }\textbf {\bibinfo {volume} {601}},\ \bibinfo {pages} {151} (\bibinfo {year}
  {2004})},\ \Eprint {http://arxiv.org/abs/arXiv:astro-ph/0310158}
  {arXiv:astro-ph/0310158} \BibitemShut {NoStop}%
\bibitem [{\citenamefont {{Daniel}}\ and\ \citenamefont
  {et~al.}(2005)}]{2005ApJ...621..181D}%
  \BibitemOpen
  \bibfield  {author} {\bibinfo {author} {\bibfnamefont {M.~K.}\ \bibnamefont
  {{Daniel}}}\ and\ \bibinfo {author} {\bibnamefont {et~al.}},\ }\href
  {\doibase 10.1086/427406} {\bibfield  {journal} {\bibinfo  {journal} {\apj}\
  }\textbf {\bibinfo {volume} {621}},\ \bibinfo {pages} {181} (\bibinfo {year}
  {2005})},\ \Eprint {http://arxiv.org/abs/arXiv:astro-ph/0503085}
  {arXiv:astro-ph/0503085} \BibitemShut {NoStop}%
\bibitem [{\citenamefont {{B{\l}a{\.z}ejowski}}\ and\ \citenamefont
  {et~al.}(2005)}]{2005ApJ...630..130B}%
  \BibitemOpen
  \bibfield  {author} {\bibinfo {author} {\bibfnamefont {M.}~\bibnamefont
  {{B{\l}a{\.z}ejowski}}}\ and\ \bibinfo {author} {\bibnamefont {et~al.}},\
  }\href {\doibase 10.1086/431925} {\bibfield  {journal} {\bibinfo  {journal}
  {\apj}\ }\textbf {\bibinfo {volume} {630}},\ \bibinfo {pages} {130} (\bibinfo
  {year} {2005})},\ \Eprint {http://arxiv.org/abs/arXiv:astro-ph/0505325}
  {arXiv:astro-ph/0505325} \BibitemShut {NoStop}%
\bibitem [{\citenamefont {{Acciari}}\ and\ \citenamefont
  {et~al.}(2011{\natexlab{a}})}]{2011ApJ...738...25A}%
  \BibitemOpen
  \bibfield  {author} {\bibinfo {author} {\bibfnamefont {V.~A.}\ \bibnamefont
  {{Acciari}}}\ and\ \bibinfo {author} {\bibnamefont {et~al.}},\ }\href
  {\doibase 10.1088/0004-637X/738/1/25} {\bibfield  {journal} {\bibinfo
  {journal} {\apj}\ }\textbf {\bibinfo {volume} {738}},\ \bibinfo {eid} {25}
  (\bibinfo {year} {2011}{\natexlab{a}})},\ \Eprint
  {http://arxiv.org/abs/1106.1210} {arXiv:1106.1210 [astro-ph.HE]} \BibitemShut
  {NoStop}%
\bibitem [{\citenamefont {{Kusunose}}\ and\ \citenamefont
  {{Takahara}}(2006)}]{2006ApJ...651..113K}%
  \BibitemOpen
  \bibfield  {author} {\bibinfo {author} {\bibfnamefont {M.}~\bibnamefont
  {{Kusunose}}}\ and\ \bibinfo {author} {\bibfnamefont {F.}~\bibnamefont
  {{Takahara}}},\ }\href {\doibase 10.1086/507403} {\bibfield  {journal}
  {\bibinfo  {journal} {\apj}\ }\textbf {\bibinfo {volume} {651}},\ \bibinfo
  {pages} {113} (\bibinfo {year} {2006})},\ \Eprint
  {http://arxiv.org/abs/arXiv:astro-ph/0607063} {arXiv:astro-ph/0607063}
  \BibitemShut {NoStop}%
\bibitem [{\citenamefont {{B{\"o}ttcher}}(2005)}]{2005ApJ...621..176B}%
  \BibitemOpen
  \bibfield  {author} {\bibinfo {author} {\bibfnamefont {M.}~\bibnamefont
  {{B{\"o}ttcher}}},\ }\href {\doibase 10.1086/427430} {\bibfield  {journal}
  {\bibinfo  {journal} {\apj}\ }\textbf {\bibinfo {volume} {621}},\ \bibinfo
  {pages} {176} (\bibinfo {year} {2005})},\ \Eprint
  {http://arxiv.org/abs/arXiv:astro-ph/0411248} {arXiv:astro-ph/0411248}
  \BibitemShut {NoStop}%
\bibitem [{\citenamefont {{Sahu}}\ \emph {et~al.}(2013)\citenamefont {{Sahu}},
  \citenamefont {{Oliveros}},\ and\ \citenamefont
  {{Sanabria}}}]{2013PhRvD..87j3015S}%
  \BibitemOpen
  \bibfield  {author} {\bibinfo {author} {\bibfnamefont {S.}~\bibnamefont
  {{Sahu}}}, \bibinfo {author} {\bibfnamefont {A.~F.~O.}\ \bibnamefont
  {{Oliveros}}}, \ and\ \bibinfo {author} {\bibfnamefont {J.~C.}\ \bibnamefont
  {{Sanabria}}},\ }\href {\doibase 10.1103/PhysRevD.87.103015} {\bibfield
  {journal} {\bibinfo  {journal} {\prd}\ }\textbf {\bibinfo {volume} {87}},\
  \bibinfo {eid} {103015} (\bibinfo {year} {2013})},\ \Eprint
  {http://arxiv.org/abs/1305.4985} {arXiv:1305.4985 [hep-ph]} \BibitemShut
  {NoStop}%
\bibitem [{\citenamefont {{Reimer}}\ \emph {et~al.}(2005)\citenamefont
  {{Reimer}}, \citenamefont {{B{\"o}ttcher}},\ and\ \citenamefont
  {{Postnikov}}}]{2005ApJ...630..186R}%
  \BibitemOpen
  \bibfield  {author} {\bibinfo {author} {\bibfnamefont {A.}~\bibnamefont
  {{Reimer}}}, \bibinfo {author} {\bibfnamefont {M.}~\bibnamefont
  {{B{\"o}ttcher}}}, \ and\ \bibinfo {author} {\bibfnamefont {S.}~\bibnamefont
  {{Postnikov}}},\ }\href {\doibase 10.1086/431948} {\bibfield  {journal}
  {\bibinfo  {journal} {\apj}\ }\textbf {\bibinfo {volume} {630}},\ \bibinfo
  {pages} {186} (\bibinfo {year} {2005})},\ \Eprint
  {http://arxiv.org/abs/arXiv:astro-ph/0505233} {arXiv:astro-ph/0505233}
  \BibitemShut {NoStop}%
\bibitem [{\citenamefont {{Halzen}}\ and\ \citenamefont
  {{Hooper}}(2005)}]{2005APh....23..537H}%
  \BibitemOpen
  \bibfield  {author} {\bibinfo {author} {\bibfnamefont {F.}~\bibnamefont
  {{Halzen}}}\ and\ \bibinfo {author} {\bibfnamefont {D.}~\bibnamefont
  {{Hooper}}},\ }\href {\doibase 10.1016/j.astropartphys.2005.03.007}
  {\bibfield  {journal} {\bibinfo  {journal} {Astroparticle Physics}\ }\textbf
  {\bibinfo {volume} {23}},\ \bibinfo {pages} {537} (\bibinfo {year} {2005})},\
  \Eprint {http://arxiv.org/abs/arXiv:astro-ph/0502449}
  {arXiv:astro-ph/0502449} \BibitemShut {NoStop}%
\bibitem [{\citenamefont {{Wardle}}\ \emph {et~al.}(1998)\citenamefont
  {{Wardle}}, \citenamefont {{Homan}}, \citenamefont {{Ojha}},\ and\
  \citenamefont {{Roberts}}}]{1998Natur.395..457W}%
  \BibitemOpen
  \bibfield  {author} {\bibinfo {author} {\bibfnamefont {J.~F.~C.}\
  \bibnamefont {{Wardle}}}, \bibinfo {author} {\bibfnamefont {D.~C.}\
  \bibnamefont {{Homan}}}, \bibinfo {author} {\bibfnamefont {R.}~\bibnamefont
  {{Ojha}}}, \ and\ \bibinfo {author} {\bibfnamefont {D.~H.}\ \bibnamefont
  {{Roberts}}},\ }\href {\doibase 10.1038/26675} {\bibfield  {journal}
  {\bibinfo  {journal} {\nat}\ }\textbf {\bibinfo {volume} {395}},\ \bibinfo
  {pages} {457} (\bibinfo {year} {1998})}\BibitemShut {NoStop}%
\bibitem [{\citenamefont {{Katz}}(1996)}]{1996ApJ...463..305K}%
  \BibitemOpen
  \bibfield  {author} {\bibinfo {author} {\bibfnamefont {J.~I.}\ \bibnamefont
  {{Katz}}},\ }\href {\doibase 10.1086/177242} {\bibfield  {journal} {\bibinfo
  {journal} {\apj}\ }\textbf {\bibinfo {volume} {463}},\ \bibinfo {pages} {305}
  (\bibinfo {year} {1996})}\BibitemShut {NoStop}%
\bibitem [{\citenamefont {{Katz}}(2006)}]{2006astro.ph..3772K}%
  \BibitemOpen
  \bibfield  {author} {\bibinfo {author} {\bibfnamefont {J.~I.}\ \bibnamefont
  {{Katz}}},\ }\href@noop {} {\bibfield  {journal} {\bibinfo  {journal} {ArXiv
  Astrophysics e-prints}\ } (\bibinfo {year} {2006})},\ \Eprint
  {http://arxiv.org/abs/astro-ph/0603772} {astro-ph/0603772} \BibitemShut
  {NoStop}%
\bibitem [{\citenamefont {{Zhang}}\ and\ \citenamefont
  {{M{\'e}sz{\'a}ros}}(2004)}]{2004IJMPA..19.2385Z}%
  \BibitemOpen
  \bibfield  {author} {\bibinfo {author} {\bibfnamefont {B.}~\bibnamefont
  {{Zhang}}}\ and\ \bibinfo {author} {\bibfnamefont {P.}~\bibnamefont
  {{M{\'e}sz{\'a}ros}}},\ }\href {\doibase 10.1142/S0217751X0401746X}
  {\bibfield  {journal} {\bibinfo  {journal} {International Journal of Modern
  Physics A}\ }\textbf {\bibinfo {volume} {19}},\ \bibinfo {pages} {2385}
  (\bibinfo {year} {2004})},\ \Eprint {http://arxiv.org/abs/astro-ph/0311321}
  {astro-ph/0311321} \BibitemShut {NoStop}%
\bibitem [{\citenamefont {{Piran}}(1999)}]{1999PhR...314..575P}%
  \BibitemOpen
  \bibfield  {author} {\bibinfo {author} {\bibfnamefont {T.}~\bibnamefont
  {{Piran}}},\ }\href {\doibase 10.1016/S0370-1573(98)00127-6} {\bibfield
  {journal} {\bibinfo  {journal} {\physrep}\ }\textbf {\bibinfo {volume}
  {314}},\ \bibinfo {pages} {575} (\bibinfo {year} {1999})},\ \Eprint
  {http://arxiv.org/abs/astro-ph/9810256} {astro-ph/9810256} \BibitemShut
  {NoStop}%
\bibitem [{\citenamefont {{Iwamoto}}\ and\ \citenamefont
  {{Takahara}}(2004)}]{2004ApJ...601...78I}%
  \BibitemOpen
  \bibfield  {author} {\bibinfo {author} {\bibfnamefont {S.}~\bibnamefont
  {{Iwamoto}}}\ and\ \bibinfo {author} {\bibfnamefont {F.}~\bibnamefont
  {{Takahara}}},\ }\href {\doibase 10.1086/380300} {\bibfield  {journal}
  {\bibinfo  {journal} {\apj}\ }\textbf {\bibinfo {volume} {601}},\ \bibinfo
  {pages} {78} (\bibinfo {year} {2004})},\ \Eprint
  {http://arxiv.org/abs/arXiv:astro-ph/0309782} {arXiv:astro-ph/0309782}
  \BibitemShut {NoStop}%
\bibitem [{\citenamefont {{Iwamoto}}\ and\ \citenamefont
  {{Takahara}}(2002)}]{2002ApJ...565..163I}%
  \BibitemOpen
  \bibfield  {author} {\bibinfo {author} {\bibfnamefont {S.}~\bibnamefont
  {{Iwamoto}}}\ and\ \bibinfo {author} {\bibfnamefont {F.}~\bibnamefont
  {{Takahara}}},\ }\href {\doibase 10.1086/324480} {\bibfield  {journal}
  {\bibinfo  {journal} {\apj}\ }\textbf {\bibinfo {volume} {565}},\ \bibinfo
  {pages} {163} (\bibinfo {year} {2002})},\ \Eprint
  {http://arxiv.org/abs/arXiv:astro-ph/0204223} {arXiv:astro-ph/0204223}
  \BibitemShut {NoStop}%
\bibitem [{\citenamefont {{Asano}}\ and\ \citenamefont
  {{Takahara}}(2007)}]{2007ApJ...655..762A}%
  \BibitemOpen
  \bibfield  {author} {\bibinfo {author} {\bibfnamefont {K.}~\bibnamefont
  {{Asano}}}\ and\ \bibinfo {author} {\bibfnamefont {F.}~\bibnamefont
  {{Takahara}}},\ }\href {\doibase 10.1086/509756} {\bibfield  {journal}
  {\bibinfo  {journal} {\apj}\ }\textbf {\bibinfo {volume} {655}},\ \bibinfo
  {pages} {762} (\bibinfo {year} {2007})},\ \Eprint
  {http://arxiv.org/abs/arXiv:astro-ph/0611142} {arXiv:astro-ph/0611142}
  \BibitemShut {NoStop}%
\bibitem [{\citenamefont {{Asano}}\ and\ \citenamefont
  {{Takahara}}(2009)}]{2009ApJ...690L..81A}%
  \BibitemOpen
  \bibfield  {author} {\bibinfo {author} {\bibfnamefont {K.}~\bibnamefont
  {{Asano}}}\ and\ \bibinfo {author} {\bibfnamefont {F.}~\bibnamefont
  {{Takahara}}},\ }\href {\doibase 10.1088/0004-637X/690/1/L81} {\bibfield
  {journal} {\bibinfo  {journal} {\apjl}\ }\textbf {\bibinfo {volume} {690}},\
  \bibinfo {pages} {L81} (\bibinfo {year} {2009})},\ \Eprint
  {http://arxiv.org/abs/0812.2102} {arXiv:0812.2102} \BibitemShut {NoStop}%
\bibitem [{\citenamefont {{Schwinger}}(1951)}]{1951PhRv...82..664S}%
  \BibitemOpen
  \bibfield  {author} {\bibinfo {author} {\bibfnamefont {J.}~\bibnamefont
  {{Schwinger}}},\ }\href {\doibase 10.1103/PhysRev.82.664} {\bibfield
  {journal} {\bibinfo  {journal} {Physical Review}\ }\textbf {\bibinfo {volume}
  {82}},\ \bibinfo {pages} {664} (\bibinfo {year} {1951})}\BibitemShut
  {NoStop}%
\bibitem [{\citenamefont {{Wolfenstein}}(1978)}]{1978PhRvD..17.2369W}%
  \BibitemOpen
  \bibfield  {author} {\bibinfo {author} {\bibfnamefont {L.}~\bibnamefont
  {{Wolfenstein}}},\ }\href {\doibase 10.1103/PhysRevD.17.2369} {\bibfield
  {journal} {\bibinfo  {journal} {\prd}\ }\textbf {\bibinfo {volume} {17}},\
  \bibinfo {pages} {2369} (\bibinfo {year} {1978})}\BibitemShut {NoStop}%
\bibitem [{\citenamefont {{Sahu}}\ \emph
  {et~al.}(2009{\natexlab{a}})\citenamefont {{Sahu}}, \citenamefont
  {{Fraija}},\ and\ \citenamefont {{Keum}}}]{2009PhRvD..80c3009S}%
  \BibitemOpen
  \bibfield  {author} {\bibinfo {author} {\bibfnamefont {S.}~\bibnamefont
  {{Sahu}}}, \bibinfo {author} {\bibfnamefont {N.}~\bibnamefont {{Fraija}}}, \
  and\ \bibinfo {author} {\bibfnamefont {Y.-Y.}\ \bibnamefont {{Keum}}},\
  }\href {\doibase 10.1103/PhysRevD.80.033009} {\bibfield  {journal} {\bibinfo
  {journal} {\prd}\ }\textbf {\bibinfo {volume} {80}},\ \bibinfo {eid} {033009}
  (\bibinfo {year} {2009}{\natexlab{a}})},\ \Eprint
  {http://arxiv.org/abs/0904.0138} {arXiv:0904.0138 [hep-ph]} \BibitemShut
  {NoStop}%
\bibitem [{\citenamefont {{Sahu}}\ \emph
  {et~al.}(2009{\natexlab{b}})\citenamefont {{Sahu}}, \citenamefont
  {{Fraija}},\ and\ \citenamefont {{Keum}}}]{2009JCAP...11..024S}%
  \BibitemOpen
  \bibfield  {author} {\bibinfo {author} {\bibfnamefont {S.}~\bibnamefont
  {{Sahu}}}, \bibinfo {author} {\bibfnamefont {N.}~\bibnamefont {{Fraija}}}, \
  and\ \bibinfo {author} {\bibfnamefont {Y.-Y.}\ \bibnamefont {{Keum}}},\
  }\href {\doibase 10.1088/1475-7516/2009/11/024} {\bibfield  {journal}
  {\bibinfo  {journal} {\jcap}\ }\textbf {\bibinfo {volume} {11}},\ \bibinfo
  {eid} {024} (\bibinfo {year} {2009}{\natexlab{b}})},\ \Eprint
  {http://arxiv.org/abs/0909.3003} {arXiv:0909.3003 [hep-ph]} \BibitemShut
  {NoStop}%
\bibitem [{\citenamefont {{Fraija}}(2014{\natexlab{a}})}]{2014ApJ...787..140F}%
  \BibitemOpen
  \bibfield  {author} {\bibinfo {author} {\bibfnamefont {N.}~\bibnamefont
  {{Fraija}}},\ }\href {\doibase 10.1088/0004-637X/787/2/140} {\bibfield
  {journal} {\bibinfo  {journal} {\apj}\ }\textbf {\bibinfo {volume} {787}},\
  \bibinfo {eid} {140} (\bibinfo {year} {2014}{\natexlab{a}})},\ \Eprint
  {http://arxiv.org/abs/1401.1581} {arXiv:1401.1581 [astro-ph.HE]} \BibitemShut
  {NoStop}%
\bibitem [{\citenamefont {{Fraija}}\ and\ \citenamefont
  {{Marinelli}}(2014)}]{2014arXiv1411.7354F}%
  \BibitemOpen
  \bibfield  {author} {\bibinfo {author} {\bibfnamefont {N.}~\bibnamefont
  {{Fraija}}}\ and\ \bibinfo {author} {\bibfnamefont {A.}~\bibnamefont
  {{Marinelli}}},\ }\href@noop {} {\bibfield  {journal} {\bibinfo  {journal}
  {ArXiv e-prints}\ } (\bibinfo {year} {2014})},\ \Eprint
  {http://arxiv.org/abs/1411.7354} {arXiv:1411.7354 [astro-ph.HE]} \BibitemShut
  {NoStop}%
\bibitem [{\citenamefont {{IceCube Collaboration}}\ \emph
  {et~al.}(2013{\natexlab{a}})\citenamefont {{IceCube Collaboration}},
  \citenamefont {{Aartsen}}, \citenamefont {{Abbasi}}, \citenamefont {{Abdou}},
  \citenamefont {{Ackermann}}, \citenamefont {{Adams}}, \citenamefont
  {{Aguilar}}, \citenamefont {{Ahlers}}, \citenamefont {{Altmann}},
  \citenamefont {{Auffenberg}},\ and\ \citenamefont
  {et~al.}}]{2013arXiv1311.5238I}%
  \BibitemOpen
  \bibfield  {author} {\bibinfo {author} {\bibnamefont {{IceCube
  Collaboration}}}, \bibinfo {author} {\bibfnamefont {M.~G.}\ \bibnamefont
  {{Aartsen}}}, \bibinfo {author} {\bibfnamefont {R.}~\bibnamefont {{Abbasi}}},
  \bibinfo {author} {\bibfnamefont {Y.}~\bibnamefont {{Abdou}}}, \bibinfo
  {author} {\bibfnamefont {M.}~\bibnamefont {{Ackermann}}}, \bibinfo {author}
  {\bibfnamefont {J.}~\bibnamefont {{Adams}}}, \bibinfo {author} {\bibfnamefont
  {J.~A.}\ \bibnamefont {{Aguilar}}}, \bibinfo {author} {\bibfnamefont
  {M.}~\bibnamefont {{Ahlers}}}, \bibinfo {author} {\bibfnamefont
  {D.}~\bibnamefont {{Altmann}}}, \bibinfo {author} {\bibfnamefont
  {J.}~\bibnamefont {{Auffenberg}}}, \ and\ \bibinfo {author} {\bibnamefont
  {et~al.}},\ }\href@noop {} {\bibfield  {journal} {\bibinfo  {journal} {ArXiv
  e-prints}\ } (\bibinfo {year} {2013}{\natexlab{a}})},\ \Eprint
  {http://arxiv.org/abs/1311.5238} {arXiv:1311.5238 [astro-ph.HE]} \BibitemShut
  {NoStop}%
\bibitem [{\citenamefont {{IceCube Collaboration}}\ \emph
  {et~al.}(2013{\natexlab{b}})\citenamefont {{IceCube Collaboration}},
  \citenamefont {{Aartsen}}, \citenamefont {{Abbasi}}, \citenamefont {{Abdou}},
  \citenamefont {{Ackermann}}, \citenamefont {{Adams}}, \citenamefont
  {{Aguilar}}, \citenamefont {{Ahlers}}, \citenamefont {{Altmann}},
  \citenamefont {{Auffenberg}},\ and\ \citenamefont
  {et~al.}}]{2013arXiv1304.5356I}%
  \BibitemOpen
  \bibfield  {author} {\bibinfo {author} {\bibnamefont {{IceCube
  Collaboration}}}, \bibinfo {author} {\bibfnamefont {M.~G.}\ \bibnamefont
  {{Aartsen}}}, \bibinfo {author} {\bibfnamefont {R.}~\bibnamefont {{Abbasi}}},
  \bibinfo {author} {\bibfnamefont {Y.}~\bibnamefont {{Abdou}}}, \bibinfo
  {author} {\bibfnamefont {M.}~\bibnamefont {{Ackermann}}}, \bibinfo {author}
  {\bibfnamefont {J.}~\bibnamefont {{Adams}}}, \bibinfo {author} {\bibfnamefont
  {J.~A.}\ \bibnamefont {{Aguilar}}}, \bibinfo {author} {\bibfnamefont
  {M.}~\bibnamefont {{Ahlers}}}, \bibinfo {author} {\bibfnamefont
  {D.}~\bibnamefont {{Altmann}}}, \bibinfo {author} {\bibfnamefont
  {J.}~\bibnamefont {{Auffenberg}}}, \ and\ \bibinfo {author} {\bibnamefont
  {et~al.}},\ }\href@noop {} {\bibfield  {journal} {\bibinfo  {journal} {ArXiv
  e-prints}\ } (\bibinfo {year} {2013}{\natexlab{b}})},\ \Eprint
  {http://arxiv.org/abs/1304.5356} {arXiv:1304.5356 [astro-ph.HE]} \BibitemShut
  {NoStop}%
\bibitem [{\citenamefont {{Mannheim}}(1993)}]{1993A&A...269...67M}%
  \BibitemOpen
  \bibfield  {author} {\bibinfo {author} {\bibfnamefont {K.}~\bibnamefont
  {{Mannheim}}},\ }\href@noop {} {\bibfield  {journal} {\bibinfo  {journal}
  {\aap}\ }\textbf {\bibinfo {volume} {269}},\ \bibinfo {pages} {67} (\bibinfo
  {year} {1993})},\ \Eprint {http://arxiv.org/abs/astro-ph/9302006}
  {astro-ph/9302006} \BibitemShut {NoStop}%
\bibitem [{\citenamefont {{Sikora}}\ and\ \citenamefont
  {{Madejski}}(2000)}]{2000ApJ...534..109S}%
  \BibitemOpen
  \bibfield  {author} {\bibinfo {author} {\bibfnamefont {M.}~\bibnamefont
  {{Sikora}}}\ and\ \bibinfo {author} {\bibfnamefont {G.}~\bibnamefont
  {{Madejski}}},\ }\href {\doibase 10.1086/308756} {\bibfield  {journal}
  {\bibinfo  {journal} {\apj}\ }\textbf {\bibinfo {volume} {534}},\ \bibinfo
  {pages} {109} (\bibinfo {year} {2000})},\ \Eprint
  {http://arxiv.org/abs/astro-ph/9912335} {astro-ph/9912335} \BibitemShut
  {NoStop}%
\bibitem [{\citenamefont {{Aharonian}}(2002)}]{2002MNRAS.332..215A}%
  \BibitemOpen
  \bibfield  {author} {\bibinfo {author} {\bibfnamefont {F.~A.}\ \bibnamefont
  {{Aharonian}}},\ }\href {\doibase 10.1046/j.1365-8711.2002.05292.x}
  {\bibfield  {journal} {\bibinfo  {journal} {\mnras}\ }\textbf {\bibinfo
  {volume} {332}},\ \bibinfo {pages} {215} (\bibinfo {year} {2002})},\ \Eprint
  {http://arxiv.org/abs/arXiv:astro-ph/0106037} {arXiv:astro-ph/0106037}
  \BibitemShut {NoStop}%
\bibitem [{\citenamefont {{Aharonian}}(2000)}]{2000NewA....5..377A}%
  \BibitemOpen
  \bibfield  {author} {\bibinfo {author} {\bibfnamefont {F.~A.}\ \bibnamefont
  {{Aharonian}}},\ }\href {\doibase 10.1016/S1384-1076(00)00039-7} {\bibfield
  {journal} {\bibinfo  {journal} {\na}\ }\textbf {\bibinfo {volume} {5}},\
  \bibinfo {pages} {377} (\bibinfo {year} {2000})},\ \Eprint
  {http://arxiv.org/abs/astro-ph/0003159} {astro-ph/0003159} \BibitemShut
  {NoStop}%
\bibitem [{\citenamefont {{Svensson}}(1984)}]{1984MNRAS.209..175S}%
  \BibitemOpen
  \bibfield  {author} {\bibinfo {author} {\bibfnamefont {R.}~\bibnamefont
  {{Svensson}}},\ }\href@noop {} {\bibfield  {journal} {\bibinfo  {journal}
  {\mnras}\ }\textbf {\bibinfo {volume} {209}},\ \bibinfo {pages} {175}
  (\bibinfo {year} {1984})}\BibitemShut {NoStop}%
\bibitem [{\citenamefont {{Fragile}}\ \emph {et~al.}(2004)\citenamefont
  {{Fragile}}, \citenamefont {{Mathews}}, \citenamefont {{Poirier}},\ and\
  \citenamefont {{Totani}}}]{2004APh....20..591F}%
  \BibitemOpen
  \bibfield  {author} {\bibinfo {author} {\bibfnamefont {P.~C.}\ \bibnamefont
  {{Fragile}}}, \bibinfo {author} {\bibfnamefont {G.~J.}\ \bibnamefont
  {{Mathews}}}, \bibinfo {author} {\bibfnamefont {J.}~\bibnamefont
  {{Poirier}}}, \ and\ \bibinfo {author} {\bibfnamefont {T.}~\bibnamefont
  {{Totani}}},\ }\href {\doibase 10.1016/j.astropartphys.2003.08.005}
  {\bibfield  {journal} {\bibinfo  {journal} {Astroparticle Physics}\ }\textbf
  {\bibinfo {volume} {20}},\ \bibinfo {pages} {591} (\bibinfo {year} {2004})},\
  \Eprint {http://arxiv.org/abs/arXiv:astro-ph/0206383}
  {arXiv:astro-ph/0206383} \BibitemShut {NoStop}%
\bibitem [{\citenamefont {{Peterson}}(1997)}]{1997iagn.book.....P}%
  \BibitemOpen
  \bibfield  {author} {\bibinfo {author} {\bibfnamefont {B.~M.}\ \bibnamefont
  {{Peterson}}},\ }\href@noop {} {\emph {\bibinfo {title} {An introduction to
  active galactic nuclei}}}\ (\bibinfo {year} {1997})\BibitemShut {NoStop}%
\bibitem [{\citenamefont {{Ghisellini}}\ and\ \citenamefont
  {{Madau}}(1996)}]{1996MNRAS.280...67G}%
  \BibitemOpen
  \bibfield  {author} {\bibinfo {author} {\bibfnamefont {G.}~\bibnamefont
  {{Ghisellini}}}\ and\ \bibinfo {author} {\bibfnamefont {P.}~\bibnamefont
  {{Madau}}},\ }\href@noop {} {\bibfield  {journal} {\bibinfo  {journal}
  {\mnras}\ }\textbf {\bibinfo {volume} {280}},\ \bibinfo {pages} {67}
  (\bibinfo {year} {1996})}\BibitemShut {NoStop}%
\bibitem [{\citenamefont {{B{\"o}ttcher}}\ and\ \citenamefont
  {{Dermer}}(2002)}]{2002ApJ...564...86B}%
  \BibitemOpen
  \bibfield  {author} {\bibinfo {author} {\bibfnamefont {M.}~\bibnamefont
  {{B{\"o}ttcher}}}\ and\ \bibinfo {author} {\bibfnamefont {C.~D.}\
  \bibnamefont {{Dermer}}},\ }\href {\doibase 10.1086/324134} {\bibfield
  {journal} {\bibinfo  {journal} {\apj}\ }\textbf {\bibinfo {volume} {564}},\
  \bibinfo {pages} {86} (\bibinfo {year} {2002})},\ \Eprint
  {http://arxiv.org/abs/astro-ph/0106395} {astro-ph/0106395} \BibitemShut
  {NoStop}%
\bibitem [{\citenamefont {{Abdo}}\ \emph {et~al.}(2011)\citenamefont {{Abdo}},
  \citenamefont {{Ackermann}}, \citenamefont {{Ajello}}, \citenamefont
  {{Baldini}}, \citenamefont {{Ballet}}, \citenamefont {{Barbiellini}},
  \citenamefont {{Bastieri}}, \citenamefont {{Bechtol}}, \citenamefont
  {{Bellazzini}}, \citenamefont {{Berenji}},\ and\ \citenamefont
  {et~al.}}]{2011ApJ...736..131A}%
  \BibitemOpen
  \bibfield  {author} {\bibinfo {author} {\bibfnamefont {A.~A.}\ \bibnamefont
  {{Abdo}}}, \bibinfo {author} {\bibfnamefont {M.}~\bibnamefont {{Ackermann}}},
  \bibinfo {author} {\bibfnamefont {M.}~\bibnamefont {{Ajello}}}, \bibinfo
  {author} {\bibfnamefont {L.}~\bibnamefont {{Baldini}}}, \bibinfo {author}
  {\bibfnamefont {J.}~\bibnamefont {{Ballet}}}, \bibinfo {author}
  {\bibfnamefont {G.}~\bibnamefont {{Barbiellini}}}, \bibinfo {author}
  {\bibfnamefont {D.}~\bibnamefont {{Bastieri}}}, \bibinfo {author}
  {\bibfnamefont {K.}~\bibnamefont {{Bechtol}}}, \bibinfo {author}
  {\bibfnamefont {R.}~\bibnamefont {{Bellazzini}}}, \bibinfo {author}
  {\bibfnamefont {B.}~\bibnamefont {{Berenji}}}, \ and\ \bibinfo {author}
  {\bibnamefont {et~al.}},\ }\href {\doibase 10.1088/0004-637X/736/2/131}
  {\bibfield  {journal} {\bibinfo  {journal} {\apj}\ }\textbf {\bibinfo
  {volume} {736}},\ \bibinfo {eid} {131} (\bibinfo {year} {2011})},\ \Eprint
  {http://arxiv.org/abs/1106.1348} {arXiv:1106.1348 [astro-ph.HE]} \BibitemShut
  {NoStop}%
\bibitem [{\citenamefont {{Kino}}\ and\ \citenamefont
  {{Takahara}}(2004)}]{2004MNRAS.349..336K}%
  \BibitemOpen
  \bibfield  {author} {\bibinfo {author} {\bibfnamefont {M.}~\bibnamefont
  {{Kino}}}\ and\ \bibinfo {author} {\bibfnamefont {F.}~\bibnamefont
  {{Takahara}}},\ }\href {\doibase 10.1111/j.1365-2966.2004.07511.x} {\bibfield
   {journal} {\bibinfo  {journal} {\mnras}\ }\textbf {\bibinfo {volume}
  {349}},\ \bibinfo {pages} {336} (\bibinfo {year} {2004})},\ \Eprint
  {http://arxiv.org/abs/astro-ph/0312530} {astro-ph/0312530} \BibitemShut
  {NoStop}%
\bibitem [{\citenamefont {{Fraija}}\ \emph {et~al.}(2012)\citenamefont
  {{Fraija}}, \citenamefont {{Gonz{\'a}lez}}, \citenamefont {{Perez}},\ and\
  \citenamefont {{Marinelli}}}]{2012ApJ...753...40F}%
  \BibitemOpen
  \bibfield  {author} {\bibinfo {author} {\bibfnamefont {N.}~\bibnamefont
  {{Fraija}}}, \bibinfo {author} {\bibfnamefont {M.~M.}\ \bibnamefont
  {{Gonz{\'a}lez}}}, \bibinfo {author} {\bibfnamefont {M.}~\bibnamefont
  {{Perez}}}, \ and\ \bibinfo {author} {\bibfnamefont {A.}~\bibnamefont
  {{Marinelli}}},\ }\href {\doibase 10.1088/0004-637X/753/1/40} {\bibfield
  {journal} {\bibinfo  {journal} {\apj}\ }\textbf {\bibinfo {volume} {753}},\
  \bibinfo {eid} {40} (\bibinfo {year} {2012})},\ \Eprint
  {http://arxiv.org/abs/1204.4500} {arXiv:1204.4500 [astro-ph.HE]} \BibitemShut
  {NoStop}%
\bibitem [{\citenamefont {{Fraija}}(2014{\natexlab{b}})}]{2014ApJ...783...44F}%
  \BibitemOpen
  \bibfield  {author} {\bibinfo {author} {\bibfnamefont {N.}~\bibnamefont
  {{Fraija}}},\ }\href {\doibase 10.1088/0004-637X/783/1/44} {\bibfield
  {journal} {\bibinfo  {journal} {\apj}\ }\textbf {\bibinfo {volume} {783}},\
  \bibinfo {eid} {44} (\bibinfo {year} {2014}{\natexlab{b}})},\ \Eprint
  {http://arxiv.org/abs/1312.6944} {arXiv:1312.6944 [astro-ph.HE]} \BibitemShut
  {NoStop}%
\bibitem [{\citenamefont {{Dermer}}\ and\ \citenamefont
  {{Menon}}(2009)}]{der09}%
  \BibitemOpen
  \bibfield  {author} {\bibinfo {author} {\bibfnamefont {C.~D.}\ \bibnamefont
  {{Dermer}}}\ and\ \bibinfo {author} {\bibfnamefont {G.}~\bibnamefont
  {{Menon}}},\ }\href@noop {} {\emph {\bibinfo {title} {High Energy Radiation
  from Black Holes.}}}\ (\bibinfo {year} {2009})\BibitemShut {NoStop}%
\bibitem [{\citenamefont {{M{\"u}cke}}\ \emph {et~al.}(2000)\citenamefont
  {{M{\"u}cke}}, \citenamefont {{Engel}}, \citenamefont {{Rachen}},
  \citenamefont {{Protheroe}},\ and\ \citenamefont
  {{Stanev}}}]{2000CoPhC.124..290M}%
  \BibitemOpen
  \bibfield  {author} {\bibinfo {author} {\bibfnamefont {A.}~\bibnamefont
  {{M{\"u}cke}}}, \bibinfo {author} {\bibfnamefont {R.}~\bibnamefont
  {{Engel}}}, \bibinfo {author} {\bibfnamefont {J.~P.}\ \bibnamefont
  {{Rachen}}}, \bibinfo {author} {\bibfnamefont {R.~J.}\ \bibnamefont
  {{Protheroe}}}, \ and\ \bibinfo {author} {\bibfnamefont {T.}~\bibnamefont
  {{Stanev}}},\ }\href {\doibase 10.1016/S0010-4655(99)00446-4} {\bibfield
  {journal} {\bibinfo  {journal} {Computer Physics Communications}\ }\textbf
  {\bibinfo {volume} {124}},\ \bibinfo {pages} {290} (\bibinfo {year}
  {2000})},\ \Eprint {http://arxiv.org/abs/astro-ph/9903478} {astro-ph/9903478}
  \BibitemShut {NoStop}%
\bibitem [{\citenamefont {{Stecker}}(1968)}]{1968PhRvL..21.1016S}%
  \BibitemOpen
  \bibfield  {author} {\bibinfo {author} {\bibfnamefont {F.~W.}\ \bibnamefont
  {{Stecker}}},\ }\href {\doibase 10.1103/PhysRevLett.21.1016} {\bibfield
  {journal} {\bibinfo  {journal} {Physical Review Letters}\ }\textbf {\bibinfo
  {volume} {21}},\ \bibinfo {pages} {1016} (\bibinfo {year}
  {1968})}\BibitemShut {NoStop}%
\bibitem [{\citenamefont {Waxman}\ and\ \citenamefont
  {Bahcall}(1997)}]{PhysRevLett.78.2292}%
  \BibitemOpen
  \bibfield  {author} {\bibinfo {author} {\bibfnamefont {E.}~\bibnamefont
  {Waxman}}\ and\ \bibinfo {author} {\bibfnamefont {J.}~\bibnamefont
  {Bahcall}},\ }\href {\doibase 10.1103/PhysRevLett.78.2292} {\bibfield
  {journal} {\bibinfo  {journal} {Phys. Rev. Lett.}\ }\textbf {\bibinfo
  {volume} {78}},\ \bibinfo {pages} {2292} (\bibinfo {year}
  {1997})}\BibitemShut {NoStop}%
\bibitem [{\citenamefont {{Dermer}}(2013)}]{2013avhe.book..225D}%
  \BibitemOpen
  \bibfield  {author} {\bibinfo {author} {\bibfnamefont {C.~D.}\ \bibnamefont
  {{Dermer}}},\ }\enquote {\bibinfo {title} {{Sources of GeV Photons and the
  Fermi Results}},}\ in\ \href {\doibase 10.1007/978-3-642-36134-0_3} {\emph
  {\bibinfo {booktitle} {Astrophysics at Very High Energies, Saas-Fee Advanced
  Course}}},\ \bibinfo {editor} {edited by\ \bibinfo {editor} {\bibfnamefont
  {F.}~\bibnamefont {{Aharonian}}}, \bibinfo {editor} {\bibfnamefont
  {L.}~\bibnamefont {{Bergstr{\"o}m}}}, \ and\ \bibinfo {editor} {\bibfnamefont
  {C.}~\bibnamefont {{Dermer}}}}\ (\bibinfo {year} {2013})\ p.\ \bibinfo
  {pages} {225}\BibitemShut {NoStop}%
\bibitem [{\citenamefont {{Murase}}\ \emph {et~al.}(2014)\citenamefont
  {{Murase}}, \citenamefont {{Inoue}},\ and\ \citenamefont
  {{Dermer}}}]{2014arXiv1403.4089M}%
  \BibitemOpen
  \bibfield  {author} {\bibinfo {author} {\bibfnamefont {K.}~\bibnamefont
  {{Murase}}}, \bibinfo {author} {\bibfnamefont {Y.}~\bibnamefont {{Inoue}}}, \
  and\ \bibinfo {author} {\bibfnamefont {C.~D.}\ \bibnamefont {{Dermer}}},\
  }\href@noop {} {\bibfield  {journal} {\bibinfo  {journal} {ArXiv e-prints}\ }
  (\bibinfo {year} {2014})},\ \Eprint {http://arxiv.org/abs/1403.4089}
  {arXiv:1403.4089 [astro-ph.HE]} \BibitemShut {NoStop}%
\bibitem [{\citenamefont {{Rachen}}\ and\ \citenamefont
  {{M{\'e}sz{\'a}ros}}(1998)}]{1998PhRvD..58l3005R}%
  \BibitemOpen
  \bibfield  {author} {\bibinfo {author} {\bibfnamefont {J.~P.}\ \bibnamefont
  {{Rachen}}}\ and\ \bibinfo {author} {\bibfnamefont {P.}~\bibnamefont
  {{M{\'e}sz{\'a}ros}}},\ }\href {\doibase 10.1103/PhysRevD.58.123005}
  {\bibfield  {journal} {\bibinfo  {journal} {\prd}\ }\textbf {\bibinfo
  {volume} {58}},\ \bibinfo {eid} {123005} (\bibinfo {year} {1998})},\ \Eprint
  {http://arxiv.org/abs/astro-ph/9802280} {astro-ph/9802280} \BibitemShut
  {NoStop}%
\bibitem [{\citenamefont {{Longair}}(1994)}]{1994hea2.book.....L}%
  \BibitemOpen
  \bibfield  {author} {\bibinfo {author} {\bibfnamefont {M.~S.}\ \bibnamefont
  {{Longair}}},\ }\href@noop {} {\emph {\bibinfo {title} {High energy
  astrophysics.~Volume 2}}}\ (\bibinfo {year} {1994})\BibitemShut {NoStop}%
\bibitem [{\citenamefont {{Rybicki}}\ and\ \citenamefont
  {{Lightman}}(1986)}]{1986rpa..book.....R}%
  \BibitemOpen
  \bibfield  {author} {\bibinfo {author} {\bibfnamefont {G.~B.}\ \bibnamefont
  {{Rybicki}}}\ and\ \bibinfo {author} {\bibfnamefont {A.~P.}\ \bibnamefont
  {{Lightman}}},\ }\href@noop {} {\emph {\bibinfo {title} {Radiative Processes
  in Astrophysics}}}\ (\bibinfo {year} {1986})\BibitemShut {NoStop}%
\bibitem [{\citenamefont {{Mohapatra}}\ and\ \citenamefont
  {{Pal}}(2004)}]{2004mnpa.book.....M}%
  \BibitemOpen
  \bibfield  {author} {\bibinfo {author} {\bibfnamefont {R.~N.}\ \bibnamefont
  {{Mohapatra}}}\ and\ \bibinfo {author} {\bibfnamefont {P.~B.}\ \bibnamefont
  {{Pal}}},\ }\href@noop {} {\emph {\bibinfo {title} {Massive neutrinos in
  physics and astrophysics.}}}\ (\bibinfo {year} {2004})\BibitemShut {NoStop}%
\bibitem [{\citenamefont {{Bahcall}}(1989)}]{1989neas.book.....B}%
  \BibitemOpen
  \bibfield  {author} {\bibinfo {author} {\bibfnamefont {J.~N.}\ \bibnamefont
  {{Bahcall}}},\ }\href@noop {} {\emph {\bibinfo {title} {Neutrino
  astrophysics}}}\ (\bibinfo {year} {1989})\BibitemShut {NoStop}%
\bibitem [{\citenamefont {{Giunti}}\ and\ \citenamefont
  {{Chung}}(2007)}]{2007fnpa.book.....G}%
  \BibitemOpen
  \bibfield  {author} {\bibinfo {author} {\bibfnamefont {C.}~\bibnamefont
  {{Giunti}}}\ and\ \bibinfo {author} {\bibfnamefont {W.~K.}\ \bibnamefont
  {{Chung}}},\ }\href@noop {} {\emph {\bibinfo {title} {Fundamentals of
  Neutrino Physics and Astrophysics.}}}\ (\bibinfo  {publisher} {Oxford
  University Press},\ \bibinfo {year} {2007})\BibitemShut {NoStop}%
\bibitem [{\citenamefont {{Fraija}}\ \emph {et~al.}(2014)\citenamefont
  {{Fraija}}, \citenamefont {{Bernal}},\ and\ \citenamefont
  {{Hidalgo-Gam{\'e}z}}}]{2014MNRAS.442..239F}%
  \BibitemOpen
  \bibfield  {author} {\bibinfo {author} {\bibfnamefont {N.}~\bibnamefont
  {{Fraija}}}, \bibinfo {author} {\bibfnamefont {C.~G.}\ \bibnamefont
  {{Bernal}}}, \ and\ \bibinfo {author} {\bibfnamefont {A.~M.}\ \bibnamefont
  {{Hidalgo-Gam{\'e}z}}},\ }\href {\doibase 10.1093/mnras/stu872} {\bibfield
  {journal} {\bibinfo  {journal} {\mnras}\ }\textbf {\bibinfo {volume} {442}},\
  \bibinfo {pages} {239} (\bibinfo {year} {2014})},\ \Eprint
  {http://arxiv.org/abs/1402.6292} {arXiv:1402.6292 [astro-ph.HE]} \BibitemShut
  {NoStop}%
\bibitem [{\citenamefont {{Fraija}}(2014{\natexlab{c}})}]{2014MNRAS.437.2187F}%
  \BibitemOpen
  \bibfield  {author} {\bibinfo {author} {\bibfnamefont {N.}~\bibnamefont
  {{Fraija}}},\ }\href {\doibase 10.1093/mnras/stt2036} {\bibfield  {journal}
  {\bibinfo  {journal} {\mnras}\ }\textbf {\bibinfo {volume} {437}},\ \bibinfo
  {pages} {2187} (\bibinfo {year} {2014}{\natexlab{c}})},\ \Eprint
  {http://arxiv.org/abs/1310.7061} {arXiv:1310.7061 [astro-ph.HE]} \BibitemShut
  {NoStop}%
\bibitem [{\citenamefont {{Aharmim}}\ and\ \citenamefont
  {et~al.}(2011)}]{aha11}%
  \BibitemOpen
  \bibfield  {author} {\bibinfo {author} {\bibfnamefont {B.}~\bibnamefont
  {{Aharmim}}}\ and\ \bibinfo {author} {\bibnamefont {et~al.}},\ }\href@noop {}
  {\bibfield  {journal} {\bibinfo  {journal} {ArXiv e-prints}\ } (\bibinfo
  {year} {2011})},\ \Eprint {http://arxiv.org/abs/1109.0763} {arXiv:1109.0763
  [nucl-ex]} \BibitemShut {NoStop}%
\bibitem [{\citenamefont {{Abe}}\ and\ \citenamefont {et~al.}(2011)}]{abe11a}%
  \BibitemOpen
  \bibfield  {author} {\bibinfo {author} {\bibfnamefont {K.}~\bibnamefont
  {{Abe}}}\ and\ \bibinfo {author} {\bibnamefont {et~al.}},\ }\href {\doibase
  10.1103/PhysRevLett.107.241801} {\bibfield  {journal} {\bibinfo  {journal}
  {Physical Review Letters}\ }\textbf {\bibinfo {volume} {107}},\ \bibinfo
  {eid} {241801} (\bibinfo {year} {2011})},\ \Eprint
  {http://arxiv.org/abs/1109.1621} {arXiv:1109.1621 [hep-ex]} \BibitemShut
  {NoStop}%
\bibitem [{\citenamefont {{Athanassopoulos}}\ and\ \citenamefont
  {et~al.}(1996)}]{ath96}%
  \BibitemOpen
  \bibfield  {author} {\bibinfo {author} {\bibfnamefont {C.}~\bibnamefont
  {{Athanassopoulos}}}\ and\ \bibinfo {author} {\bibnamefont {et~al.}},\ }\href
  {\doibase 10.1103/PhysRevLett.77.3082} {\bibfield  {journal} {\bibinfo
  {journal} {Physical Review Letters}\ }\textbf {\bibinfo {volume} {77}},\
  \bibinfo {pages} {3082} (\bibinfo {year} {1996})},\ \Eprint
  {http://arxiv.org/abs/arXiv:nucl-ex/9605003} {arXiv:nucl-ex/9605003}
  \BibitemShut {NoStop}%
\bibitem [{\citenamefont {{Athanassopoulos}}\ and\ \citenamefont
  {et~al.}(1998)}]{ath98}%
  \BibitemOpen
  \bibfield  {author} {\bibinfo {author} {\bibfnamefont {C.}~\bibnamefont
  {{Athanassopoulos}}}\ and\ \bibinfo {author} {\bibnamefont {et~al.}},\ }\href
  {\doibase 10.1103/PhysRevLett.81.1774} {\bibfield  {journal} {\bibinfo
  {journal} {Physical Review Letters}\ }\textbf {\bibinfo {volume} {81}},\
  \bibinfo {pages} {1774} (\bibinfo {year} {1998})},\ \Eprint
  {http://arxiv.org/abs/arXiv:nucl-ex/9709006} {arXiv:nucl-ex/9709006}
  \BibitemShut {NoStop}%
\bibitem [{\citenamefont {{Gonzalez-Garcia}}\ and\ \citenamefont
  {{Nir}}(2003)}]{gon03}%
  \BibitemOpen
  \bibfield  {author} {\bibinfo {author} {\bibfnamefont {M.~C.}\ \bibnamefont
  {{Gonzalez-Garcia}}}\ and\ \bibinfo {author} {\bibfnamefont {Y.}~\bibnamefont
  {{Nir}}},\ }\href {\doibase 10.1103/RevModPhys.75.345} {\bibfield  {journal}
  {\bibinfo  {journal} {Reviews of Modern Physics}\ }\textbf {\bibinfo {volume}
  {75}},\ \bibinfo {pages} {345} (\bibinfo {year} {2003})},\ \Eprint
  {http://arxiv.org/abs/arXiv:hep-ph/0202058} {arXiv:hep-ph/0202058}
  \BibitemShut {NoStop}%
\bibitem [{\citenamefont {{Akhmedov}}\ \emph {et~al.}(2004)\citenamefont
  {{Akhmedov}}, \citenamefont {{Johansson}}, \citenamefont {{Lindner}},
  \citenamefont {{Ohlsson}},\ and\ \citenamefont {{Schwetz}}}]{akh04}%
  \BibitemOpen
  \bibfield  {author} {\bibinfo {author} {\bibfnamefont {E.~K.}\ \bibnamefont
  {{Akhmedov}}}, \bibinfo {author} {\bibfnamefont {R.}~\bibnamefont
  {{Johansson}}}, \bibinfo {author} {\bibfnamefont {M.}~\bibnamefont
  {{Lindner}}}, \bibinfo {author} {\bibfnamefont {T.}~\bibnamefont
  {{Ohlsson}}}, \ and\ \bibinfo {author} {\bibfnamefont {T.}~\bibnamefont
  {{Schwetz}}},\ }\href {\doibase 10.1088/1126-6708/2004/04/078} {\bibfield
  {journal} {\bibinfo  {journal} {Journal of High Energy Physics}\ }\textbf
  {\bibinfo {volume} {4}},\ \bibinfo {eid} {078} (\bibinfo {year} {2004})},\
  \Eprint {http://arxiv.org/abs/arXiv:hep-ph/0402175} {arXiv:hep-ph/0402175}
  \BibitemShut {NoStop}%
\bibitem [{\citenamefont {{Gonzalez-Garcia}}\ and\ \citenamefont
  {{Maltoni}}(2008)}]{gon08}%
  \BibitemOpen
  \bibfield  {author} {\bibinfo {author} {\bibfnamefont {M.~C.}\ \bibnamefont
  {{Gonzalez-Garcia}}}\ and\ \bibinfo {author} {\bibfnamefont {M.}~\bibnamefont
  {{Maltoni}}},\ }\href {\doibase 10.1016/j.physrep.2007.12.004} {\bibfield
  {journal} {\bibinfo  {journal} {\physrep}\ }\textbf {\bibinfo {volume}
  {460}},\ \bibinfo {pages} {1} (\bibinfo {year} {2008})},\ \Eprint
  {http://arxiv.org/abs/0704.1800} {arXiv:0704.1800 [hep-ph]} \BibitemShut
  {NoStop}%
\bibitem [{\citenamefont {{Gonzalez-Garcia}}(2011)}]{gon11}%
  \BibitemOpen
  \bibfield  {author} {\bibinfo {author} {\bibfnamefont {M.~C.}\ \bibnamefont
  {{Gonzalez-Garcia}}},\ }\href {\doibase 10.1134/S106377961104006X} {\bibfield
   {journal} {\bibinfo  {journal} {Physics of Particles and Nuclei}\ }\textbf
  {\bibinfo {volume} {42}},\ \bibinfo {pages} {577} (\bibinfo {year}
  {2011})}\BibitemShut {NoStop}%
\bibitem [{\citenamefont {{Wendell}}\ and\ \citenamefont
  {et~al.}(2010)}]{wen10}%
  \BibitemOpen
  \bibfield  {author} {\bibinfo {author} {\bibfnamefont {R.}~\bibnamefont
  {{Wendell}}}\ and\ \bibinfo {author} {\bibnamefont {et~al.}},\ }\href
  {\doibase 10.1103/PhysRevD.81.092004} {\bibfield  {journal} {\bibinfo
  {journal} {\prd}\ }\textbf {\bibinfo {volume} {81}},\ \bibinfo {eid} {092004}
  (\bibinfo {year} {2010})},\ \Eprint {http://arxiv.org/abs/1002.3471}
  {arXiv:1002.3471 [hep-ex]} \BibitemShut {NoStop}%
\bibitem [{\citenamefont {{Learned}}\ and\ \citenamefont
  {{Pakvasa}}(1995)}]{lea95}%
  \BibitemOpen
  \bibfield  {author} {\bibinfo {author} {\bibfnamefont {J.~G.}\ \bibnamefont
  {{Learned}}}\ and\ \bibinfo {author} {\bibfnamefont {S.}~\bibnamefont
  {{Pakvasa}}},\ }\href {\doibase 10.1016/0927-6505(94)00043-3} {\bibfield
  {journal} {\bibinfo  {journal} {Astroparticle Physics}\ }\textbf {\bibinfo
  {volume} {3}},\ \bibinfo {pages} {267} (\bibinfo {year} {1995})},\ \Eprint
  {http://arxiv.org/abs/arXiv:hep-ph/9405296} {arXiv:hep-ph/9405296}
  \BibitemShut {NoStop}%
\bibitem [{\citenamefont {{Andres}}\ and\ \citenamefont
  {et~al.}(2000)}]{2000APh....13....1A}%
  \BibitemOpen
  \bibfield  {author} {\bibinfo {author} {\bibfnamefont {E.}~\bibnamefont
  {{Andres}}}\ and\ \bibinfo {author} {\bibnamefont {et~al.}},\ }\href
  {\doibase 10.1016/S0927-6505(99)00092-4} {\bibfield  {journal} {\bibinfo
  {journal} {Astroparticle Physics}\ }\textbf {\bibinfo {volume} {13}},\
  \bibinfo {pages} {1} (\bibinfo {year} {2000})},\ \Eprint
  {http://arxiv.org/abs/astro-ph/9906203} {astro-ph/9906203} \BibitemShut
  {NoStop}%
\bibitem [{\citenamefont {{Andr{\'e}s}}\ and\ \citenamefont
  {et~al.}(2001)}]{2001Natur.410..441A}%
  \BibitemOpen
  \bibfield  {author} {\bibinfo {author} {\bibfnamefont {E.}~\bibnamefont
  {{Andr{\'e}s}}}\ and\ \bibinfo {author} {\bibnamefont {et~al.}},\ }\href
  {\doibase 10.1038/410441A0} {\bibfield  {journal} {\bibinfo  {journal}
  {\nat}\ }\textbf {\bibinfo {volume} {410}},\ \bibinfo {pages} {441} (\bibinfo
  {year} {2001})}\BibitemShut {NoStop}%
\bibitem [{\citenamefont {{Andres}}\ and\ \citenamefont
  {et~al.}(2001)}]{2001NuPhS..91..423A}%
  \BibitemOpen
  \bibfield  {author} {\bibinfo {author} {\bibfnamefont {E.}~\bibnamefont
  {{Andres}}}\ and\ \bibinfo {author} {\bibnamefont {et~al.}},\ }\href
  {\doibase 10.1016/S0920-5632(00)00971-3} {\bibfield  {journal} {\bibinfo
  {journal} {Nuclear Physics B Proceedings Supplements}\ }\textbf {\bibinfo
  {volume} {91}},\ \bibinfo {pages} {423} (\bibinfo {year} {2001})},\ \Eprint
  {http://arxiv.org/abs/astro-ph/0009242} {astro-ph/0009242} \BibitemShut
  {NoStop}%
\bibitem [{\citenamefont {{Halzen}}(2007)}]{2007Ap&SS.309..407H}%
  \BibitemOpen
  \bibfield  {author} {\bibinfo {author} {\bibfnamefont {F.}~\bibnamefont
  {{Halzen}}},\ }\href {\doibase 10.1007/s10509-007-9434-7} {\bibfield
  {journal} {\bibinfo  {journal} {\apss}\ }\textbf {\bibinfo {volume} {309}},\
  \bibinfo {pages} {407} (\bibinfo {year} {2007})},\ \Eprint
  {http://arxiv.org/abs/arXiv:astro-ph/0611915} {arXiv:astro-ph/0611915}
  \BibitemShut {NoStop}%
\bibitem [{\citenamefont {{Becker}}(2008)}]{2008PhR...458..173B}%
  \BibitemOpen
  \bibfield  {author} {\bibinfo {author} {\bibfnamefont {J.~K.}\ \bibnamefont
  {{Becker}}},\ }\href {\doibase 10.1016/j.physrep.2007.10.006} {\bibfield
  {journal} {\bibinfo  {journal} {\physrep}\ }\textbf {\bibinfo {volume}
  {458}},\ \bibinfo {pages} {173} (\bibinfo {year} {2008})},\ \Eprint
  {http://arxiv.org/abs/0710.1557} {arXiv:0710.1557} \BibitemShut {NoStop}%
\bibitem [{\citenamefont {Halzen}(2013)}]{Halzen:2013bta}%
  \BibitemOpen
  \bibfield  {author} {\bibinfo {author} {\bibfnamefont {F.}~\bibnamefont
  {Halzen}},\ }\href {\doibase 10.1393/ncr/i2013-10086-y} {\bibfield  {journal}
  {\bibinfo  {journal} {Riv.Nuovo Cim.}\ }\textbf {\bibinfo {volume} {036}},\
  \bibinfo {pages} {81} (\bibinfo {year} {2013})}\BibitemShut {NoStop}%
\bibitem [{\citenamefont {{Gandhi}}\ \emph {et~al.}(1998)\citenamefont
  {{Gandhi}}, \citenamefont {{Quigg}}, \citenamefont {{Reno}},\ and\
  \citenamefont {{Sarcevic}}}]{1998PhRvD..58i3009G}%
  \BibitemOpen
  \bibfield  {author} {\bibinfo {author} {\bibfnamefont {R.}~\bibnamefont
  {{Gandhi}}}, \bibinfo {author} {\bibfnamefont {C.}~\bibnamefont {{Quigg}}},
  \bibinfo {author} {\bibfnamefont {M.~H.}\ \bibnamefont {{Reno}}}, \ and\
  \bibinfo {author} {\bibfnamefont {I.}~\bibnamefont {{Sarcevic}}},\ }\href
  {\doibase 10.1103/PhysRevD.58.093009} {\bibfield  {journal} {\bibinfo
  {journal} {\prd}\ }\textbf {\bibinfo {volume} {58}},\ \bibinfo {eid} {093009}
  (\bibinfo {year} {1998})},\ \Eprint
  {http://arxiv.org/abs/arXiv:hep-ph/9807264} {arXiv:hep-ph/9807264}
  \BibitemShut {NoStop}%
\bibitem [{\citenamefont {{Murase}}\ \emph {et~al.}(2012)\citenamefont
  {{Murase}}, \citenamefont {{Dermer}}, \citenamefont {{Takami}},\ and\
  \citenamefont {{Migliori}}}]{2012ApJ...749...63M}%
  \BibitemOpen
  \bibfield  {author} {\bibinfo {author} {\bibfnamefont {K.}~\bibnamefont
  {{Murase}}}, \bibinfo {author} {\bibfnamefont {C.~D.}\ \bibnamefont
  {{Dermer}}}, \bibinfo {author} {\bibfnamefont {H.}~\bibnamefont {{Takami}}},
  \ and\ \bibinfo {author} {\bibfnamefont {G.}~\bibnamefont {{Migliori}}},\
  }\href {\doibase 10.1088/0004-637X/749/1/63} {\bibfield  {journal} {\bibinfo
  {journal} {\apj}\ }\textbf {\bibinfo {volume} {749}},\ \bibinfo {eid} {63}
  (\bibinfo {year} {2012})},\ \Eprint {http://arxiv.org/abs/1107.5576}
  {arXiv:1107.5576 [astro-ph.HE]} \BibitemShut {NoStop}%
\bibitem [{\citenamefont {{Razzaque}}\ \emph {et~al.}(2012)\citenamefont
  {{Razzaque}}, \citenamefont {{Dermer}},\ and\ \citenamefont
  {{Finke}}}]{2012ApJ...745..196R}%
  \BibitemOpen
  \bibfield  {author} {\bibinfo {author} {\bibfnamefont {S.}~\bibnamefont
  {{Razzaque}}}, \bibinfo {author} {\bibfnamefont {C.~D.}\ \bibnamefont
  {{Dermer}}}, \ and\ \bibinfo {author} {\bibfnamefont {J.~D.}\ \bibnamefont
  {{Finke}}},\ }\href {\doibase 10.1088/0004-637X/745/2/196} {\bibfield
  {journal} {\bibinfo  {journal} {\apj}\ }\textbf {\bibinfo {volume} {745}},\
  \bibinfo {eid} {196} (\bibinfo {year} {2012})},\ \Eprint
  {http://arxiv.org/abs/1110.0853} {arXiv:1110.0853 [astro-ph.HE]} \BibitemShut
  {NoStop}%
\bibitem [{\citenamefont {{Hillas}}(1984)}]{1984ARA&A..22..425H}%
  \BibitemOpen
  \bibfield  {author} {\bibinfo {author} {\bibfnamefont {A.~M.}\ \bibnamefont
  {{Hillas}}},\ }\href {\doibase 10.1146/annurev.aa.22.090184.002233}
  {\bibfield  {journal} {\bibinfo  {journal} {\araa}\ }\textbf {\bibinfo
  {volume} {22}},\ \bibinfo {pages} {425} (\bibinfo {year} {1984})}\BibitemShut
  {NoStop}%
\bibitem [{\citenamefont {{Jiang}}\ \emph {et~al.}(2010)\citenamefont
  {{Jiang}}, \citenamefont {{Hou}}, \citenamefont {{Han}}, \citenamefont
  {{Sun}},\ and\ \citenamefont {{Wang}}}]{2010ApJ...719..459J}%
  \BibitemOpen
  \bibfield  {author} {\bibinfo {author} {\bibfnamefont {Y.-Y.}\ \bibnamefont
  {{Jiang}}}, \bibinfo {author} {\bibfnamefont {L.~G.}\ \bibnamefont {{Hou}}},
  \bibinfo {author} {\bibfnamefont {J.~L.}\ \bibnamefont {{Han}}}, \bibinfo
  {author} {\bibfnamefont {X.~H.}\ \bibnamefont {{Sun}}}, \ and\ \bibinfo
  {author} {\bibfnamefont {W.}~\bibnamefont {{Wang}}},\ }\href {\doibase
  10.1088/0004-637X/719/1/459} {\bibfield  {journal} {\bibinfo  {journal}
  {\apj}\ }\textbf {\bibinfo {volume} {719}},\ \bibinfo {pages} {459} (\bibinfo
  {year} {2010})},\ \Eprint {http://arxiv.org/abs/1004.1877} {arXiv:1004.1877
  [astro-ph.HE]} \BibitemShut {NoStop}%
\bibitem [{\citenamefont {{Lemoine}}\ and\ \citenamefont
  {{Waxman}}(2009)}]{2009JCAP...11..009L}%
  \BibitemOpen
  \bibfield  {author} {\bibinfo {author} {\bibfnamefont {M.}~\bibnamefont
  {{Lemoine}}}\ and\ \bibinfo {author} {\bibfnamefont {E.}~\bibnamefont
  {{Waxman}}},\ }\href {\doibase 10.1088/1475-7516/2009/11/009} {\bibfield
  {journal} {\bibinfo  {journal} {\jcap}\ }\textbf {\bibinfo {volume} {11}},\
  \bibinfo {eid} {009} (\bibinfo {year} {2009})},\ \Eprint
  {http://arxiv.org/abs/0907.1354} {arXiv:0907.1354 [astro-ph.HE]} \BibitemShut
  {NoStop}%
\bibitem [{\citenamefont {{Dermer}}\ \emph {et~al.}(2009)\citenamefont
  {{Dermer}}, \citenamefont {{Razzaque}}, \citenamefont {{Finke}},\ and\
  \citenamefont {{Atoyan}}}]{2009NJPh...11f5016D}%
  \BibitemOpen
  \bibfield  {author} {\bibinfo {author} {\bibfnamefont {C.~D.}\ \bibnamefont
  {{Dermer}}}, \bibinfo {author} {\bibfnamefont {S.}~\bibnamefont
  {{Razzaque}}}, \bibinfo {author} {\bibfnamefont {J.~D.}\ \bibnamefont
  {{Finke}}}, \ and\ \bibinfo {author} {\bibfnamefont {A.}~\bibnamefont
  {{Atoyan}}},\ }\href {\doibase 10.1088/1367-2630/11/6/065016} {\bibfield
  {journal} {\bibinfo  {journal} {New Journal of Physics}\ }\textbf {\bibinfo
  {volume} {11}},\ \bibinfo {eid} {065016} (\bibinfo {year} {2009})},\ \Eprint
  {http://arxiv.org/abs/0811.1160} {arXiv:0811.1160} \BibitemShut {NoStop}%
\bibitem [{\citenamefont {{Stanev}}(1997)}]{1997ApJ...479..290S}%
  \BibitemOpen
  \bibfield  {author} {\bibinfo {author} {\bibfnamefont {T.}~\bibnamefont
  {{Stanev}}},\ }\href@noop {} {\bibfield  {journal} {\bibinfo  {journal}
  {\apj}\ }\textbf {\bibinfo {volume} {479}},\ \bibinfo {pages} {290} (\bibinfo
  {year} {1997})},\ \Eprint {http://arxiv.org/abs/astro-ph/9607086}
  {astro-ph/9607086} \BibitemShut {NoStop}%
\bibitem [{\citenamefont {{Abu-Zayyad}}\ and\ \citenamefont
  {et~al.}(2012)}]{2012NIMPA.689...87A}%
  \BibitemOpen
  \bibfield  {author} {\bibinfo {author} {\bibfnamefont {T.}~\bibnamefont
  {{Abu-Zayyad}}}\ and\ \bibinfo {author} {\bibnamefont {et~al.}},\ }\href
  {\doibase 10.1016/j.nima.2012.05.079} {\bibfield  {journal} {\bibinfo
  {journal} {Nuclear Instruments and Methods in Physics Research A}\ }\textbf
  {\bibinfo {volume} {689}},\ \bibinfo {pages} {87} (\bibinfo {year} {2012})},\
  \Eprint {http://arxiv.org/abs/1201.4964} {arXiv:1201.4964 [astro-ph.IM]}
  \BibitemShut {NoStop}%
\bibitem [{\citenamefont {{Sommers}}(2001)}]{2001APh....14..271S}%
  \BibitemOpen
  \bibfield  {author} {\bibinfo {author} {\bibfnamefont {P.}~\bibnamefont
  {{Sommers}}},\ }\href {\doibase 10.1016/S0927-6505(00)00130-4} {\bibfield
  {journal} {\bibinfo  {journal} {Astroparticle Physics}\ }\textbf {\bibinfo
  {volume} {14}},\ \bibinfo {pages} {271} (\bibinfo {year} {2001})},\ \Eprint
  {http://arxiv.org/abs/astro-ph/0004016} {astro-ph/0004016} \BibitemShut
  {NoStop}%
\bibitem [{\citenamefont {{Kotera}}\ and\ \citenamefont
  {{Olinto}}(2011)}]{2011ARA&A..49..119K}%
  \BibitemOpen
  \bibfield  {author} {\bibinfo {author} {\bibfnamefont {K.}~\bibnamefont
  {{Kotera}}}\ and\ \bibinfo {author} {\bibfnamefont {A.~V.}\ \bibnamefont
  {{Olinto}}},\ }\href {\doibase 10.1146/annurev-astro-081710-102620}
  {\bibfield  {journal} {\bibinfo  {journal} {\araa}\ }\textbf {\bibinfo
  {volume} {49}},\ \bibinfo {pages} {119} (\bibinfo {year} {2011})},\ \Eprint
  {http://arxiv.org/abs/1101.4256} {arXiv:1101.4256 [astro-ph.HE]} \BibitemShut
  {NoStop}%
\bibitem [{\citenamefont {{Brun}}\ and\ \citenamefont
  {{Rademakers}}(1997)}]{1997NIMPA.389...81B}%
  \BibitemOpen
  \bibfield  {author} {\bibinfo {author} {\bibfnamefont {R.}~\bibnamefont
  {{Brun}}}\ and\ \bibinfo {author} {\bibfnamefont {F.}~\bibnamefont
  {{Rademakers}}},\ }\href {\doibase 10.1016/S0168-9002(97)00048-X} {\bibfield
  {journal} {\bibinfo  {journal} {Nuclear Instruments and Methods in Physics
  Research A}\ }\textbf {\bibinfo {volume} {389}},\ \bibinfo {pages} {81}
  (\bibinfo {year} {1997})}\BibitemShut {NoStop}%
\bibitem [{\citenamefont {{Fraija}}(2014{\natexlab{d}})}]{2014MNRAS.441.1209F}%
  \BibitemOpen
  \bibfield  {author} {\bibinfo {author} {\bibfnamefont {N.}~\bibnamefont
  {{Fraija}}},\ }\href {\doibase 10.1093/mnras/stu652} {\bibfield  {journal}
  {\bibinfo  {journal} {\mnras}\ }\textbf {\bibinfo {volume} {441}},\ \bibinfo
  {pages} {1209} (\bibinfo {year} {2014}{\natexlab{d}})},\ \Eprint
  {http://arxiv.org/abs/1402.4558} {arXiv:1402.4558 [astro-ph.HE]} \BibitemShut
  {NoStop}%
\bibitem [{\citenamefont {{Tagliaferri}}\ and\ \citenamefont
  {et~al.}(2008)}]{2008ApJ...679.1029T}%
  \BibitemOpen
  \bibfield  {author} {\bibinfo {author} {\bibfnamefont {G.}~\bibnamefont
  {{Tagliaferri}}}\ and\ \bibinfo {author} {\bibnamefont {et~al.}},\ }\href
  {\doibase 10.1086/586731} {\bibfield  {journal} {\bibinfo  {journal} {\apj}\
  }\textbf {\bibinfo {volume} {679}},\ \bibinfo {pages} {1029} (\bibinfo {year}
  {2008})},\ \Eprint {http://arxiv.org/abs/0801.4029} {arXiv:0801.4029}
  \BibitemShut {NoStop}%
\bibitem [{\citenamefont {{V{\'e}ron-Cetty}}\ and\ \citenamefont
  {{V{\'e}ron}}(2006)}]{2006A&A...455..773V}%
  \BibitemOpen
  \bibfield  {author} {\bibinfo {author} {\bibfnamefont {M.-P.}\ \bibnamefont
  {{V{\'e}ron-Cetty}}}\ and\ \bibinfo {author} {\bibfnamefont {P.}~\bibnamefont
  {{V{\'e}ron}}},\ }\href {\doibase 10.1051/0004-6361:20065177} {\bibfield
  {journal} {\bibinfo  {journal} {\aap}\ }\textbf {\bibinfo {volume} {455}},\
  \bibinfo {pages} {773} (\bibinfo {year} {2006})}\BibitemShut {NoStop}%
\bibitem [{\citenamefont {{Aliu}}\ and\ \citenamefont
  {et~al.}(2014)}]{2014ApJ...797...89A}%
  \BibitemOpen
  \bibfield  {author} {\bibinfo {author} {\bibfnamefont {E.}~\bibnamefont
  {{Aliu}}}\ and\ \bibinfo {author} {\bibnamefont {et~al.}},\ }\href {\doibase
  10.1088/0004-637X/797/2/89} {\bibfield  {journal} {\bibinfo  {journal}
  {\apj}\ }\textbf {\bibinfo {volume} {797}},\ \bibinfo {eid} {89} (\bibinfo
  {year} {2014})},\ \Eprint {http://arxiv.org/abs/1412.1031} {arXiv:1412.1031
  [astro-ph.HE]} \BibitemShut {NoStop}%
\bibitem [{\citenamefont {{Falomo}}\ \emph {et~al.}(2003)\citenamefont
  {{Falomo}}, \citenamefont {{Kotilainen}}, \citenamefont {{Carangelo}},\ and\
  \citenamefont {{Treves}}}]{2003ApJ...595..624F}%
  \BibitemOpen
  \bibfield  {author} {\bibinfo {author} {\bibfnamefont {R.}~\bibnamefont
  {{Falomo}}}, \bibinfo {author} {\bibfnamefont {J.~K.}\ \bibnamefont
  {{Kotilainen}}}, \bibinfo {author} {\bibfnamefont {N.}~\bibnamefont
  {{Carangelo}}}, \ and\ \bibinfo {author} {\bibfnamefont {A.}~\bibnamefont
  {{Treves}}},\ }\href {\doibase 10.1086/377432} {\bibfield  {journal}
  {\bibinfo  {journal} {\apj}\ }\textbf {\bibinfo {volume} {595}},\ \bibinfo
  {pages} {624} (\bibinfo {year} {2003})},\ \Eprint
  {http://arxiv.org/abs/astro-ph/0306163} {astro-ph/0306163} \BibitemShut
  {NoStop}%
\bibitem [{\citenamefont {{Elvis}}\ \emph {et~al.}(1992)\citenamefont
  {{Elvis}}, \citenamefont {{Plummer}}, \citenamefont {{Schachter}},\ and\
  \citenamefont {{Fabbiano}}}]{1992ApJS...80..257E}%
  \BibitemOpen
  \bibfield  {author} {\bibinfo {author} {\bibfnamefont {M.}~\bibnamefont
  {{Elvis}}}, \bibinfo {author} {\bibfnamefont {D.}~\bibnamefont {{Plummer}}},
  \bibinfo {author} {\bibfnamefont {J.}~\bibnamefont {{Schachter}}}, \ and\
  \bibinfo {author} {\bibfnamefont {G.}~\bibnamefont {{Fabbiano}}},\ }\href
  {\doibase 10.1086/191665} {\bibfield  {journal} {\bibinfo  {journal} {\apjs}\
  }\textbf {\bibinfo {volume} {80}},\ \bibinfo {pages} {257} (\bibinfo {year}
  {1992})}\BibitemShut {NoStop}%
\bibitem [{\citenamefont {{Schachter}}\ \emph {et~al.}(1993)\citenamefont
  {{Schachter}}, \citenamefont {{Stocke}}, \citenamefont {{Perlman}},
  \citenamefont {{Elvis}}, \citenamefont {{Remillard}}, \citenamefont
  {{Granados}}, \citenamefont {{Luu}}, \citenamefont {{Huchra}}, \citenamefont
  {{Humphreys}}, \citenamefont {{Urry}},\ and\ \citenamefont
  {{Wallin}}}]{1993ApJ...412..541S}%
  \BibitemOpen
  \bibfield  {author} {\bibinfo {author} {\bibfnamefont {J.~F.}\ \bibnamefont
  {{Schachter}}}, \bibinfo {author} {\bibfnamefont {J.~T.}\ \bibnamefont
  {{Stocke}}}, \bibinfo {author} {\bibfnamefont {E.}~\bibnamefont {{Perlman}}},
  \bibinfo {author} {\bibfnamefont {M.}~\bibnamefont {{Elvis}}}, \bibinfo
  {author} {\bibfnamefont {R.}~\bibnamefont {{Remillard}}}, \bibinfo {author}
  {\bibfnamefont {A.}~\bibnamefont {{Granados}}}, \bibinfo {author}
  {\bibfnamefont {J.}~\bibnamefont {{Luu}}}, \bibinfo {author} {\bibfnamefont
  {J.~P.}\ \bibnamefont {{Huchra}}}, \bibinfo {author} {\bibfnamefont
  {R.}~\bibnamefont {{Humphreys}}}, \bibinfo {author} {\bibfnamefont {C.~M.}\
  \bibnamefont {{Urry}}}, \ and\ \bibinfo {author} {\bibfnamefont
  {J.}~\bibnamefont {{Wallin}}},\ }\href {\doibase 10.1086/172942} {\bibfield
  {journal} {\bibinfo  {journal} {\apj}\ }\textbf {\bibinfo {volume} {412}},\
  \bibinfo {pages} {541} (\bibinfo {year} {1993})}\BibitemShut {NoStop}%
\bibitem [{\citenamefont {{Beckmann}}\ \emph {et~al.}(2002)\citenamefont
  {{Beckmann}}, \citenamefont {{Wolter}}, \citenamefont {{Celotti}},
  \citenamefont {{Costamante}}, \citenamefont {{Ghisellini}}, \citenamefont
  {{Maccacaro}},\ and\ \citenamefont {{Tagliaferri}}}]{2002A&A...383..410B}%
  \BibitemOpen
  \bibfield  {author} {\bibinfo {author} {\bibfnamefont {V.}~\bibnamefont
  {{Beckmann}}}, \bibinfo {author} {\bibfnamefont {A.}~\bibnamefont
  {{Wolter}}}, \bibinfo {author} {\bibfnamefont {A.}~\bibnamefont {{Celotti}}},
  \bibinfo {author} {\bibfnamefont {L.}~\bibnamefont {{Costamante}}}, \bibinfo
  {author} {\bibfnamefont {G.}~\bibnamefont {{Ghisellini}}}, \bibinfo {author}
  {\bibfnamefont {T.}~\bibnamefont {{Maccacaro}}}, \ and\ \bibinfo {author}
  {\bibfnamefont {G.}~\bibnamefont {{Tagliaferri}}},\ }\href {\doibase
  10.1051/0004-6361:20011752} {\bibfield  {journal} {\bibinfo  {journal}
  {\aap}\ }\textbf {\bibinfo {volume} {383}},\ \bibinfo {pages} {410} (\bibinfo
  {year} {2002})},\ \Eprint {http://arxiv.org/abs/arXiv:astro-ph/0112311}
  {arXiv:astro-ph/0112311} \BibitemShut {NoStop}%
\bibitem [{\citenamefont {{Aharonian}}\ and\ \citenamefont
  {et~al.}(2003{\natexlab{a}})}]{2003A&A...406L...9A}%
  \BibitemOpen
  \bibfield  {author} {\bibinfo {author} {\bibfnamefont {F.}~\bibnamefont
  {{Aharonian}}}\ and\ \bibinfo {author} {\bibnamefont {et~al.}},\ }\href
  {\doibase 10.1051/0004-6361:20030838} {\bibfield  {journal} {\bibinfo
  {journal} {\aap}\ }\textbf {\bibinfo {volume} {406}},\ \bibinfo {pages} {L9}
  (\bibinfo {year} {2003}{\natexlab{a}})}\BibitemShut {NoStop}%
\bibitem [{\citenamefont {{Ahrens}}\ and\ \citenamefont
  {et~al.}(2003)}]{2003PhRvD..67a2003A}%
  \BibitemOpen
  \bibfield  {author} {\bibinfo {author} {\bibfnamefont {J.}~\bibnamefont
  {{Ahrens}}}\ and\ \bibinfo {author} {\bibnamefont {et~al.}},\ }\href
  {\doibase 10.1103/PhysRevD.67.012003} {\bibfield  {journal} {\bibinfo
  {journal} {\prd}\ }\textbf {\bibinfo {volume} {67}},\ \bibinfo {eid} {012003}
  (\bibinfo {year} {2003})},\ \Eprint {http://arxiv.org/abs/astro-ph/0206487}
  {astro-ph/0206487} \BibitemShut {NoStop}%
\bibitem [{\citenamefont {{Donnarumma}}\ \emph {et~al.}(2009)\citenamefont
  {{Donnarumma}}, \citenamefont {{Vittorini}}, \citenamefont {{Vercellone}},
  \citenamefont {{del Monte}}, \citenamefont {{Feroci}}, \citenamefont
  {{D'Ammando}}, \citenamefont {{Pacciani}}, \citenamefont {{Chen}},
  \citenamefont {{Tavani}}, \citenamefont {{Bulgarelli}},\ and\ \citenamefont
  {et~al.}}]{2009ApJ...691L..13D}%
  \BibitemOpen
  \bibfield  {author} {\bibinfo {author} {\bibfnamefont {I.}~\bibnamefont
  {{Donnarumma}}}, \bibinfo {author} {\bibfnamefont {V.}~\bibnamefont
  {{Vittorini}}}, \bibinfo {author} {\bibfnamefont {S.}~\bibnamefont
  {{Vercellone}}}, \bibinfo {author} {\bibfnamefont {E.}~\bibnamefont {{del
  Monte}}}, \bibinfo {author} {\bibfnamefont {M.}~\bibnamefont {{Feroci}}},
  \bibinfo {author} {\bibfnamefont {F.}~\bibnamefont {{D'Ammando}}}, \bibinfo
  {author} {\bibfnamefont {L.}~\bibnamefont {{Pacciani}}}, \bibinfo {author}
  {\bibfnamefont {A.~W.}\ \bibnamefont {{Chen}}}, \bibinfo {author}
  {\bibfnamefont {M.}~\bibnamefont {{Tavani}}}, \bibinfo {author}
  {\bibfnamefont {A.}~\bibnamefont {{Bulgarelli}}}, \ and\ \bibinfo {author}
  {\bibnamefont {et~al.}},\ }\href {\doibase 10.1088/0004-637X/691/1/L13}
  {\bibfield  {journal} {\bibinfo  {journal} {\apjl}\ }\textbf {\bibinfo
  {volume} {691}},\ \bibinfo {pages} {L13} (\bibinfo {year} {2009})},\ \Eprint
  {http://arxiv.org/abs/0812.1500} {arXiv:0812.1500} \BibitemShut {NoStop}%
\bibitem [{\citenamefont {{Barth}}\ \emph {et~al.}(2003)\citenamefont
  {{Barth}}, \citenamefont {{Ho}},\ and\ \citenamefont
  {{Sargent}}}]{2003ApJ...583..134B}%
  \BibitemOpen
  \bibfield  {author} {\bibinfo {author} {\bibfnamefont {A.~J.}\ \bibnamefont
  {{Barth}}}, \bibinfo {author} {\bibfnamefont {L.~C.}\ \bibnamefont {{Ho}}}, \
  and\ \bibinfo {author} {\bibfnamefont {W.~L.~W.}\ \bibnamefont {{Sargent}}},\
  }\href {\doibase 10.1086/345083} {\bibfield  {journal} {\bibinfo  {journal}
  {\apj}\ }\textbf {\bibinfo {volume} {583}},\ \bibinfo {pages} {134} (\bibinfo
  {year} {2003})},\ \Eprint {http://arxiv.org/abs/astro-ph/0209562}
  {astro-ph/0209562} \BibitemShut {NoStop}%
\bibitem [{\citenamefont {{Fossati}}\ and\ \citenamefont
  {et~al}(2008)}]{2008ApJ...677..906F}%
  \BibitemOpen
  \bibfield  {author} {\bibinfo {author} {\bibfnamefont {G.}~\bibnamefont
  {{Fossati}}}\ and\ \bibinfo {author} {\bibnamefont {et~al}},\ }\href
  {\doibase 10.1086/527311} {\bibfield  {journal} {\bibinfo  {journal} {\apj}\
  }\textbf {\bibinfo {volume} {677}},\ \bibinfo {pages} {906} (\bibinfo {year}
  {2008})},\ \Eprint {http://arxiv.org/abs/0710.4138} {arXiv:0710.4138}
  \BibitemShut {NoStop}%
\bibitem [{\citenamefont {{Punch}}\ \emph {et~al.}(1992)\citenamefont
  {{Punch}}, \citenamefont {{Akerlof}}, \citenamefont {{Cawley}}, \citenamefont
  {{Chantell}}, \citenamefont {{Fegan}}, \citenamefont {{Fennell}},
  \citenamefont {{Gaidos}}, \citenamefont {{Hagan}}, \citenamefont {{Hillas}},
  \citenamefont {{Jiang}}, \citenamefont {{Kerrick}}, \citenamefont {{Lamb}},
  \citenamefont {{Lawrence}}, \citenamefont {{Lewis}}, \citenamefont {{Meyer}},
  \citenamefont {{Mohanty}}, \citenamefont {{O'Flaherty}}, \citenamefont
  {{Reynolds}}, \citenamefont {{Rovero}}, \citenamefont {{Schubnell}},
  \citenamefont {{Sembroski}}, \citenamefont {{Weekes}},\ and\ \citenamefont
  {{Wilson}}}]{1992Natur.358..477P}%
  \BibitemOpen
  \bibfield  {author} {\bibinfo {author} {\bibfnamefont {M.}~\bibnamefont
  {{Punch}}}, \bibinfo {author} {\bibfnamefont {C.~W.}\ \bibnamefont
  {{Akerlof}}}, \bibinfo {author} {\bibfnamefont {M.~F.}\ \bibnamefont
  {{Cawley}}}, \bibinfo {author} {\bibfnamefont {M.}~\bibnamefont
  {{Chantell}}}, \bibinfo {author} {\bibfnamefont {D.~J.}\ \bibnamefont
  {{Fegan}}}, \bibinfo {author} {\bibfnamefont {S.}~\bibnamefont {{Fennell}}},
  \bibinfo {author} {\bibfnamefont {J.~A.}\ \bibnamefont {{Gaidos}}}, \bibinfo
  {author} {\bibfnamefont {J.}~\bibnamefont {{Hagan}}}, \bibinfo {author}
  {\bibfnamefont {A.~M.}\ \bibnamefont {{Hillas}}}, \bibinfo {author}
  {\bibfnamefont {Y.}~\bibnamefont {{Jiang}}}, \bibinfo {author} {\bibfnamefont
  {A.~D.}\ \bibnamefont {{Kerrick}}}, \bibinfo {author} {\bibfnamefont {R.~C.}\
  \bibnamefont {{Lamb}}}, \bibinfo {author} {\bibfnamefont {M.~A.}\
  \bibnamefont {{Lawrence}}}, \bibinfo {author} {\bibfnamefont {D.~A.}\
  \bibnamefont {{Lewis}}}, \bibinfo {author} {\bibfnamefont {D.~I.}\
  \bibnamefont {{Meyer}}}, \bibinfo {author} {\bibfnamefont {G.}~\bibnamefont
  {{Mohanty}}}, \bibinfo {author} {\bibfnamefont {K.~S.}\ \bibnamefont
  {{O'Flaherty}}}, \bibinfo {author} {\bibfnamefont {P.~T.}\ \bibnamefont
  {{Reynolds}}}, \bibinfo {author} {\bibfnamefont {A.~C.}\ \bibnamefont
  {{Rovero}}}, \bibinfo {author} {\bibfnamefont {M.~S.}\ \bibnamefont
  {{Schubnell}}}, \bibinfo {author} {\bibfnamefont {G.}~\bibnamefont
  {{Sembroski}}}, \bibinfo {author} {\bibfnamefont {T.~C.}\ \bibnamefont
  {{Weekes}}}, \ and\ \bibinfo {author} {\bibfnamefont {C.}~\bibnamefont
  {{Wilson}}},\ }\href {\doibase 10.1038/358477a0} {\bibfield  {journal}
  {\bibinfo  {journal} {\nat}\ }\textbf {\bibinfo {volume} {358}},\ \bibinfo
  {pages} {477} (\bibinfo {year} {1992})}\BibitemShut {NoStop}%
\bibitem [{\citenamefont {{Gaidos}}\ \emph {et~al.}(1996)\citenamefont
  {{Gaidos}}, \citenamefont {{Akerlof}}, \citenamefont {{Biller}},
  \citenamefont {{Boyle}}, \citenamefont {{Breslin}}, \citenamefont
  {{Buckley}}, \citenamefont {{Carter-Lewis}}, \citenamefont {{Catanese}},
  \citenamefont {{Cawley}}, \citenamefont {{Fegan}}, \citenamefont {{Finley}},
  \citenamefont {{Gordo}}, \citenamefont {{Hillas}}, \citenamefont
  {{Krennrich}}, \citenamefont {{Lamb}}, \citenamefont {{Lessard}},
  \citenamefont {{McEnery}}, \citenamefont {{Masterson}}, \citenamefont
  {{Mohanty}}, \citenamefont {{Moriarty}}, \citenamefont {{Quinn}},
  \citenamefont {{Rodgers}}, \citenamefont {{Rose}}, \citenamefont
  {{Samuelson}}, \citenamefont {{Schubnell}}, \citenamefont {{Sembroski}},
  \citenamefont {{Srinivasan}}, \citenamefont {{Weekes}}, \citenamefont
  {{Wilson}},\ and\ \citenamefont {{Zweerink}}}]{1996Natur.383..319G}%
  \BibitemOpen
  \bibfield  {author} {\bibinfo {author} {\bibfnamefont {J.~A.}\ \bibnamefont
  {{Gaidos}}}, \bibinfo {author} {\bibfnamefont {C.~W.}\ \bibnamefont
  {{Akerlof}}}, \bibinfo {author} {\bibfnamefont {S.}~\bibnamefont {{Biller}}},
  \bibinfo {author} {\bibfnamefont {P.~J.}\ \bibnamefont {{Boyle}}}, \bibinfo
  {author} {\bibfnamefont {A.~C.}\ \bibnamefont {{Breslin}}}, \bibinfo {author}
  {\bibfnamefont {J.~H.}\ \bibnamefont {{Buckley}}}, \bibinfo {author}
  {\bibfnamefont {D.~A.}\ \bibnamefont {{Carter-Lewis}}}, \bibinfo {author}
  {\bibfnamefont {M.}~\bibnamefont {{Catanese}}}, \bibinfo {author}
  {\bibfnamefont {M.~F.}\ \bibnamefont {{Cawley}}}, \bibinfo {author}
  {\bibfnamefont {D.~J.}\ \bibnamefont {{Fegan}}}, \bibinfo {author}
  {\bibfnamefont {J.~P.}\ \bibnamefont {{Finley}}}, \bibinfo {author}
  {\bibfnamefont {J.~B.}\ \bibnamefont {{Gordo}}}, \bibinfo {author}
  {\bibfnamefont {A.~M.}\ \bibnamefont {{Hillas}}}, \bibinfo {author}
  {\bibfnamefont {F.}~\bibnamefont {{Krennrich}}}, \bibinfo {author}
  {\bibfnamefont {R.~C.}\ \bibnamefont {{Lamb}}}, \bibinfo {author}
  {\bibfnamefont {R.~W.}\ \bibnamefont {{Lessard}}}, \bibinfo {author}
  {\bibfnamefont {J.~E.}\ \bibnamefont {{McEnery}}}, \bibinfo {author}
  {\bibfnamefont {C.}~\bibnamefont {{Masterson}}}, \bibinfo {author}
  {\bibfnamefont {G.}~\bibnamefont {{Mohanty}}}, \bibinfo {author}
  {\bibfnamefont {P.}~\bibnamefont {{Moriarty}}}, \bibinfo {author}
  {\bibfnamefont {J.}~\bibnamefont {{Quinn}}}, \bibinfo {author} {\bibfnamefont
  {A.~J.}\ \bibnamefont {{Rodgers}}}, \bibinfo {author} {\bibfnamefont {H.~J.}\
  \bibnamefont {{Rose}}}, \bibinfo {author} {\bibfnamefont {F.}~\bibnamefont
  {{Samuelson}}}, \bibinfo {author} {\bibfnamefont {M.~S.}\ \bibnamefont
  {{Schubnell}}}, \bibinfo {author} {\bibfnamefont {G.~H.}\ \bibnamefont
  {{Sembroski}}}, \bibinfo {author} {\bibfnamefont {R.}~\bibnamefont
  {{Srinivasan}}}, \bibinfo {author} {\bibfnamefont {T.~C.}\ \bibnamefont
  {{Weekes}}}, \bibinfo {author} {\bibfnamefont {C.~L.}\ \bibnamefont
  {{Wilson}}}, \ and\ \bibinfo {author} {\bibfnamefont {J.}~\bibnamefont
  {{Zweerink}}},\ }\href {\doibase 10.1038/383319a0} {\bibfield  {journal}
  {\bibinfo  {journal} {\nat}\ }\textbf {\bibinfo {volume} {383}},\ \bibinfo
  {pages} {319} (\bibinfo {year} {1996})}\BibitemShut {NoStop}%
\bibitem [{\citenamefont {{Krennrich}}\ and\ \citenamefont
  {et~al.}(2002)}]{2002ApJ...575L...9K}%
  \BibitemOpen
  \bibfield  {author} {\bibinfo {author} {\bibfnamefont {F.}~\bibnamefont
  {{Krennrich}}}\ and\ \bibinfo {author} {\bibnamefont {et~al.}},\ }\href
  {\doibase 10.1086/342700} {\bibfield  {journal} {\bibinfo  {journal} {\apjl}\
  }\textbf {\bibinfo {volume} {575}},\ \bibinfo {pages} {L9} (\bibinfo {year}
  {2002})},\ \Eprint {http://arxiv.org/abs/arXiv:astro-ph/0207184}
  {arXiv:astro-ph/0207184} \BibitemShut {NoStop}%
\bibitem [{\citenamefont {{Albert}}\ and\ \citenamefont
  {et~al.}(2007)}]{2007ApJ...663..125A}%
  \BibitemOpen
  \bibfield  {author} {\bibinfo {author} {\bibfnamefont {J.}~\bibnamefont
  {{Albert}}}\ and\ \bibinfo {author} {\bibnamefont {et~al.}},\ }\href
  {\doibase 10.1086/518221} {\bibfield  {journal} {\bibinfo  {journal} {\apj}\
  }\textbf {\bibinfo {volume} {663}},\ \bibinfo {pages} {125} (\bibinfo {year}
  {2007})},\ \Eprint {http://arxiv.org/abs/arXiv:astro-ph/0603478}
  {arXiv:astro-ph/0603478} \BibitemShut {NoStop}%
\bibitem [{\citenamefont {{Aharonian}}\ and\ \citenamefont
  {et~al.}(2005)}]{2005A&A...437...95A}%
  \BibitemOpen
  \bibfield  {author} {\bibinfo {author} {\bibfnamefont {F.}~\bibnamefont
  {{Aharonian}}}\ and\ \bibinfo {author} {\bibnamefont {et~al.}},\ }\href
  {\doibase 10.1051/0004-6361:20053050} {\bibfield  {journal} {\bibinfo
  {journal} {\aap}\ }\textbf {\bibinfo {volume} {437}},\ \bibinfo {pages} {95}
  (\bibinfo {year} {2005})},\ \Eprint
  {http://arxiv.org/abs/arXiv:astro-ph/0506319} {arXiv:astro-ph/0506319}
  \BibitemShut {NoStop}%
\bibitem [{\citenamefont {{Aharonian}}\ and\ \citenamefont
  {et~al.}(2002)}]{2002A&A...393...89A}%
  \BibitemOpen
  \bibfield  {author} {\bibinfo {author} {\bibfnamefont {F.}~\bibnamefont
  {{Aharonian}}}\ and\ \bibinfo {author} {\bibnamefont {et~al.}},\ }\href
  {\doibase 10.1051/0004-6361:20021005} {\bibfield  {journal} {\bibinfo
  {journal} {\aap}\ }\textbf {\bibinfo {volume} {393}},\ \bibinfo {pages} {89}
  (\bibinfo {year} {2002})},\ \Eprint
  {http://arxiv.org/abs/arXiv:astro-ph/0205499} {arXiv:astro-ph/0205499}
  \BibitemShut {NoStop}%
\bibitem [{\citenamefont {{Aharonian}}\ and\ \citenamefont
  {et~al.}(2003{\natexlab{b}})}]{2003A&A...410..813A}%
  \BibitemOpen
  \bibfield  {author} {\bibinfo {author} {\bibfnamefont {F.}~\bibnamefont
  {{Aharonian}}}\ and\ \bibinfo {author} {\bibnamefont {et~al.}},\ }\href
  {\doibase 10.1051/0004-6361:20031265} {\bibfield  {journal} {\bibinfo
  {journal} {\aap}\ }\textbf {\bibinfo {volume} {410}},\ \bibinfo {pages} {813}
  (\bibinfo {year} {2003}{\natexlab{b}})}\BibitemShut {NoStop}%
\bibitem [{\citenamefont {{Carson}}\ \emph {et~al.}(2007)\citenamefont
  {{Carson}}, \citenamefont {{Kildea}}, \citenamefont {{Ong}}, \citenamefont
  {{Ball}}, \citenamefont {{Bramel}}, \citenamefont {{Covault}}, \citenamefont
  {{Driscoll}}, \citenamefont {{Fortin}}, \citenamefont {{Gingrich}},
  \citenamefont {{Hanna}}, \citenamefont {{Lindner}}, \citenamefont
  {{Mueller}}, \citenamefont {{Jarvis}}, \citenamefont {{Mukherjee}},
  \citenamefont {{Ragan}}, \citenamefont {{Scalzo}}, \citenamefont
  {{Williams}},\ and\ \citenamefont {{Zweerink}}}]{2007ApJ...662..199C}%
  \BibitemOpen
  \bibfield  {author} {\bibinfo {author} {\bibfnamefont {J.~E.}\ \bibnamefont
  {{Carson}}}, \bibinfo {author} {\bibfnamefont {J.}~\bibnamefont {{Kildea}}},
  \bibinfo {author} {\bibfnamefont {R.~A.}\ \bibnamefont {{Ong}}}, \bibinfo
  {author} {\bibfnamefont {J.}~\bibnamefont {{Ball}}}, \bibinfo {author}
  {\bibfnamefont {D.~A.}\ \bibnamefont {{Bramel}}}, \bibinfo {author}
  {\bibfnamefont {C.~E.}\ \bibnamefont {{Covault}}}, \bibinfo {author}
  {\bibfnamefont {D.}~\bibnamefont {{Driscoll}}}, \bibinfo {author}
  {\bibfnamefont {P.}~\bibnamefont {{Fortin}}}, \bibinfo {author}
  {\bibfnamefont {D.~M.}\ \bibnamefont {{Gingrich}}}, \bibinfo {author}
  {\bibfnamefont {D.~S.}\ \bibnamefont {{Hanna}}}, \bibinfo {author}
  {\bibfnamefont {T.}~\bibnamefont {{Lindner}}}, \bibinfo {author}
  {\bibfnamefont {C.}~\bibnamefont {{Mueller}}}, \bibinfo {author}
  {\bibfnamefont {A.}~\bibnamefont {{Jarvis}}}, \bibinfo {author}
  {\bibfnamefont {R.}~\bibnamefont {{Mukherjee}}}, \bibinfo {author}
  {\bibfnamefont {K.}~\bibnamefont {{Ragan}}}, \bibinfo {author} {\bibfnamefont
  {R.~A.}\ \bibnamefont {{Scalzo}}}, \bibinfo {author} {\bibfnamefont {D.~A.}\
  \bibnamefont {{Williams}}}, \ and\ \bibinfo {author} {\bibfnamefont
  {J.}~\bibnamefont {{Zweerink}}},\ }\href {\doibase 10.1086/516818} {\bibfield
   {journal} {\bibinfo  {journal} {\apj}\ }\textbf {\bibinfo {volume} {662}},\
  \bibinfo {pages} {199} (\bibinfo {year} {2007})},\ \Eprint
  {http://arxiv.org/abs/astro-ph/0612562} {astro-ph/0612562} \BibitemShut
  {NoStop}%
\bibitem [{\citenamefont {{Abdo}}\ and\ \citenamefont
  {et~al.}(2014)}]{2014ApJ...782..110A}%
  \BibitemOpen
  \bibfield  {author} {\bibinfo {author} {\bibfnamefont {A.~A.}\ \bibnamefont
  {{Abdo}}}\ and\ \bibinfo {author} {\bibnamefont {et~al.}},\ }\href {\doibase
  10.1088/0004-637X/782/2/110} {\bibfield  {journal} {\bibinfo  {journal}
  {\apj}\ }\textbf {\bibinfo {volume} {782}},\ \bibinfo {eid} {110} (\bibinfo
  {year} {2014})},\ \Eprint {http://arxiv.org/abs/1401.2161} {arXiv:1401.2161
  [astro-ph.HE]} \BibitemShut {NoStop}%
\bibitem [{\citenamefont {{Isobe}}\ \emph {et~al.}(2010)\citenamefont
  {{Isobe}}, \citenamefont {{Sugimori}}, \citenamefont {{Kawai}}, \citenamefont
  {{Ueda}}, \citenamefont {{Negoro}}, \citenamefont {{Sugizaki}}, \citenamefont
  {{Matsuoka}}, \citenamefont {{Daikyuji}}, \citenamefont {{Eguchi}},
  \citenamefont {{Hiroi}}, \citenamefont {{Ishikawa}}, \citenamefont
  {{Ishiwata}}, \citenamefont {{Kawasaki}}, \citenamefont {{Kimura}},
  \citenamefont {{Kohama}}, \citenamefont {{Mihara}}, \citenamefont
  {{Miyoshi}}, \citenamefont {{Morii}}, \citenamefont {{Nakagawa}},
  \citenamefont {{Nakahira}}, \citenamefont {{Nakajima}}, \citenamefont
  {{Ozawa}}, \citenamefont {{Sootome}}, \citenamefont {{Suzuki}}, \citenamefont
  {{Tomida}}, \citenamefont {{Tsunemi}}, \citenamefont {{Ueno}}, \citenamefont
  {{Yamamoto}}, \citenamefont {{Yamaoka}}, \citenamefont {{Yoshida}},\ and\
  \citenamefont {{MAXI Team}}}]{2010PASJ...62L..55I}%
  \BibitemOpen
  \bibfield  {author} {\bibinfo {author} {\bibfnamefont {N.}~\bibnamefont
  {{Isobe}}}, \bibinfo {author} {\bibfnamefont {K.}~\bibnamefont {{Sugimori}}},
  \bibinfo {author} {\bibfnamefont {N.}~\bibnamefont {{Kawai}}}, \bibinfo
  {author} {\bibfnamefont {Y.}~\bibnamefont {{Ueda}}}, \bibinfo {author}
  {\bibfnamefont {H.}~\bibnamefont {{Negoro}}}, \bibinfo {author}
  {\bibfnamefont {M.}~\bibnamefont {{Sugizaki}}}, \bibinfo {author}
  {\bibfnamefont {M.}~\bibnamefont {{Matsuoka}}}, \bibinfo {author}
  {\bibfnamefont {A.}~\bibnamefont {{Daikyuji}}}, \bibinfo {author}
  {\bibfnamefont {S.}~\bibnamefont {{Eguchi}}}, \bibinfo {author}
  {\bibfnamefont {K.}~\bibnamefont {{Hiroi}}}, \bibinfo {author} {\bibfnamefont
  {M.}~\bibnamefont {{Ishikawa}}}, \bibinfo {author} {\bibfnamefont
  {R.}~\bibnamefont {{Ishiwata}}}, \bibinfo {author} {\bibfnamefont
  {K.}~\bibnamefont {{Kawasaki}}}, \bibinfo {author} {\bibfnamefont
  {M.}~\bibnamefont {{Kimura}}}, \bibinfo {author} {\bibfnamefont
  {M.}~\bibnamefont {{Kohama}}}, \bibinfo {author} {\bibfnamefont
  {T.}~\bibnamefont {{Mihara}}}, \bibinfo {author} {\bibfnamefont
  {S.}~\bibnamefont {{Miyoshi}}}, \bibinfo {author} {\bibfnamefont
  {M.}~\bibnamefont {{Morii}}}, \bibinfo {author} {\bibfnamefont {Y.~E.}\
  \bibnamefont {{Nakagawa}}}, \bibinfo {author} {\bibfnamefont
  {S.}~\bibnamefont {{Nakahira}}}, \bibinfo {author} {\bibfnamefont
  {M.}~\bibnamefont {{Nakajima}}}, \bibinfo {author} {\bibfnamefont
  {H.}~\bibnamefont {{Ozawa}}}, \bibinfo {author} {\bibfnamefont
  {T.}~\bibnamefont {{Sootome}}}, \bibinfo {author} {\bibfnamefont
  {M.}~\bibnamefont {{Suzuki}}}, \bibinfo {author} {\bibfnamefont
  {H.}~\bibnamefont {{Tomida}}}, \bibinfo {author} {\bibfnamefont
  {H.}~\bibnamefont {{Tsunemi}}}, \bibinfo {author} {\bibfnamefont
  {S.}~\bibnamefont {{Ueno}}}, \bibinfo {author} {\bibfnamefont
  {T.}~\bibnamefont {{Yamamoto}}}, \bibinfo {author} {\bibfnamefont
  {K.}~\bibnamefont {{Yamaoka}}}, \bibinfo {author} {\bibfnamefont
  {A.}~\bibnamefont {{Yoshida}}}, \ and\ \bibinfo {author} {\bibnamefont {{MAXI
  Team}}},\ }\href {\doibase 10.1093/pasj/62.6.L55} {\bibfield  {journal}
  {\bibinfo  {journal} {\pasj}\ }\textbf {\bibinfo {volume} {62}},\ \bibinfo
  {pages} {L55} (\bibinfo {year} {2010})},\ \Eprint
  {http://arxiv.org/abs/1010.1003} {arXiv:1010.1003 [astro-ph.HE]} \BibitemShut
  {NoStop}%
\bibitem [{\citenamefont {{Niinuma}}\ \emph {et~al.}(2012)\citenamefont
  {{Niinuma}}, \citenamefont {{Kino}}, \citenamefont {{Nagai}}, \citenamefont
  {{Isobe}}, \citenamefont {{Gabanyi}}, \citenamefont {{Hada}}, \citenamefont
  {{Koyama}}, \citenamefont {{Asada}}, \citenamefont {{Oyama}},\ and\
  \citenamefont {{Fujisawa}}}]{2012ApJ...759...84N}%
  \BibitemOpen
  \bibfield  {author} {\bibinfo {author} {\bibfnamefont {K.}~\bibnamefont
  {{Niinuma}}}, \bibinfo {author} {\bibfnamefont {M.}~\bibnamefont {{Kino}}},
  \bibinfo {author} {\bibfnamefont {H.}~\bibnamefont {{Nagai}}}, \bibinfo
  {author} {\bibfnamefont {N.}~\bibnamefont {{Isobe}}}, \bibinfo {author}
  {\bibfnamefont {K.~E.}\ \bibnamefont {{Gabanyi}}}, \bibinfo {author}
  {\bibfnamefont {K.}~\bibnamefont {{Hada}}}, \bibinfo {author} {\bibfnamefont
  {S.}~\bibnamefont {{Koyama}}}, \bibinfo {author} {\bibfnamefont
  {K.}~\bibnamefont {{Asada}}}, \bibinfo {author} {\bibfnamefont
  {T.}~\bibnamefont {{Oyama}}}, \ and\ \bibinfo {author} {\bibfnamefont
  {K.}~\bibnamefont {{Fujisawa}}},\ }\href {\doibase
  10.1088/0004-637X/759/2/84} {\bibfield  {journal} {\bibinfo  {journal}
  {\apj}\ }\textbf {\bibinfo {volume} {759}},\ \bibinfo {eid} {84} (\bibinfo
  {year} {2012})},\ \Eprint {http://arxiv.org/abs/1209.2466} {arXiv:1209.2466
  [astro-ph.HE]} \BibitemShut {NoStop}%
\bibitem [{\citenamefont {{Krimm}}\ \emph {et~al.}(2009)\citenamefont
  {{Krimm}}, \citenamefont {{Barthelmy}}, \citenamefont {{Baumgartner}},
  \citenamefont {{Cummings}}, \citenamefont {{Fenimore}}, \citenamefont
  {{Gehrels}}, \citenamefont {{Markwardt}}, \citenamefont {{Palmer}},
  \citenamefont {{Sakamoto}}, \citenamefont {{Skinner}}, \citenamefont
  {{Stamatikos}}, \citenamefont {{Tueller}},\ and\ \citenamefont
  {{Ukwatta}}}]{2009ATel.2292....1K}%
  \BibitemOpen
  \bibfield  {author} {\bibinfo {author} {\bibfnamefont {H.~A.}\ \bibnamefont
  {{Krimm}}}, \bibinfo {author} {\bibfnamefont {S.~D.}\ \bibnamefont
  {{Barthelmy}}}, \bibinfo {author} {\bibfnamefont {W.}~\bibnamefont
  {{Baumgartner}}}, \bibinfo {author} {\bibfnamefont {J.}~\bibnamefont
  {{Cummings}}}, \bibinfo {author} {\bibfnamefont {E.}~\bibnamefont
  {{Fenimore}}}, \bibinfo {author} {\bibfnamefont {N.}~\bibnamefont
  {{Gehrels}}}, \bibinfo {author} {\bibfnamefont {C.~B.}\ \bibnamefont
  {{Markwardt}}}, \bibinfo {author} {\bibfnamefont {D.}~\bibnamefont
  {{Palmer}}}, \bibinfo {author} {\bibfnamefont {T.}~\bibnamefont
  {{Sakamoto}}}, \bibinfo {author} {\bibfnamefont {G.}~\bibnamefont
  {{Skinner}}}, \bibinfo {author} {\bibfnamefont {M.}~\bibnamefont
  {{Stamatikos}}}, \bibinfo {author} {\bibfnamefont {J.}~\bibnamefont
  {{Tueller}}}, \ and\ \bibinfo {author} {\bibfnamefont {T.}~\bibnamefont
  {{Ukwatta}}},\ }\href@noop {} {\bibfield  {journal} {\bibinfo  {journal} {The
  Astronomer's Telegram}\ }\textbf {\bibinfo {volume} {2292}},\ \bibinfo
  {pages} {1} (\bibinfo {year} {2009})}\BibitemShut {NoStop}%
\bibitem [{\citenamefont {{Costa}}\ \emph {et~al.}(2008)\citenamefont
  {{Costa}}, \citenamefont {{Del Monte}}, \citenamefont {{Donnarumma}},
  \citenamefont {{Evangelista}}, \citenamefont {{Feroci}}, \citenamefont
  {{Lapshov}}, \citenamefont {{Lazzarotto}}, \citenamefont {{Pacciani}},
  \citenamefont {{Rapisarda}}, \citenamefont {{Soffitta}}, \citenamefont
  {{Argan}}, \citenamefont {{Trois}}, \citenamefont {{Tavani}}, \citenamefont
  {{Pucella}}, \citenamefont {{D'Ammando}}, \citenamefont {{Vittorini}},
  \citenamefont {{Chen}}, \citenamefont {{Vercellone}}, \citenamefont
  {{Giuliani}}, \citenamefont {{Mereghetti}}, \citenamefont {{Pellizzoni}},
  \citenamefont {{Perotti}}, \citenamefont {{Fornari}}, \citenamefont
  {{Fiorini}}, \citenamefont {{Caraveo}}, \citenamefont {{Bulgarelli}},
  \citenamefont {{Gianotti}}, \citenamefont {{Trifoglio}}, \citenamefont {{Di
  Cocco}}, \citenamefont {{Labanti}}, \citenamefont {{Fuschino}}, \citenamefont
  {{Marisaldi}}, \citenamefont {{Galli}}, \citenamefont {{Barbiellini}},
  \citenamefont {{Longo}}, \citenamefont {{Vallazza}}, \citenamefont
  {{Picozza}}, \citenamefont {{Morselli}}, \citenamefont {{Prest}},
  \citenamefont {{Lipari}}, \citenamefont {{Zanello}}, \citenamefont
  {{Cattaneo}}, \citenamefont {{Pittori}}, \citenamefont {{Verrecchia}},
  \citenamefont {{Preger}}, \citenamefont {{Giommi}},\ and\ \citenamefont
  {{Salotti}}}]{2008ATel.1574....1C}%
  \BibitemOpen
  \bibfield  {author} {\bibinfo {author} {\bibfnamefont {E.}~\bibnamefont
  {{Costa}}}, \bibinfo {author} {\bibfnamefont {E.}~\bibnamefont {{Del
  Monte}}}, \bibinfo {author} {\bibfnamefont {I.}~\bibnamefont {{Donnarumma}}},
  \bibinfo {author} {\bibfnamefont {Y.}~\bibnamefont {{Evangelista}}}, \bibinfo
  {author} {\bibfnamefont {M.}~\bibnamefont {{Feroci}}}, \bibinfo {author}
  {\bibfnamefont {I.}~\bibnamefont {{Lapshov}}}, \bibinfo {author}
  {\bibfnamefont {F.}~\bibnamefont {{Lazzarotto}}}, \bibinfo {author}
  {\bibfnamefont {L.}~\bibnamefont {{Pacciani}}}, \bibinfo {author}
  {\bibfnamefont {M.}~\bibnamefont {{Rapisarda}}}, \bibinfo {author}
  {\bibfnamefont {P.}~\bibnamefont {{Soffitta}}}, \bibinfo {author}
  {\bibfnamefont {A.}~\bibnamefont {{Argan}}}, \bibinfo {author} {\bibfnamefont
  {A.}~\bibnamefont {{Trois}}}, \bibinfo {author} {\bibfnamefont
  {M.}~\bibnamefont {{Tavani}}}, \bibinfo {author} {\bibfnamefont
  {G.}~\bibnamefont {{Pucella}}}, \bibinfo {author} {\bibfnamefont
  {F.}~\bibnamefont {{D'Ammando}}}, \bibinfo {author} {\bibfnamefont
  {V.}~\bibnamefont {{Vittorini}}}, \bibinfo {author} {\bibfnamefont
  {A.}~\bibnamefont {{Chen}}}, \bibinfo {author} {\bibfnamefont
  {S.}~\bibnamefont {{Vercellone}}}, \bibinfo {author} {\bibfnamefont
  {A.}~\bibnamefont {{Giuliani}}}, \bibinfo {author} {\bibfnamefont
  {S.}~\bibnamefont {{Mereghetti}}}, \bibinfo {author} {\bibfnamefont
  {A.}~\bibnamefont {{Pellizzoni}}}, \bibinfo {author} {\bibfnamefont
  {F.}~\bibnamefont {{Perotti}}}, \bibinfo {author} {\bibfnamefont
  {F.}~\bibnamefont {{Fornari}}}, \bibinfo {author} {\bibfnamefont
  {M.}~\bibnamefont {{Fiorini}}}, \bibinfo {author} {\bibfnamefont
  {P.}~\bibnamefont {{Caraveo}}}, \bibinfo {author} {\bibfnamefont
  {A.}~\bibnamefont {{Bulgarelli}}}, \bibinfo {author} {\bibfnamefont
  {F.}~\bibnamefont {{Gianotti}}}, \bibinfo {author} {\bibfnamefont
  {M.}~\bibnamefont {{Trifoglio}}}, \bibinfo {author} {\bibfnamefont
  {G.}~\bibnamefont {{Di Cocco}}}, \bibinfo {author} {\bibfnamefont
  {C.}~\bibnamefont {{Labanti}}}, \bibinfo {author} {\bibfnamefont
  {F.}~\bibnamefont {{Fuschino}}}, \bibinfo {author} {\bibfnamefont
  {M.}~\bibnamefont {{Marisaldi}}}, \bibinfo {author} {\bibfnamefont
  {M.}~\bibnamefont {{Galli}}}, \bibinfo {author} {\bibfnamefont
  {G.}~\bibnamefont {{Barbiellini}}}, \bibinfo {author} {\bibfnamefont
  {F.}~\bibnamefont {{Longo}}}, \bibinfo {author} {\bibfnamefont
  {E.}~\bibnamefont {{Vallazza}}}, \bibinfo {author} {\bibfnamefont
  {P.}~\bibnamefont {{Picozza}}}, \bibinfo {author} {\bibfnamefont
  {A.}~\bibnamefont {{Morselli}}}, \bibinfo {author} {\bibfnamefont
  {M.}~\bibnamefont {{Prest}}}, \bibinfo {author} {\bibfnamefont
  {P.}~\bibnamefont {{Lipari}}}, \bibinfo {author} {\bibfnamefont
  {D.}~\bibnamefont {{Zanello}}}, \bibinfo {author} {\bibfnamefont
  {P.}~\bibnamefont {{Cattaneo}}}, \bibinfo {author} {\bibfnamefont
  {C.}~\bibnamefont {{Pittori}}}, \bibinfo {author} {\bibfnamefont
  {F.}~\bibnamefont {{Verrecchia}}}, \bibinfo {author} {\bibfnamefont
  {B.}~\bibnamefont {{Preger}}}, \bibinfo {author} {\bibfnamefont
  {P.}~\bibnamefont {{Giommi}}}, \ and\ \bibinfo {author} {\bibfnamefont
  {L.}~\bibnamefont {{Salotti}}},\ }\href@noop {} {\bibfield  {journal}
  {\bibinfo  {journal} {The Astronomer's Telegram}\ }\textbf {\bibinfo {volume}
  {1574}},\ \bibinfo {pages} {1} (\bibinfo {year} {2008})}\BibitemShut
  {NoStop}%
\bibitem [{\citenamefont {{Lichti}}\ \emph {et~al.}(2006)\citenamefont
  {{Lichti}}, \citenamefont {{Paltani}}, \citenamefont {{Mowlavi}},
  \citenamefont {{Ajello}}, \citenamefont {{Collmar}}, \citenamefont {{von
  Kienlin}}, \citenamefont {{Beckmann}}, \citenamefont {{Boisson}},
  \citenamefont {{Sol}}, \citenamefont {{Buckley}}, \citenamefont {{Charlot}},
  \citenamefont {{Degrange}}, \citenamefont {{Djannati-Atai}}, \citenamefont
  {{Punch}}, \citenamefont {{Falcone}}, \citenamefont {{Finley}}, \citenamefont
  {{Fossati}}, \citenamefont {{Henri}}, \citenamefont {{Sauge}}, \citenamefont
  {{Katarzynski}}, \citenamefont {{Kieda}}, \citenamefont
  {{L{\"a}hteenm{\"a}ki}}, \citenamefont {{Tornikoski}}, \citenamefont
  {{Mannheim}}, \citenamefont {{Marcowith}}, \citenamefont {{Saggione}},
  \citenamefont {{Sillanpaa}}, \citenamefont {{Takalo}}, \citenamefont
  {{Smith}},\ and\ \citenamefont {{Weekes}}}]{2006ATel..848....1L}%
  \BibitemOpen
  \bibfield  {author} {\bibinfo {author} {\bibfnamefont {G.~G.}\ \bibnamefont
  {{Lichti}}}, \bibinfo {author} {\bibfnamefont {S.}~\bibnamefont {{Paltani}}},
  \bibinfo {author} {\bibfnamefont {M.}~\bibnamefont {{Mowlavi}}}, \bibinfo
  {author} {\bibfnamefont {M.}~\bibnamefont {{Ajello}}}, \bibinfo {author}
  {\bibfnamefont {W.}~\bibnamefont {{Collmar}}}, \bibinfo {author}
  {\bibfnamefont {A.}~\bibnamefont {{von Kienlin}}}, \bibinfo {author}
  {\bibfnamefont {V.}~\bibnamefont {{Beckmann}}}, \bibinfo {author}
  {\bibfnamefont {C.}~\bibnamefont {{Boisson}}}, \bibinfo {author}
  {\bibfnamefont {H.}~\bibnamefont {{Sol}}}, \bibinfo {author} {\bibfnamefont
  {J.}~\bibnamefont {{Buckley}}}, \bibinfo {author} {\bibfnamefont
  {P.}~\bibnamefont {{Charlot}}}, \bibinfo {author} {\bibfnamefont
  {B.}~\bibnamefont {{Degrange}}}, \bibinfo {author} {\bibfnamefont
  {A.}~\bibnamefont {{Djannati-Atai}}}, \bibinfo {author} {\bibfnamefont
  {M.}~\bibnamefont {{Punch}}}, \bibinfo {author} {\bibfnamefont
  {A.}~\bibnamefont {{Falcone}}}, \bibinfo {author} {\bibfnamefont
  {J.}~\bibnamefont {{Finley}}}, \bibinfo {author} {\bibfnamefont
  {G.}~\bibnamefont {{Fossati}}}, \bibinfo {author} {\bibfnamefont
  {G.}~\bibnamefont {{Henri}}}, \bibinfo {author} {\bibfnamefont
  {L.}~\bibnamefont {{Sauge}}}, \bibinfo {author} {\bibfnamefont
  {K.}~\bibnamefont {{Katarzynski}}}, \bibinfo {author} {\bibfnamefont
  {D.}~\bibnamefont {{Kieda}}}, \bibinfo {author} {\bibfnamefont
  {A.}~\bibnamefont {{L{\"a}hteenm{\"a}ki}}}, \bibinfo {author} {\bibfnamefont
  {M.}~\bibnamefont {{Tornikoski}}}, \bibinfo {author} {\bibfnamefont
  {K.}~\bibnamefont {{Mannheim}}}, \bibinfo {author} {\bibfnamefont
  {A.}~\bibnamefont {{Marcowith}}}, \bibinfo {author} {\bibfnamefont
  {A.}~\bibnamefont {{Saggione}}}, \bibinfo {author} {\bibfnamefont
  {A.}~\bibnamefont {{Sillanpaa}}}, \bibinfo {author} {\bibfnamefont
  {L.}~\bibnamefont {{Takalo}}}, \bibinfo {author} {\bibfnamefont
  {D.}~\bibnamefont {{Smith}}}, \ and\ \bibinfo {author} {\bibfnamefont
  {T.}~\bibnamefont {{Weekes}}},\ }\href@noop {} {\bibfield  {journal}
  {\bibinfo  {journal} {The Astronomer's Telegram}\ }\textbf {\bibinfo {volume}
  {848}},\ \bibinfo {pages} {1} (\bibinfo {year} {2006})}\BibitemShut {NoStop}%
\bibitem [{\citenamefont {{Balokovic}}\ \emph {et~al.}(2013)\citenamefont
  {{Balokovic}}, \citenamefont {{Furniss}}, \citenamefont {{Madejski}},\ and\
  \citenamefont {{Harrison}}}]{2013ATel.4974....1B}%
  \BibitemOpen
  \bibfield  {author} {\bibinfo {author} {\bibfnamefont {M.}~\bibnamefont
  {{Balokovic}}}, \bibinfo {author} {\bibfnamefont {A.}~\bibnamefont
  {{Furniss}}}, \bibinfo {author} {\bibfnamefont {G.}~\bibnamefont
  {{Madejski}}}, \ and\ \bibinfo {author} {\bibfnamefont {F.}~\bibnamefont
  {{Harrison}}},\ }\href@noop {} {\bibfield  {journal} {\bibinfo  {journal}
  {The Astronomer's Telegram}\ }\textbf {\bibinfo {volume} {4974}},\ \bibinfo
  {pages} {1} (\bibinfo {year} {2013})}\BibitemShut {NoStop}%
\bibitem [{\citenamefont {{Acciari}}\ and\ \citenamefont
  {et~al.}(2011{\natexlab{b}})}]{2011ApJ...743...62A}%
  \BibitemOpen
  \bibfield  {author} {\bibinfo {author} {\bibfnamefont {V.~A.}\ \bibnamefont
  {{Acciari}}}\ and\ \bibinfo {author} {\bibnamefont {et~al.}},\ }\href
  {\doibase 10.1088/0004-637X/743/1/62} {\bibfield  {journal} {\bibinfo
  {journal} {\apj}\ }\textbf {\bibinfo {volume} {743}},\ \bibinfo {eid} {62}
  (\bibinfo {year} {2011}{\natexlab{b}})},\ \Eprint
  {http://arxiv.org/abs/1109.0050} {arXiv:1109.0050 [astro-ph.HE]} \BibitemShut
  {NoStop}%
\bibitem [{\citenamefont {{Abe}}\ \emph {et~al.}(2011)\citenamefont {{Abe}},
  \citenamefont {{Abe}}, \citenamefont {{Aihara}}, \citenamefont {{Fukuda}},
  \citenamefont {{Hayato}}, \citenamefont {{Huang}}, \citenamefont
  {{Ichikawa}}, \citenamefont {{Ikeda}}, \citenamefont {{Inoue}}, \citenamefont
  {{Ishino}}, \citenamefont {{Itow}}, \citenamefont {{Kajita}}, \citenamefont
  {{Kameda}}, \citenamefont {{Kishimoto}}, \citenamefont {{Koga}},
  \citenamefont {{Koshio}}, \citenamefont {{Lee}}, \citenamefont {{Minamino}},
  \citenamefont {{Miura}}, \citenamefont {{Moriyama}}, \citenamefont
  {{Nakahata}}, \citenamefont {{Nakamura}}, \citenamefont {{Nakaya}},
  \citenamefont {{Nakayama}}, \citenamefont {{Nishijima}}, \citenamefont
  {{Nishimura}}, \citenamefont {{Obayashi}}, \citenamefont {{Okumura}},
  \citenamefont {{Sakuda}}, \citenamefont {{Sekiya}}, \citenamefont
  {{Shiozawa}}, \citenamefont {{Suzuki}}, \citenamefont {{Suzuki}},
  \citenamefont {{Takeda}}, \citenamefont {{Takeuchi}}, \citenamefont
  {{Tanaka}}, \citenamefont {{Tasaka}}, \citenamefont {{Tomura}}, \citenamefont
  {{Vagins}}, \citenamefont {{Wang}},\ and\ \citenamefont
  {{Yokoyama}}}]{2011arXiv1109.3262A}%
  \BibitemOpen
  \bibfield  {author} {\bibinfo {author} {\bibfnamefont {K.}~\bibnamefont
  {{Abe}}}, \bibinfo {author} {\bibfnamefont {T.}~\bibnamefont {{Abe}}},
  \bibinfo {author} {\bibfnamefont {H.}~\bibnamefont {{Aihara}}}, \bibinfo
  {author} {\bibfnamefont {Y.}~\bibnamefont {{Fukuda}}}, \bibinfo {author}
  {\bibfnamefont {Y.}~\bibnamefont {{Hayato}}}, \bibinfo {author}
  {\bibfnamefont {K.}~\bibnamefont {{Huang}}}, \bibinfo {author} {\bibfnamefont
  {A.~K.}\ \bibnamefont {{Ichikawa}}}, \bibinfo {author} {\bibfnamefont
  {M.}~\bibnamefont {{Ikeda}}}, \bibinfo {author} {\bibfnamefont
  {K.}~\bibnamefont {{Inoue}}}, \bibinfo {author} {\bibfnamefont
  {H.}~\bibnamefont {{Ishino}}}, \bibinfo {author} {\bibfnamefont
  {Y.}~\bibnamefont {{Itow}}}, \bibinfo {author} {\bibfnamefont
  {T.}~\bibnamefont {{Kajita}}}, \bibinfo {author} {\bibfnamefont
  {J.}~\bibnamefont {{Kameda}}}, \bibinfo {author} {\bibfnamefont
  {Y.}~\bibnamefont {{Kishimoto}}}, \bibinfo {author} {\bibfnamefont
  {M.}~\bibnamefont {{Koga}}}, \bibinfo {author} {\bibfnamefont
  {Y.}~\bibnamefont {{Koshio}}}, \bibinfo {author} {\bibfnamefont {K.~P.}\
  \bibnamefont {{Lee}}}, \bibinfo {author} {\bibfnamefont {A.}~\bibnamefont
  {{Minamino}}}, \bibinfo {author} {\bibfnamefont {M.}~\bibnamefont {{Miura}}},
  \bibinfo {author} {\bibfnamefont {S.}~\bibnamefont {{Moriyama}}}, \bibinfo
  {author} {\bibfnamefont {M.}~\bibnamefont {{Nakahata}}}, \bibinfo {author}
  {\bibfnamefont {K.}~\bibnamefont {{Nakamura}}}, \bibinfo {author}
  {\bibfnamefont {T.}~\bibnamefont {{Nakaya}}}, \bibinfo {author}
  {\bibfnamefont {S.}~\bibnamefont {{Nakayama}}}, \bibinfo {author}
  {\bibfnamefont {K.}~\bibnamefont {{Nishijima}}}, \bibinfo {author}
  {\bibfnamefont {Y.}~\bibnamefont {{Nishimura}}}, \bibinfo {author}
  {\bibfnamefont {Y.}~\bibnamefont {{Obayashi}}}, \bibinfo {author}
  {\bibfnamefont {K.}~\bibnamefont {{Okumura}}}, \bibinfo {author}
  {\bibfnamefont {M.}~\bibnamefont {{Sakuda}}}, \bibinfo {author}
  {\bibfnamefont {H.}~\bibnamefont {{Sekiya}}}, \bibinfo {author}
  {\bibfnamefont {M.}~\bibnamefont {{Shiozawa}}}, \bibinfo {author}
  {\bibfnamefont {A.~T.}\ \bibnamefont {{Suzuki}}}, \bibinfo {author}
  {\bibfnamefont {Y.}~\bibnamefont {{Suzuki}}}, \bibinfo {author}
  {\bibfnamefont {A.}~\bibnamefont {{Takeda}}}, \bibinfo {author}
  {\bibfnamefont {Y.}~\bibnamefont {{Takeuchi}}}, \bibinfo {author}
  {\bibfnamefont {H.~K.~M.}\ \bibnamefont {{Tanaka}}}, \bibinfo {author}
  {\bibfnamefont {S.}~\bibnamefont {{Tasaka}}}, \bibinfo {author}
  {\bibfnamefont {T.}~\bibnamefont {{Tomura}}}, \bibinfo {author}
  {\bibfnamefont {M.~R.}\ \bibnamefont {{Vagins}}}, \bibinfo {author}
  {\bibfnamefont {J.}~\bibnamefont {{Wang}}}, \ and\ \bibinfo {author}
  {\bibfnamefont {M.}~\bibnamefont {{Yokoyama}}},\ }\href@noop {} {\bibfield
  {journal} {\bibinfo  {journal} {ArXiv e-prints}\ } (\bibinfo {year}
  {2011})},\ \Eprint {http://arxiv.org/abs/1109.3262} {arXiv:1109.3262
  [hep-ex]} \BibitemShut {NoStop}%
\bibitem [{\citenamefont {{Junor}}\ \emph {et~al.}(1999)\citenamefont
  {{Junor}}, \citenamefont {{Biretta}},\ and\ \citenamefont
  {{Livio}}}]{1999Natur.401..891J}%
  \BibitemOpen
  \bibfield  {author} {\bibinfo {author} {\bibfnamefont {W.}~\bibnamefont
  {{Junor}}}, \bibinfo {author} {\bibfnamefont {J.~A.}\ \bibnamefont
  {{Biretta}}}, \ and\ \bibinfo {author} {\bibfnamefont {M.}~\bibnamefont
  {{Livio}}},\ }\href {\doibase 10.1038/44780} {\bibfield  {journal} {\bibinfo
  {journal} {\nat}\ }\textbf {\bibinfo {volume} {401}},\ \bibinfo {pages} {891}
  (\bibinfo {year} {1999})}\BibitemShut {NoStop}%
\bibitem [{\citenamefont {Sauty}\ \emph {et~al.}(2002)\citenamefont {Sauty},
  \citenamefont {Tsinganos},\ and\ \citenamefont {Trussoni}}]{Sauty:2001wx}%
  \BibitemOpen
  \bibfield  {author} {\bibinfo {author} {\bibfnamefont {C.}~\bibnamefont
  {Sauty}}, \bibinfo {author} {\bibfnamefont {K.}~\bibnamefont {Tsinganos}}, \
  and\ \bibinfo {author} {\bibfnamefont {E.}~\bibnamefont {Trussoni}},\
  }\href@noop {} {\bibfield  {journal} {\bibinfo  {journal} {Lect.Notes Phys.}\
  }\textbf {\bibinfo {volume} {589}},\ \bibinfo {pages} {41} (\bibinfo {year}
  {2002})},\ \Eprint {http://arxiv.org/abs/astro-ph/0108509}
  {arXiv:astro-ph/0108509 [astro-ph]} \BibitemShut {NoStop}%
\bibitem [{\citenamefont {{Kumar}}\ and\ \citenamefont
  {{Zhang}}(2014)}]{2014arXiv1410.0679K}%
  \BibitemOpen
  \bibfield  {author} {\bibinfo {author} {\bibfnamefont {P.}~\bibnamefont
  {{Kumar}}}\ and\ \bibinfo {author} {\bibfnamefont {B.}~\bibnamefont
  {{Zhang}}},\ }\href@noop {} {\bibfield  {journal} {\bibinfo  {journal} {ArXiv
  e-prints}\ } (\bibinfo {year} {2014})},\ \Eprint
  {http://arxiv.org/abs/1410.0679} {arXiv:1410.0679 [astro-ph.HE]} \BibitemShut
  {NoStop}%
\bibitem [{\citenamefont {{Greisen}}(1966)}]{1966PhRvL..16..748G}%
  \BibitemOpen
  \bibfield  {author} {\bibinfo {author} {\bibfnamefont {K.}~\bibnamefont
  {{Greisen}}},\ }\href {\doibase 10.1103/PhysRevLett.16.748} {\bibfield
  {journal} {\bibinfo  {journal} {Physical Review Letters}\ }\textbf {\bibinfo
  {volume} {16}},\ \bibinfo {pages} {748} (\bibinfo {year} {1966})}\BibitemShut
  {NoStop}%
\bibitem [{\citenamefont {{Zatsepin}}\ and\ \citenamefont
  {{Kuz'min}}(1966)}]{1966JETPL...4...78Z}%
  \BibitemOpen
  \bibfield  {author} {\bibinfo {author} {\bibfnamefont {G.~T.}\ \bibnamefont
  {{Zatsepin}}}\ and\ \bibinfo {author} {\bibfnamefont {V.~A.}\ \bibnamefont
  {{Kuz'min}}},\ }\href@noop {} {\bibfield  {journal} {\bibinfo  {journal}
  {Soviet Journal of Experimental and Theoretical Physics Letters}\ }\textbf
  {\bibinfo {volume} {4}},\ \bibinfo {pages} {78} (\bibinfo {year}
  {1966})}\BibitemShut {NoStop}%
\bibitem [{\citenamefont {{Raue}}\ and\ \citenamefont
  {{Mazin}}(2008)}]{2008IJMPD..17.1515R}%
  \BibitemOpen
  \bibfield  {author} {\bibinfo {author} {\bibfnamefont {M.}~\bibnamefont
  {{Raue}}}\ and\ \bibinfo {author} {\bibfnamefont {D.}~\bibnamefont
  {{Mazin}}},\ }\href {\doibase 10.1142/S0218271808013091} {\bibfield
  {journal} {\bibinfo  {journal} {International Journal of Modern Physics D}\
  }\textbf {\bibinfo {volume} {17}},\ \bibinfo {pages} {1515} (\bibinfo {year}
  {2008})},\ \Eprint {http://arxiv.org/abs/0802.0129} {arXiv:0802.0129}
  \BibitemShut {NoStop}%
\bibitem [{\citenamefont {{Abeysekara}}\ and\ \citenamefont
  {et~al.}(2013)}]{2013APh....50...26A}%
  \BibitemOpen
  \bibfield  {author} {\bibinfo {author} {\bibfnamefont {A.~U.}\ \bibnamefont
  {{Abeysekara}}}\ and\ \bibinfo {author} {\bibnamefont {et~al.}},\ }\href
  {\doibase 10.1016/j.astropartphys.2013.08.002} {\bibfield  {journal}
  {\bibinfo  {journal} {Astroparticle Physics}\ }\textbf {\bibinfo {volume}
  {50}},\ \bibinfo {pages} {26} (\bibinfo {year} {2013})},\ \Eprint
  {http://arxiv.org/abs/1306.5800} {arXiv:1306.5800 [astro-ph.HE]} \BibitemShut
  {NoStop}%
\bibitem [{\citenamefont {{Harrison}}\ and\ \citenamefont
  {et~al.}(2013)}]{2013ApJ...770..103H}%
  \BibitemOpen
  \bibfield  {author} {\bibinfo {author} {\bibfnamefont {F.~A.}\ \bibnamefont
  {{Harrison}}}\ and\ \bibinfo {author} {\bibnamefont {et~al.}},\ }\href
  {\doibase 10.1088/0004-637X/770/2/103} {\bibfield  {journal} {\bibinfo
  {journal} {\apj}\ }\textbf {\bibinfo {volume} {770}},\ \bibinfo {eid} {103}
  (\bibinfo {year} {2013})},\ \Eprint {http://arxiv.org/abs/1301.7307}
  {arXiv:1301.7307 [astro-ph.IM]} \BibitemShut {NoStop}%
\bibitem [{\citenamefont {{Evans}}\ \emph {et~al.}(2014)\citenamefont
  {{Evans}}, \citenamefont {{Osborne}}, \citenamefont {{Beardmore}},
  \citenamefont {{Page}}, \citenamefont {{Willingale}}, \citenamefont
  {{Mountford}}, \citenamefont {{Pagani}}, \citenamefont {{Burrows}},
  \citenamefont {{Kennea}}, \citenamefont {{Perri}}, \citenamefont
  {{Tagliaferri}},\ and\ \citenamefont {{Gehrels}}}]{2014ApJS..210....8E}%
  \BibitemOpen
  \bibfield  {author} {\bibinfo {author} {\bibfnamefont {P.~A.}\ \bibnamefont
  {{Evans}}}, \bibinfo {author} {\bibfnamefont {J.~P.}\ \bibnamefont
  {{Osborne}}}, \bibinfo {author} {\bibfnamefont {A.~P.}\ \bibnamefont
  {{Beardmore}}}, \bibinfo {author} {\bibfnamefont {K.~L.}\ \bibnamefont
  {{Page}}}, \bibinfo {author} {\bibfnamefont {R.}~\bibnamefont
  {{Willingale}}}, \bibinfo {author} {\bibfnamefont {C.~J.}\ \bibnamefont
  {{Mountford}}}, \bibinfo {author} {\bibfnamefont {C.}~\bibnamefont
  {{Pagani}}}, \bibinfo {author} {\bibfnamefont {D.~N.}\ \bibnamefont
  {{Burrows}}}, \bibinfo {author} {\bibfnamefont {J.~A.}\ \bibnamefont
  {{Kennea}}}, \bibinfo {author} {\bibfnamefont {M.}~\bibnamefont {{Perri}}},
  \bibinfo {author} {\bibfnamefont {G.}~\bibnamefont {{Tagliaferri}}}, \ and\
  \bibinfo {author} {\bibfnamefont {N.}~\bibnamefont {{Gehrels}}},\ }\href
  {\doibase 10.1088/0067-0049/210/1/8} {\bibfield  {journal} {\bibinfo
  {journal} {\apjs}\ }\textbf {\bibinfo {volume} {210}},\ \bibinfo {eid} {8}
  (\bibinfo {year} {2014})},\ \Eprint {http://arxiv.org/abs/1311.5368}
  {arXiv:1311.5368 [astro-ph.HE]} \BibitemShut {NoStop}%
\end{thebibliography}
%
%

\clearpage
\clearpage
\clearpage
\clearpage
\begin{figure}
\vspace{0.5cm}
{\centering
\resizebox*{1.1\textwidth}{0.33\textheight}
{\includegraphics{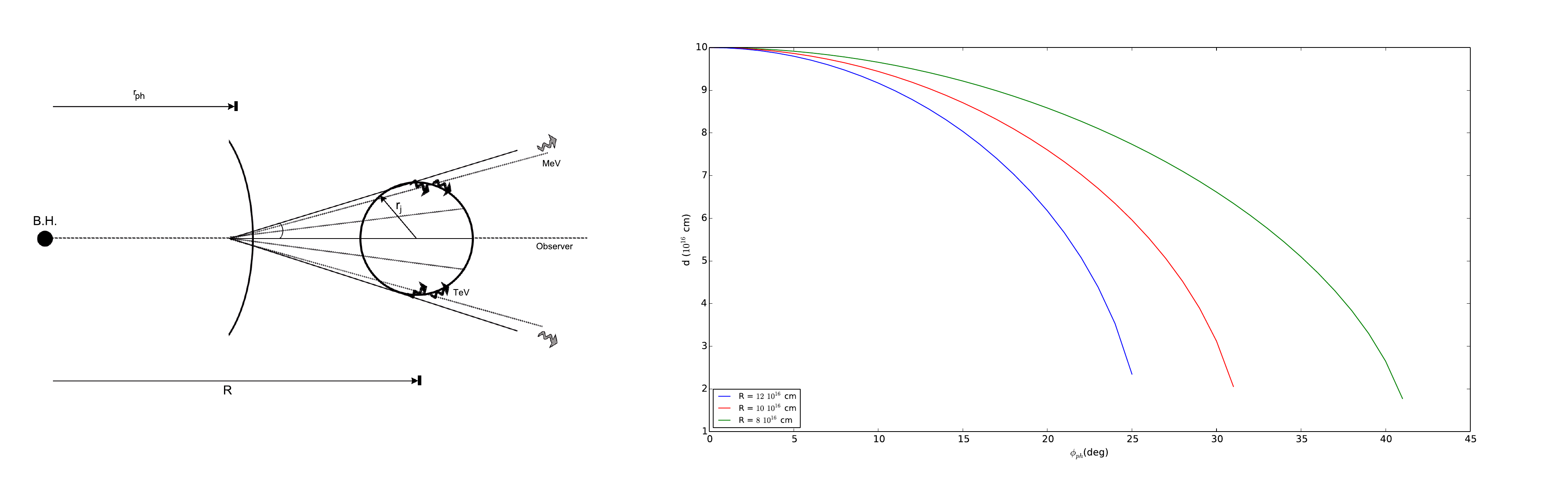}}
}
\caption{In the left-hand figure is shown a sketch of different  MeV - photon paths through the emitting region whereas in the right-hand figure, we plot the distance of the MeV photons as a function of angle $\phi_{ph}$.}
\label{sketch}
\end{figure} 
\begin{figure}
\vspace{0.5cm}
{\centering
\resizebox*{0.7\textwidth}{0.35\textheight}
{\includegraphics{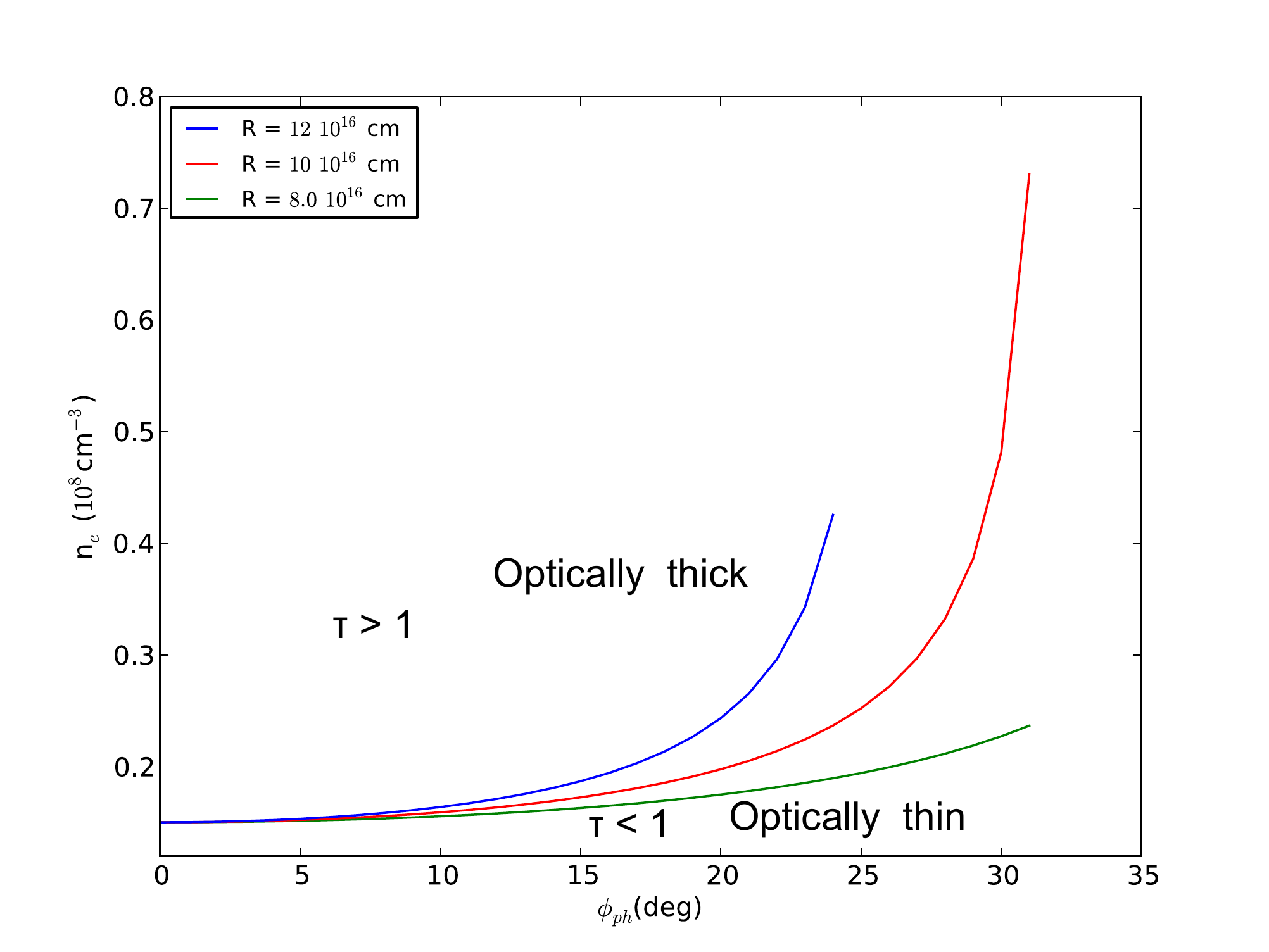}}
}
\caption{Electron density as a function of angle $\phi_{ph}$ for $\tau=1$. In this figure, one can see that the optically thick and thin zones.}
\label{optical}
\end{figure} 
\clearpage
\begin{figure}
\vspace{0.5cm}
{\centering
\resizebox*{1.0\textwidth}{0.3\textheight}
{\includegraphics{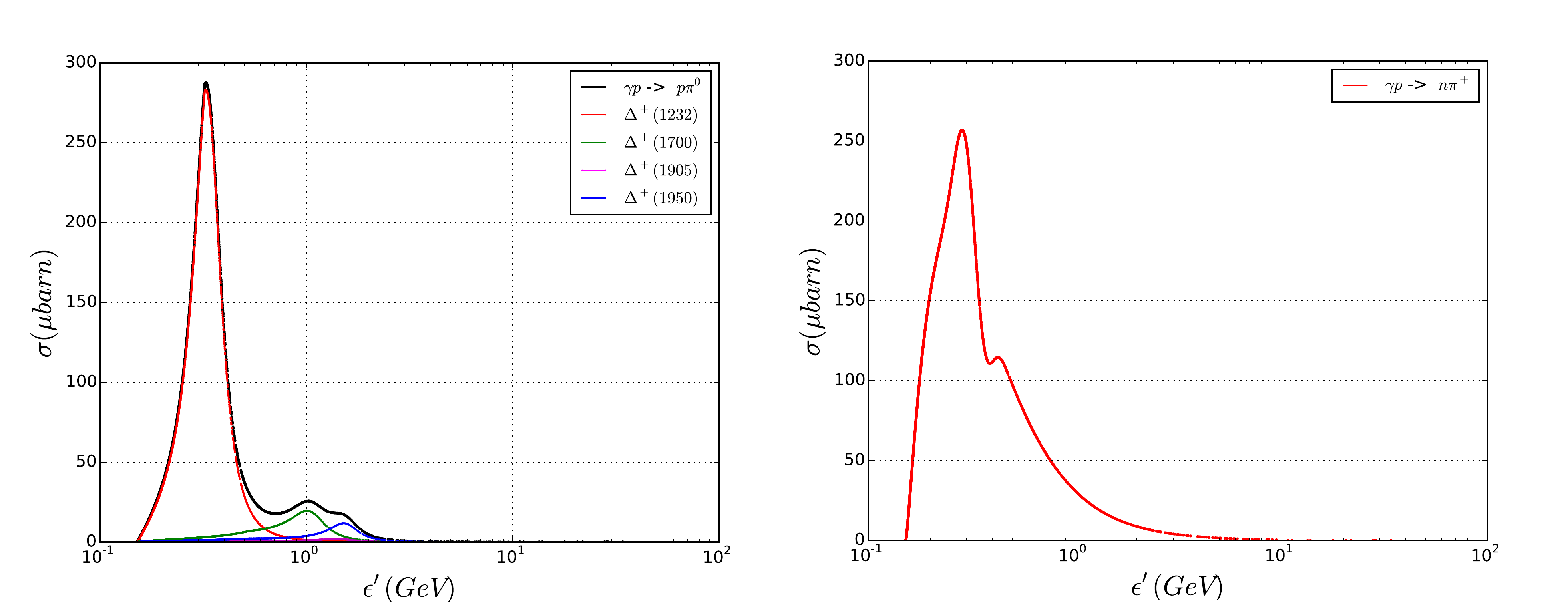}}
}
\caption{Total cross section of  the resonance production ($p\gamma \longrightarrow\Delta^{+}\longrightarrow p\pi^0$) (left) and the direct production ($p\gamma\to n\pi^+$) (right).}
\label{cross}
\end{figure} 
\clearpage
\begin{figure}
\vspace{0.5cm}
{\centering
\resizebox*{0.8\textwidth}{0.7\textheight}
{\includegraphics{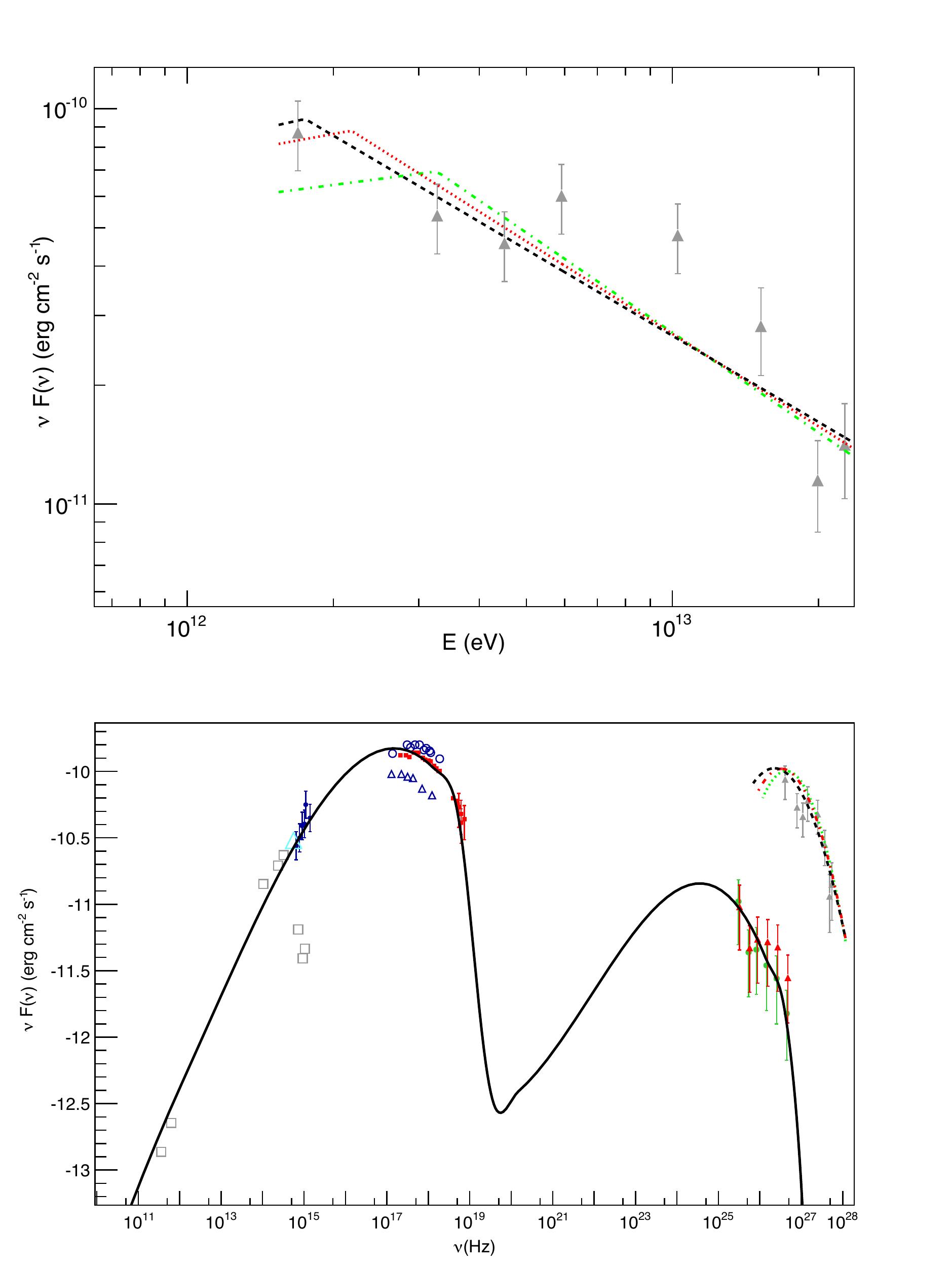}}
}
\caption{Above: Fit of the TeV flare with our hadronic model for $\Gamma=$ 8 (dashed line in black color), 12 (dotted line in red color) and 18 (dot-dashed line in green color).  Below:  SED of the multi wavelength campaign in 2006 May plus the TeV "orphan" flare in 2002 June fitted with our model.}
\label{fit_1es}
\end{figure} 
\clearpage
\begin{figure}
\vspace{0.5cm}
{\centering
\resizebox*{1.03\textwidth}{0.6\textheight}
{\includegraphics{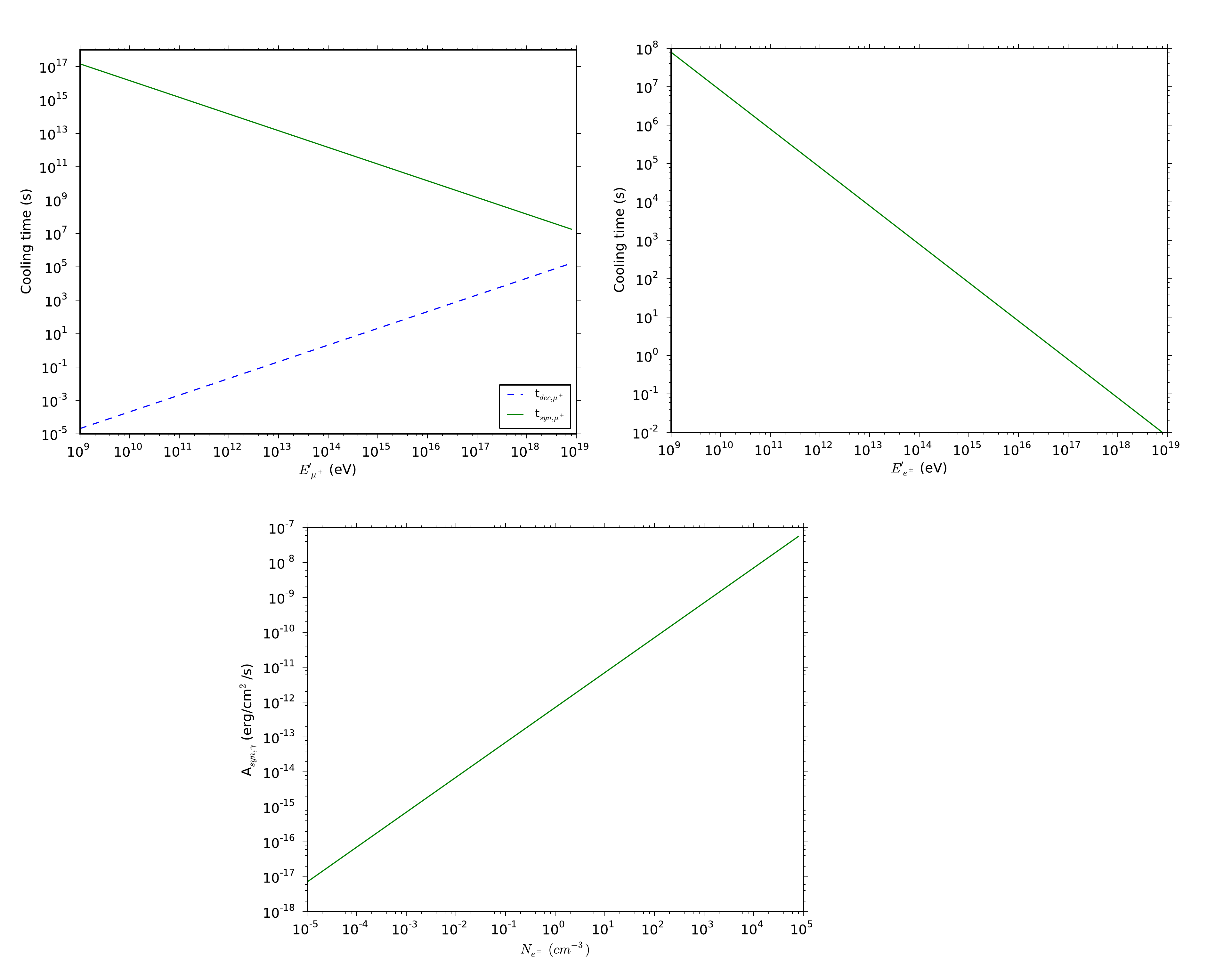}}
}
\caption{Time scale of muon and e$^\pm$ secondary synchrotron radiation presented in blazar 1ES 1959+650.     Left-hand figure above:  Cooling time scales for muons. Continuos line in green color is  synchrotron radiation scale and  the dotted line in blue color is decay scale.  Left-hand figure above:  Cooling time scale for electron/positron synchrotron radiation.   Figure below:  Proportionality constant of synchrotron radiation spectrum as a function of number density of electrons/positrons.}
\label{sync_1es}
\end{figure} 
\clearpage
\begin{figure}
\vspace{0.5cm}
{\centering
\resizebox*{1.03\textwidth}{0.7\textheight}
{\includegraphics{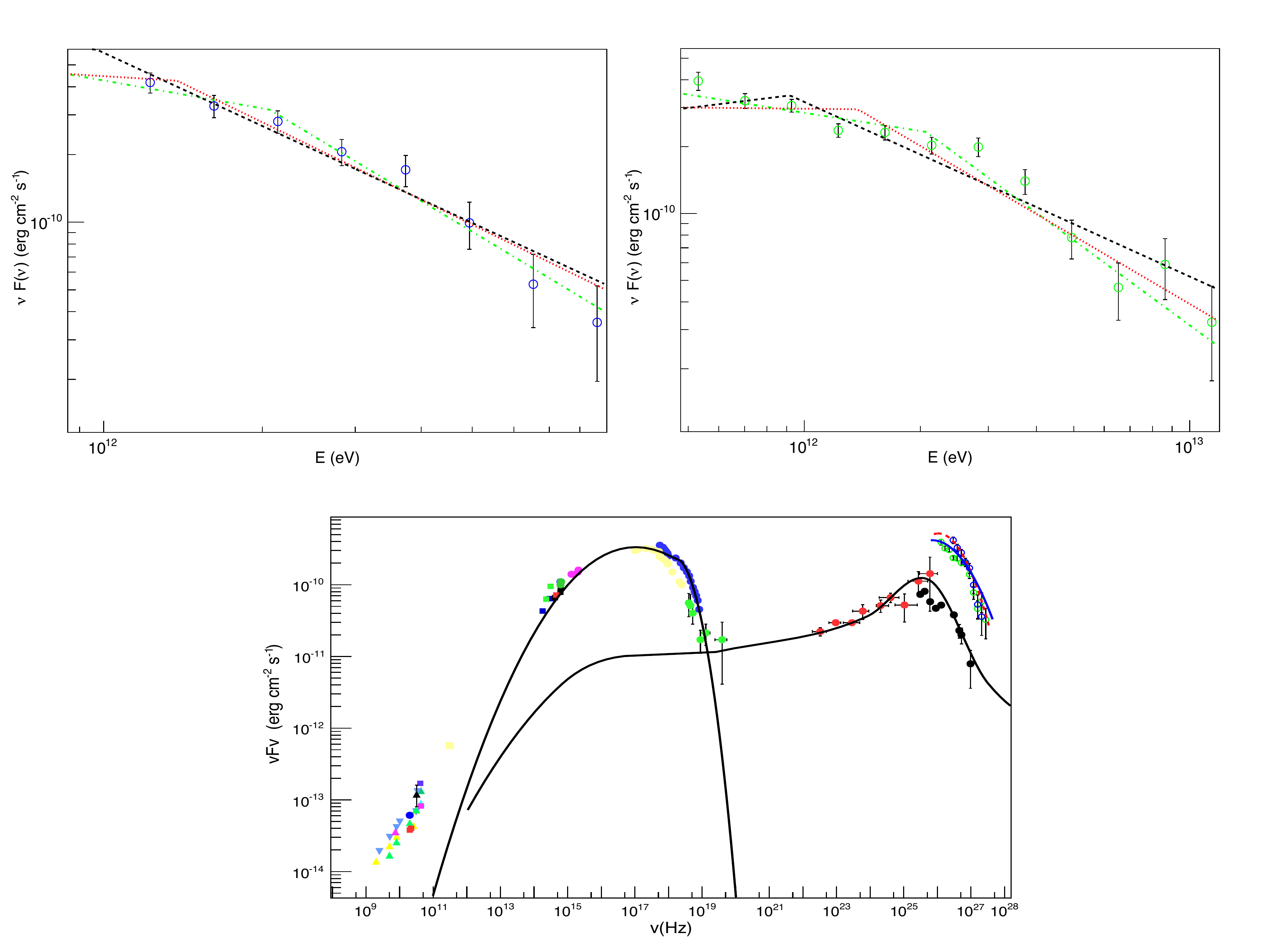}}
}
\caption{Left-hand figure above: Fit of the first TeV flare with our hadronic model for $\Gamma=$ 8 (dashed line in black color), 12 (dotted line in red color) and 18 (dot-dashed line in green color. Right-hand figure above: Fit of the second TeV flare with our hadronic model for $\Gamma=$ 8 (dashed line in black color), 12 (dotted line in red color) and 18 (dot-dashed line in green color). Below:  SED of the multi wavelength campaign in 2009 plus the TeV "orphan" flares during the night of MJD 54589.2 fitted with our mode for $\Gamma_j$=12.}
\label{fit_mrk}
\end{figure} 
\clearpage
\begin{figure}
\vspace{0.5cm}
{\centering
\resizebox*{1.03\textwidth}{0.6\textheight}
{\includegraphics{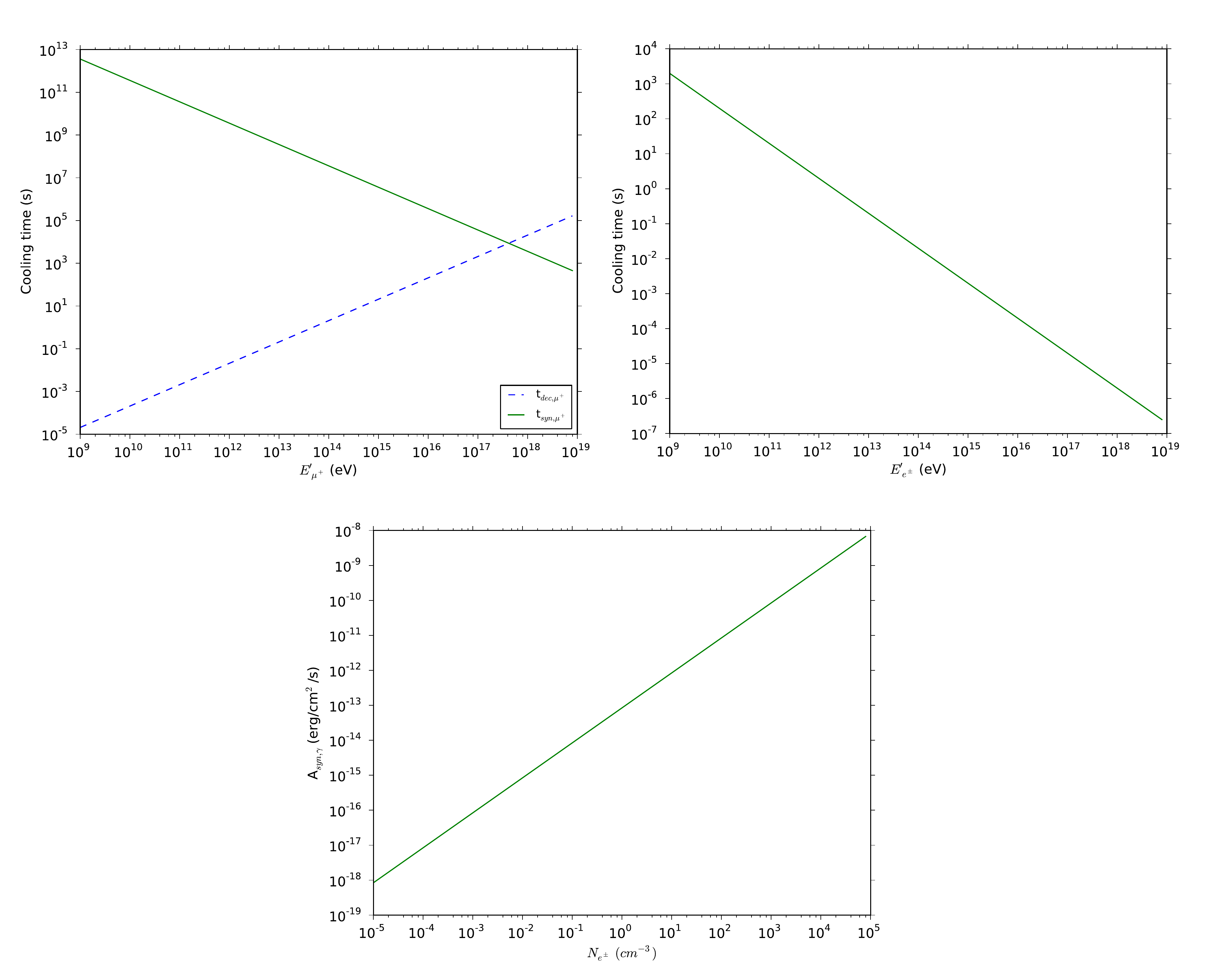}}
}
\caption{Time scale of muon and e$^\pm$ secondary synchrotron radiation presented in blazar Mrk 421.     Left-hand figure above:  Cooling time scales for muons. Continuos line in green color is  synchrotron radiation scale and  the dotted line in blue color is decay scale.  Left-hand figure above:  Cooling time scale for electron/positron synchrotron radiation.   Figure below:  Proportionality constant of synchrotron radiation spectrum as a function of number density of electrons/positrons.}
\label{sync_mrk}
\end{figure} 
\clearpage
\clearpage
\clearpage
\clearpage
\end{document}